\documentclass[aps,floats,amssymb,amsmath,prb,nofootinbib,twocolumn,superscriptaddress,floatfix]{revtex4-2}

\usepackage{booktabs}
\usepackage{calc}
\usepackage{psfrag}
\usepackage{graphicx}
\usepackage{color}
\usepackage{units}
\usepackage[dvipsnames]{xcolor}
\usepackage[unicode=true,pdfusetitle,
 bookmarks=true,bookmarksnumbered=false,bookmarksopen=false,
 breaklinks=false,pdfborder={0 0 0},backref=false,colorlinks=true]
 {hyperref}
\usepackage{geometry}
\usepackage{braket}
\pdfpageattr {/Group << /S /Transparency /I true /CS /DeviceRGB>>}

\newcommand{\X}[3]{X_{#1,\mathbf{#2},\mathbf{#3}}}
\newcommand{\Y}[3]{Y_{#1,\mathbf{#2},\mathbf{#3}}}
\newcommand{\W}[4]{W_{#1,#2,\mathbf{#3},\mathbf{#4}}}

\usepackage{pdfpages}

\makeatletter
\AtBeginDocument{\let\LS@rot\@undefined}
\makeatother

\begin{document}

\title{Open Quantum Cluster Embedding Theory}

\author{Petar Brini\'c}
\affiliation{Scientific Computing Laboratory, Center for the Study of Complex Systems, Institute of Physics Belgrade,
University of Belgrade, Pregrevica 118, 11080 Belgrade, Serbia}

\author{Hugo U. R. Strand}
\affiliation{School of Science and Technology, \"Orebro University, SE-70182 Örebro, Sweden}

\author{Jak\v sa Vu\v ci\v cevi\'c}
\affiliation{Scientific Computing Laboratory, Center for the Study of Complex Systems, Institute of Physics Belgrade,
University of Belgrade, Pregrevica 118, 11080 Belgrade, Serbia}

\begin{abstract}   
    The simulation of strongly correlated electron systems remains a formidable challenge. Certain experimentally relevant dynamical response functions are especially difficult to calculate, due to issues of finite-size effects and the ill posed analytic continuation. To address this we propose the quantum cluster embedding theory, an embedded cluster method aimed at computing the response of the system following an external perturbation; the frequency dependent dynamical susceptibility is obtained subsequently by means of inverse linear response theory. The embedded clusters, used within the method as representative of short range correlations, are open quantum systems governed by the Lindblad equation. The short-range correlations extracted from the clusters are used to close the equations of motion for the fermionic bilinear and the local double occupancy on the lattice. In turn, the clusters' Markovian baths are tuned to keep the bilinear and the double occupancy expectation values on the clusters and the lattice identical, throughout the concomitant evolution of the two sets of equations. The theory becomes numerically exact in the non-interacting, atomic and infinite cluster size limits, and it respects the total charge and energy conservation laws.  We show that our approach can treat very large lattices while avoiding analytic continuation through the explicit time evolution. Finally we compute the charge-charge correlation function in the square lattice Hubbard model and compare with a recent cold atom experiment, finding good qualitative agreement. 
\end{abstract}

\maketitle

\section{Introduction}

Strongly correlated electron systems feature highly complex phase diagrams, displaying a high degree of universality.\cite{Keimer2015,Georges2004,Kotliar2004,Furukawa2015,Kurosaki2005,Menke2024} In the disordered phase, the strange metallic regime, characterized by a linear-in-temperature resistivity, is of particular interest. It is associated with the region of maximum $T_c$ in high temperature superconductors\cite{Cooper2009,Ayres2021}, and is found in the vicinity of quantum critical points of some systems\cite{Vucicevic2015,Cha2020,Grigera2001}. To obtain a better understanding of strange metallic behavior, it is necessary to compute the conductivity in interacting lattice models, such as the Hubbard model.
\par However, dynamical correlation functions, such as the one relevant for the conductivity, have proven particularly challenging to compute in the Hubbard model (and other lattice models). Imaginary-time methods, such as determinant quantum Monte Carlo (DQMC)\cite{PhysRevD.24.2278,Denteneer1999,Huang2019} rely on analytic continuation, which is ill-conditioned and introduces uncontrolled error\cite{Fei2021}. Exact diagonalization methods, such as the finite-temperature Lanczos method (FTLM)\cite{Kokalj2017} can directly compute these quantities, but are limited to high temperatures and either uniform or short-range correlators due to the small size of lattices that can be treated with these methods. Development of new methods to compute the dynamical correlation functions is therefore one of the primary goals in the field. \cite{Qin2022,Brown2019,Kovacevic2025,Eom2025,Vucicevic2019,Vucicevic2021}
\par On the other hand, cold atoms in optical lattices provide a way to simulate a wide range of model Hamiltonians, allowing one to directly compare numerical results to experimental data\cite{Lewenstein2007,Gross2017}. However, the electrical resistivity $\rho$ is not yet directly measurable in these experiments, and the comparison with theoretical results has, so far, been only indirect\textemdash the resistivity in experiment was deduced from charge response measurements. These experiments probe the response to an applied electric field at long wavelengths. This response can be related to the dc-resistivity via hydrodynamic theory and the Nernst-Einstein relation \cite{Brown2019}. The dynamics at long wavelengths are inaccessible to the presently available numerical methods\textemdash imaginary-time methods run into the problem of analytic continuation, and exact diagonalization methods are generally limited to 4x4 clusters, which is far too small to observe the wavelengths of interest.
\par Dynamics of open quantum systems have been previously employed in the context of embedded cluster theories, but this approach has been limited to steady-state non-equlibrium regimes~\cite{Arrigoni2013,PhysRevB.89.165105,PhysRevB.92.125145,PhysRevB.109.075156}.
\par In this paper, we introduce the open quantum cluster embedding theory (OQCET), a real-time embedded cluster method that is able to access long-wavelength response without analytic continuation. This allows us to perform a direct comparison with the existing \emph{direct} measurements in a cold atom experiment.
\par OQCET follows the strategy of other embedded cluster methods\cite{Georges1996,Ayral2015,Ayral2016,Potthoff2003,Knizia2012}, but at the same time draws ideas from the methods based on the hierarchical equations of motion (HEOM)\cite{PhysRevE.75.031107,Tanimura2006,Tanimura2020,PhysRevB.92.235208,Jankovic2023,xxts-5x2c,f56z-h612}.
As in other embedded cluster methods, two sets of equations are solved self-consistently \textemdash the equations governing the lattice quantities are closed by short-range correlators computed in small effective clusters; in turn, the small clusters are tuned so as to mimic the dynamics of the large lattice. Namely, the small clusters are open quantum clusters, evolving per the Linblad equation \cite{GKS1976, Lindblad1976}, and their tuning is done via a time-dependent coupling to an external environment. The lattice equations are equations of motion for the fermionic bilinear and the local double occupancy, but additional quantities can also be used. The method is trivial in equilibrium: the starting point of the method is a solution for the instantaneous correlators extracted from a numerically exact lattice calculation, which requires no analytic continuation; in the absence of external fields, the OQCET equations yield no additional information. The dynamical response functions are obtained by probing the system using weak time-dependent external fields, and then inverting the linear-response theory equations \cite{PhysRevB.93.195144}. The method has certain desirable properties, such as fulfillment of the total charge and energy conservation; it becomes exact in the non-interacting and zero-hopping limits, as well in the limit of infinite cluster size.

\par The paper is organized as follows. In \ref{sec:Method} we introduce OQCET as an embedded cluster theory in analogy to DMFT, and discuss implementation details. We consider several variants of the method, with respect to a) how the initial density matrices and the Hamiltonians for the clusters are set up, b) what the cluster environment coupling is, c) what the constraining operators are (the quantities on the cluster that mimic the large lattice), d) how exactly we drive the system out of equilibrium and e) how big the cluster size is. In \ref{sec:Results} we benchmark our method against known results, and finally directly compare to the experimental findings. We observe good qualitative agreement between theory and experiment, in terms of temperature and wavelength dependence trends. In \ref{sec:Conclusions} we give concluding remarks and discuss prospects for future work. Detailed derivations of fomulas used in this paper, as well as further technical details can be found in Appendices \ref{sec:AppendixEOM}-\ref{sec:appendixbenchmark}.

\section{Methods}
\label{sec:Method}
\subsection{Embedded cluster methods}
\par To create an accurate picture of strongly correlated materials, we need to be able to treat local interacting physics as well as more delocalized coherent phenomena. By connecting lattice equations to clusters in which interactions are treated numerically exactly, embedded cluster theories provide a framework that can capture both local and non-local processes.
\par
\begin{figure}[htbp!]
    \centering
    \includegraphics[width=\columnwidth]{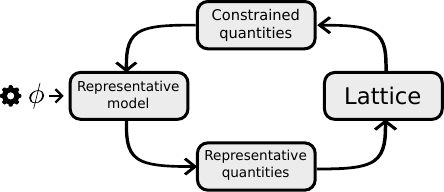}
    \caption{The self-consistency relation in embedded cluster theories. Coupling of the representative model to an effective environment is represented by the field $\phi$.}
    \label{fig:embeddedclust}
\end{figure}

 We will outline the prescription for creating an embedded cluster theory. First, we create a \textit{representative model}\cite{Georges2004} from a small subset of lattice degrees of freedom. This model should be small enough to be exactly solvable. After choosing a representative model, we choose some lattice quantities of interest. We will refer to these quantities as the \textit{constrained quantities}. Next, we write down lattice equations to express the constrained quantities. These equations will in general depend on some other quantities that are in general difficult to compute for a large lattice, but are possible to compute for the representative model; we will refer to those quantities as \textit{representative quantities}. The equations are then closed by taking the values of representative quantities from the representative model. Finally, the representative model is fully defined using a \textit{self-consistency} condition that ensures that the cluster constrained quantities match the lattice constrained quantities (Fig. \ref{fig:embeddedclust}). To satisfy the self-consistency condition, the representative model needs to have a certain degree of flexibility, which is usually achieved by coupling to an external field with as many parameters as there are constrained quantities.
\par For example, in dynamical mean-field theory (DMFT) \cite{Georges1996}, the representative model is the Anderson impurity model (AIM). The constrained quantity is the local Green's function, while the representative quantity is the self-energy. The self-consistency condition is that the impurity Green's function matches the local lattice Green's function, $G_\mathrm{loc}$.
\par Embedded cluster methods can often be rigorously derived as approximations of certain (Legendre transformed) free-energy functionals, depending precisely on the chosen representative quantities. Such methods guarantee fulfillment of the particle number and total energy conservation laws. This holds for DMFT and its cluster extensions\cite{Hettler1998,Lichtenstein2000,Kotliar2001,Maier2005,Vucicevic2018},
which can be derived as approximations of the Luttinger-Ward functional, $\Phi[G;U]\approx \Phi[G_\mathrm{loc};U]$.
In other methods such as EDMFT\cite{Sun2002,Vuicevic2017}, TRILEX\cite{Ayral2015,Vuicevic2017} and QUADRILEX\cite{Ayral2016QUADRILEX}, representative quantities include higher-order correlation functions; these methods are derived from the corresponding higher-order functionals.
There are also theories based on approximations of the self-energy functional \cite{Potthoff2003selfenergy}. However, some embedded cluster theories do not have a clear formulation in terms of a functional approximation, e.g. density matrix embedding theory (DMET)\cite{Knizia2012}. OQCET has no clear derivation in terms of a free-energy functional, but the conservation laws are still enforced, albeit in a different way.

\subsection{Open Quantum Cluster Embedding Theory}
\label{sec:OQCETembclustermethod}
\par In this section we will introduce OQCET as an embedded cluster theory in analogy to DMFT.
\par
\begin{figure}[htbp!]
    \centering
    \includegraphics[width=0.8\columnwidth]{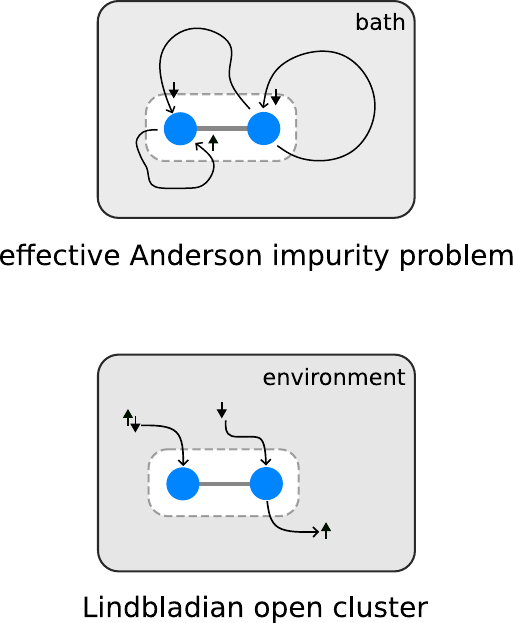}
    \caption{Schematic representation of the Anderson impurity model and a Lindbladian open quantum cluster}
    \label{fig:dmft_lindblad_cluster}
\end{figure}

 In OQCET, the role of the representative model is played by an open quantum cluster governed by the Lindblad equation \cite{GKS1976, Lindblad1976}. Like the Anderson impurity model in DMFT, Lindbladian time evolution allows for particles to enter and leave the clusters. Beyond what is usually considered in the AIM, in our open clusters we will also consider more complicated couplings with the environment.
In the AIM, the particles in the bath behave according the bare-propagator $\Delta$. In Lindbladian evolution the particles entering the cluster have no memory of their previous state (Fig. \ref{fig:dmft_lindblad_cluster}). As a result, the Lindblad equation is Markovian\textemdash the future evolution of the system only depends on its current state \cite{Manzano2020}. In practice, this means that it is only necessary to use time-local quantities, similar as in the HEOM approach\cite{PhysRevE.75.031107,Tanimura2006,Tanimura2020,PhysRevB.92.235208,Jankovic2023,xxts-5x2c,f56z-h612}. The memory requirements for non-equilibrium calculations remain constant ($O(1)$) with the number of time steps $N_t$, while the calculation time scales as $O(N_t)$. This contrasts with other methods such as non-equilibrium DMFT, where it is necessary to solve the Kadanoff-Baym equations which depend on two-time quantities. In this case the required memory scales as $O(N_t^2)$ and the computation time as $O(N_t^3)$ \cite{Aoki2014}. Significantly lower memory and processing power requirements in OQCET make it feasible to perform calculations to very long times. The coupling to the bath in (non-equilibrium) DMFT is described and tuned by the two-time hybridization function $\Delta(t,t')$, while in the Lindblad equation we tune the instantaneous (one-time dependent) coupling constants $\Gamma(t)$.
\par The constrained quantities in OQCET are instantaneous expectation values (correlators) of \textit{constraining operators}  $\{\langle A^\mathrm{latt}_\lambda (t)\rangle\}$. This is very practical, as instantaneous correlators also evolve according to equations of motion, with the same scaling as the Linblad equation for the clusters. However, the restriction that constrained quantities are instantaneous correlators
precludes the free-energy functional derivation of OQCET. In practice, this means that the choice of representative model is not unique, and different choices can lead to different results.
In this sense, OQCET is a family of methods; to fully define an OQCET method, one needs to specify the representative model in terms of its initial state, Hamiltonian and its couplings to the environment. This is unlike DMFT, where all the terms of the effective action are uniquely defined (of course, up to the parametrization of the bath, i.e. the frequency dependence of the hybridization function).
However, other embedded cluster methods also feature ambiguities such as the choice of periodization procedure in CDMFT\cite{Vucicevic2018}, the Hubbard-Stratonovich decoupling scheme (the Fierz ambiguity) in TRILEX\cite{Ayral2017}, and the choice of Lindbladian jump operators in the auxiliary master equation approach \cite{Arrigoni2013}.
Such ambiguities are not necessarily a shortcoming of a method, but can also be considered a degree of flexibility which allows one to optimize the method for a given purpose (calculation of a given quantity in a given parameter regime). If the ambiguity is resolved optimally, the results are expected to be insensitive to small changes of any free parameters\cite{Ayral2017}.

\subsubsection{Formulation - equations of motion}
\par As already noted, embedded cluster theories are formulated by coupling two sets of equations\textemdash one for the lattice, and one for the embedded clusters.

The OQCET lattice equations are equations of motion for the expectation values of constraining operators. The expectation value  of a lattice operator $A^\mathrm{latt}(t)$ evolves by the Heisenberg equation
\begin{equation} \label{eq:heisenberg_eq}
    \partial_t \langle A^\mathrm{latt} (t) \rangle = i \left\langle\left[H^\mathrm{latt}(t),A^\mathrm{latt}(t)\right]\right\rangle
\end{equation}
where $H^\mathrm{latt}(t)$ is the lattice Hamiltonian.
In general (due to interaction terms in the Hamiltonian), the set of operator expectation values $\{\langle A^\mathrm{latt}_\lambda\rangle\}$ will not yield closed equations of motion (EOM), and the commutator on the right hand side of Eq.~\ref{eq:heisenberg_eq} will also depend on some extended linear combination containing expectation values $\{\langle B^\mathrm{latt}_\lambda\rangle\}$,
\begin{equation} \label{eq:EOM}
    \partial_t \langle \hat{A}^\mathrm{latt}_\lambda (t) \rangle \equiv \sum_\mu a_\mu \langle \hat{A}^\mathrm{latt}_\mu \rangle + \sum_\nu \langle \hat{B}^\mathrm{latt}_\nu \rangle.
\end{equation}

\par
In the HEOM approach, one then writes another set of EOM's for $\{\langle \hat{B}^\mathrm{latt}_\lambda \rangle \}$, which will in turn depend on additional unknown correlators. Repeating the procedure then builds a hierarchy of EOM's, level by level. The hierarchy must eventually be truncated, and at the deepest level of HEOM, the unknown correlators need to be approximated in some way.
In OQCET, we truncate at the first level and close the EOM's for the lattice operator expectation values $\{\langle \hat{A}^\mathrm{latt}_\lambda \rangle \}$ by approximating $\{\langle \hat{B}^\mathrm{latt}_\lambda \rangle \}$: the values are constructed from the corresponding correlators on the smaller clusters (representative model), using a mapping between cluster and lattice sites. Here the 'hat' symbol denotes that $\hat{A}^\mathrm{latt}_\lambda$ and $\hat{B}^\mathrm{latt}_\lambda$ are tensors acting on the space of real space vectors $\textbf{r}$. These tensors can be of arbitrary rank, e.g. the bilinear $c^\dag_{\textbf{r}\sigma}c_{\textbf{r}'\sigma}$ is a rank two tensor and the local double occupancy operator $n_{\textbf{r}\uparrow}n_{\textbf{r}\downarrow}$ is of rank one.  
As the clusters are finite, the mapping between cluster and lattice sites is not one-to-one, and there will be some components of $\{\langle \hat{B}^\mathrm{latt}_\lambda \rangle \}$ that do not have a mapping onto the cluster. We approximate those components of $\langle \hat{B}^\mathrm{latt}_\lambda \rangle$ by setting them to zero.

\begin{table*}[htbp!]
    \begin{tabular}{l|l|l}
                     & cluster DMFT                                                                   & OQCET                                                                                                                       \\ \hline
    Rep. model       & Anderson impurity model                                                        & Open quantum cluster                                                                                                        \\
    Constr. Quantity & $G_{ij}$                                                                       & $\langle\hat{A}_\lambda \rangle$                                                                                            \\
    Rep. quantity    & $\Sigma_{ij}$                                                                  & $\langle\hat{B}_\lambda \rangle$                                                                                            \\
    Eff. field       &  $\Delta$                                                           & $\{\Gamma_l\}$                                                                                                              \\
    Latt. eq.        & $G_\mathrm{latt}= \left[G_{0,\mathrm{ latt}}-\Sigma_\mathrm{latt}\right]^{-1}$ & $\partial_t \langle \hat{A}^\mathrm{latt}_\lambda (t)\rangle \equiv \sum_\mu a_\mu \langle \hat{A}^\mathrm{latt}_\mu \rangle + \sum_\nu \langle \hat{B}^\mathrm{latt}_\nu \rangle$
    \end{tabular}
    \caption{Cluster DMFT and OQCET as embedded cluster theories.}\label{table:OQCETvscDMFT}
\end{table*}

\par The expectation value of an n-body operator can be decomposed into its connected and disconnected components
\begin{eqnarray}
    \langle \hat{O} \rangle &=& \langle \hat{O} \rangle^\mathrm{conn} + \langle \hat{O} \rangle^\mathrm{disc} \\
    \langle \hat{O} \rangle^\mathrm{disc} &=& \sum_{\mathrm{partitions}} \prod_{\alpha} \langle \hat{O}_\alpha \rangle
\end{eqnarray}
where the sum goes over all possible partitions of $\langle \hat{O} \rangle$ into lower order correlators $\langle \hat{O}_\alpha \rangle$. E.g. a three-operator expectation value $\langle \hat{M}\hat{N}\hat{O}\rangle$ can be decomposed as $\langle \hat{M}\hat{N}\hat{O}\rangle = \langle \hat{M}\hat{N}\hat{O}\rangle^\mathrm{conn} + \langle \hat{M}\hat{N}\rangle \langle \hat{O}\rangle + \langle \hat{M}\hat{O}\rangle \langle \hat{N}\rangle + \langle \hat{N}\hat{O}\rangle \langle \hat{M}\rangle - 2\langle \hat{M}\rangle \langle \hat{N}\rangle \langle \hat{O}\rangle$.
\par If the disconnected components of  $\langle \hat{B}_\lambda \rangle$ can be expressed as a product of expectation values $\{\langle \hat{A}_{\mu} \rangle \}$, we can rewrite the equations of motion Eq.~\ref{eq:EOM} such that the equations are now closed by the connected average. The time derivatives $\partial_t \langle \hat{A}^\mathrm{latt}_\lambda (t) \rangle$  can then be expressed as a function $f_\lambda \left( \{\langle \hat{A}^\mathrm{latt}_\mu \rangle\}, \{\langle \hat{B}^\mathrm{latt}_\nu \rangle^\mathrm{conn}\} \right)$ (Eq.~\ref{eq:EOMconn}). 
Analogously, we set the longer range components of $\{\langle \hat{B}^{\mathrm{\, latt}}_\lambda \rangle^\mathrm{conn} \}$ to zero. This is the approach implemented in this paper.

The method can be summarized by listing the full set of equations that are being solved, as follows
\begin{equation} \label{eq:clustmap}
    \langle \hat{B}_{\lambda}^\mathrm{latt}\rangle \approx \sum_{c \in \mathcal{C}} O(c) \langle \hat{B}_{\lambda}^{\, c, \ \mathrm{ clust}}\rangle
\end{equation}
\begin{equation}
    \partial_t \langle \hat{A}^\mathrm{latt}_\lambda (t) \rangle \equiv f_\lambda \left( \{\langle \hat{A}^\mathrm{latt}_\mu \rangle\}, \{\langle \hat{B}^\mathrm{latt}_\nu \rangle^\mathrm{conn}\} \right)
    \label{eq:EOMconn}
\end{equation}
\begin{equation}
    \begin{split}
        &\frac{d\rho_\mathrm{clust}(t)}{dt} =-i[H_\mathrm{clust},\rho_\mathrm{clust}(t)] \\&+\sum_{l} \Gamma_l(t)\left(  L_l\rho_\mathrm{clust}(t)L^\dag_l-\frac{1}{2}\left\{      L^\dag_l L_l,\rho_\mathrm{clust}(t)        \right\}\right)    
    \end{split}
    \label{eq:lindblad_clust}
\end{equation}
\begin{equation} \label{eq:clust_ev}
    \langle \hat{O}^\mathrm{clust} (t) \rangle = \mathrm{Tr} \left[\rho_{\mathrm{clust}}(t) \hat{O}^\mathrm{clust} \right]
\end{equation} 
\begin{equation} \label{eq:gamma}
    \{\Gamma_l (t)\} :  \{\langle \hat{A}^\mathrm{clust}_\lambda (t) \rangle\} = \{\langle \hat{A}^\mathrm{latt}_\lambda (t) \rangle\}
\end{equation}
The Eq.~\ref{eq:clustmap} is the essence of the approximation: it constructs the lattice correlators $\langle \hat{B}_{\lambda}^\mathrm{latt}\rangle$ from the correlators $\langle \hat{B}_{\lambda}^{\, c, \ \mathrm{ clust}}\rangle$ of all the clusters $c$ from a set of clusters $\cal C$; the coefficients $O(c)$ take into account if multiple clusters have a mapping onto the same component of $\hat{B}^\mathrm{latt}_\lambda$ (see Section \ref{sec:overlap} for details).
The self-consistency condition, Eq.~\ref{eq:gamma}, goes in the other direction - the correlators $\{\langle \hat{A}^\mathrm{clust}_\lambda (t) \rangle\}$ on any given cluster are constrained by the corresponding lattice correlators $\{\langle \hat{A}^\mathrm{latt}_\lambda (t) \rangle\}$; the constraint is satisfied by tuning the coupling to the environment $\{\Gamma_l (t)\}$, for each cluster independently. The Eqs.~\ref{eq:EOMconn} and \ref{eq:lindblad_clust} are the EOMs for the cluster and the lattice. Any correlator on the cluster can in general be obtained via Eq.~\ref{eq:clust_ev} ($\hat{O}^\mathrm{clust} \in \{ \hat{A}^\mathrm{clust}_\lambda (t), \hat{B}^\mathrm{clust}_\mu (t)\}$)

\par As already stated, the clusters are open systems, coupled to an effective Markovian environment; this allows us to fully describe each cluster using the time-dependent density-matrix $\rho_\mathrm{clust}(t)$.
The density matrix evolves per the Lindblad equation (Eq.~\ref{eq:lindblad_clust})
which has two terms.
The first term represents the unitary evolution and depends on the cluster Hamiltonian $H_\mathrm{clust}$.
(The procedure for constructing the cluster Hamiltonian $H_\mathrm{clust}$ is non-unique and it is discussed in Section \ref{sec:clust_init}.)
The second term encodes the dissipative evolution. Here $L_l$ are the \textit{jump operators}; they specify the system-environment coupling; $\Gamma_l$ are the corresponding coupling constants which are real and positive definite ($\Gamma_l \ge 0$).
For example, a jump operator $L_l=c^\dag_{i\sigma}$ ($L_l=c_{i\sigma}$) corresponds to a process where a particle is entering (leaving) the system at site $i$ with spin $\sigma$; the rate for the occurence of this process is determined by the associated $\Gamma_l$. The set of jump operators $\{L_l\}$ is in principle arbitrary and needs to be chosen by hand. However, the underlying idea is that the operators must provide sufficiently flexible coupling to the environment so that the cluster can follow the dynamics of a larger lattice. In other words, the jump operators are chosen in a way that ensures that we can reach the self-consistency between cluster and lattice equations. This means that the optimal choice of $\{L_l\}$ will in general depend on the choice of constrained quantities (i.e. the quantities we wish to calculate).
At the same time, the choice of $\Gamma_l$ to satisfy Eq.~\ref{eq:gamma} should always be unique.

\par
\begin{figure}[t]
    \centering
    \includegraphics[width=0.9\columnwidth]{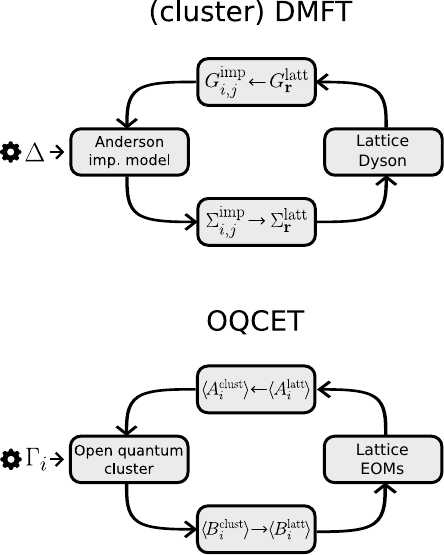}
    \caption{ Schematic comparison of the DMFT and OQCET loops. Quantities in the top boxes are imposed onto the representative model by tuning the effective field (i.e. $\Delta$ and $\Gamma_i$); quantities at the bottom are extracted from the representative model and passed onto the lattice equations.}
    \label{fig:dmft_oqcet_loop}
\end{figure}

 Finally, we illustrate the analogy with DMFT in Fig. \ref{fig:dmft_oqcet_loop} and Table \ref{table:OQCETvscDMFT}. In DMFT, the lattice equation is the Dyson equation expressing the Green's function from the self-energy and the lattice Hamiltonian; the cluster equations represent the solution for the self-energy with respect to the cluster action, which depends on the hybridization function $\Delta$. In both theories, all the equations need to be solved self-consistently; in practice this is implemented using iterative algorithms\cite{Vucicevic2018,Strand2011}.

\subsubsection{Initial conditions}
\par OQCET equations Eq~\ref{eq:clustmap}-\ref{eq:gamma} yield the time evolution of certain instantaneous correlators $\{\langle \hat{A}^\mathrm{latt}_\lambda (t) \rangle\}$ in a system.
There are no \emph{a priori} restrictions on the initial state of the system, or the external probes that might drive its dynamics.
However, here we are interested in making use of OQCET to compute the dynamical, i.e. time-dependent correlators in equilibrium.
The way to do this is to prepare the system in equilibrium, and then probe it with weak, time-dependent external fields;
as we explain in Section~\ref{sec:invlinearresponse} certain protocols allow to invert the linear response equations and extract a particular dynamical correlation function from the corresponding response of the system.

In practice, we first obtain the equilibrium expectation values for $ \{\langle \hat{A}^\mathrm{latt}_\lambda (t=0) \rangle\}$ and $ \{\langle \hat{B}^\mathrm{latt}_\lambda (t=0) \rangle\}$. This is possible to do using numerically exact methods for relatively large lattices without any analytic continuation (see Section \ref{sec:lattice_eq_avg}). Then, the equilibrium density matrix $\rho_\mathrm{clust}(t=0)$ is prepared such that $\{\langle \hat{A}^\mathrm{clust}_\lambda (t=0) \rangle\} = \{\langle \hat{A}^\mathrm{latt}_\lambda (t=0) \rangle\}$ and $\{\langle \hat{B}^\mathrm{clust}_\lambda (t=0) \rangle\} = \{\langle \hat{B}^\mathrm{latt}_\lambda (t=0) \rangle\}$. The initial cluster density matrix  $\rho_\mathrm{clust}(t=0)$ is not uniquely determined by these constraints, and several possible methods for obtaining it are discussed in Section \ref{sec:clust_init}. We note here that in the case where $\{\langle \hat{B}^\mathrm{latt}_\lambda (t=0) \rangle\}$ are zero by symmetry and the clusters inherit the relevant symmetries, the second condition is trivially satisfied.

\subsubsection{Numerical implementation}
\par In practice, the solution of OQCET is obtained by coevolving the sets of lattice and cluster equations (Eqs.~\ref{eq:EOMconn} and \ref{eq:lindblad_clust}). Lattice equations (Eq.~\ref{eq:EOMconn}) determine $ \{\langle \hat{A}^\mathrm{latt}_\lambda (t) \rangle\}$, $\Gamma_l(t)$ are tuned to satisfy the self-consistency condition $ \{\langle \hat{A}^\mathrm{clust}_\lambda (t) \rangle\} =  \{\langle \hat{A}^\mathrm{latt}_\lambda (t) \rangle\}$ (Eq.~\ref{eq:gamma}). Lattice equations for the next time step are closed by $\{\langle \hat{B}_\lambda (t) \rangle\}$ obtained from clusters (Eqs.~\ref{eq:clustmap} and~\ref{eq:clust_ev}).
One iteration of the method evolves the system in time by a step $\Delta t$. The algorithm can be described as follows (Fig.~\ref{fig:oqcet_algorithm}):
\begin{figure}[htbp!]
    \centering
    \includegraphics[width=0.9\columnwidth]{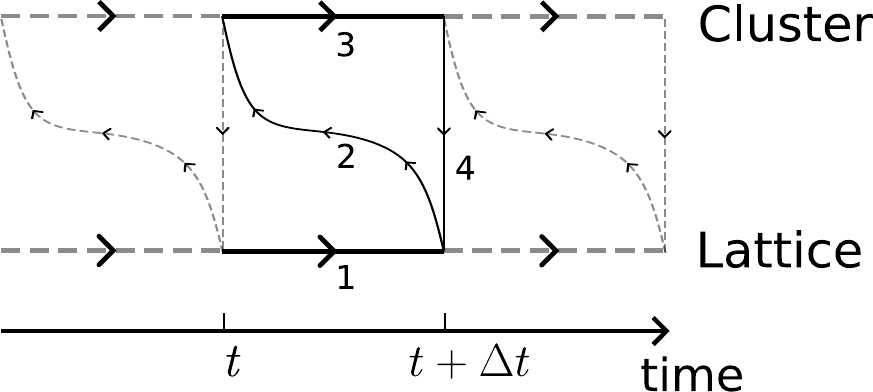}
    \caption{OQCET algorithm in time}
    \label{fig:oqcet_algorithm}
\end{figure}
\begin{enumerate}
    \item $\{\langle \hat{A}^\mathrm{latt}_\lambda (t + \Delta t) \rangle\}$ is obtained by solving the lattice differential equation by the second order Taylor method using the values of $\{\langle \hat{A}^\mathrm{latt}_\lambda (t) \rangle\}$; this requires the knowledge of the derivatives $\{\langle \hat{B}^\mathrm{latt}_\lambda (t) \rangle\}$, which are obtained by a method of finite differences using the values of $\{\langle \hat{B}^\mathrm{latt}_\lambda (t) \rangle\}$ and $\{\langle \hat{B}^\mathrm{latt}_\lambda (t- \Delta t) \rangle\}$, see Sec. \ref{sec:OQCETHubb}.
    \item On the cluster, $\Gamma_l(t)$ are chosen such that the evolution $t \rightarrow t+\Delta t$ by the Lindblad equation (Eq.~\ref{eq:lindblad_clust}) yields  $\{\langle \hat{A}^\mathrm{clust}_\lambda (t+\Delta t) \rangle\} = \{\langle \hat{A}^\mathrm{latt}_\lambda (t+\Delta t) \rangle\}$. The Lindblad equation is likewise solved by the second order Taylor method, see Sec. \ref{sec:ClusterDynamics}.
    \item The cluster density matrix is evolved to $t+\Delta t$ using $\Gamma_l(t)$. Then, cluster expectation values $\{\langle \hat{B}^\mathrm{clust}_\lambda (t+\Delta t) \rangle\}$ are calculated.
    \item The cluster expectation values $\{\langle \hat{B}^\mathrm{clust}_\lambda (t+\Delta t) \rangle\}$ are mapped onto lattice expectation values $\{\langle \hat{B}^\mathrm{latt}_\lambda (t+\Delta t) \rangle\}$.
\end{enumerate}
%
This procedure is repeated until we reach a desired final time $t_{f}$.
This can be contrasted with non-equilibrium DMFT, where the hybridization function $\Delta(t,t')$ at each time $t$ needs to be determined for all previous times $t'$ simultaneously, which is a much more computationally demanding task. \cite{Amaricci2012,Aoki2014}

The algorithm is first order accurate with a time discretization error $\sim O(\Delta t)$, despite the usage of the second order Taylor methods for the lattice and clusters. We attribute the reduction in order of their combined propagation to the optimization of $\Gamma_l(t)$ in the second step (see Appendix \ref{sec:appendixbenchmark} for details).

\subsubsection{Setting up clusters}
\par Clusters are formed in analogy to the nested cluster scheme\cite{Vucicevic2018}, which is a general approach applicable to embedded cluster methods. 
To form a cluster, the lattice Hilbert space is reduced to a subspace describing a small number of lattice sites. For example, a cluster can be formed from two neighboring lattice sites with vectors $\textbf{r}$ and $\mathbf{r+e}_x$. In principle, the clusters can be of any shape and the sites need not be nearest-neighbors. The mapping between cluster site-indices and lattice vectors is in general many-to-many: clusters can overlap, i.e. sites on different clusters can map to the same lattice vector. If there are spatial symmetries, a single cluster site can map to multiple lattice vectors.
\par To create a set of clusters $\cal{C}$ we first select the desired cluster shape (for example, 2$\times$1 nearest neighbor clusters) and we tile the lattice with all possible translations and rotations of this cluster. Next, we use lattice symmetries to select the irreducible set of non-equivalent clusters. On the clusters, $\{\langle \hat{A}^\mathrm{clust}_\lambda \rangle \}$ and $\{\langle \hat{B}^\mathrm{clust}_\lambda \rangle \}$ are tensors of cluster site indices $\{i,j,...\}$. Since the clusters are finite, long range components of $\{\langle \hat{B}^\mathrm{latt}_\lambda \rangle \}$ will not have a mapping in the space of cluster operators. For example, if we tile the lattice with two-site nearest-neighbor clusters, a rank two tensor $\langle B_{\bf r,r',\lambda}^\mathrm{latt}\rangle$ will not have a mapping if $\textbf{r}$ and $\textbf{r}'$ are not vectors of nearest-neighbor lattice sites. If that is the case, we set those tensor components of $\{\langle \hat{B}^\mathrm{latt}_\lambda \rangle \}$ to zero.
\label{sec:overlap}
\begin{figure}[htbp!]
    \centering
    \includegraphics[width=0.75\columnwidth]{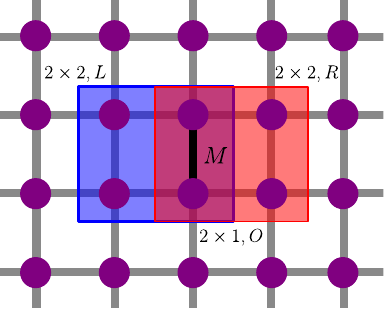}
    \caption{Two overlapping 2$\times$2 clusters, shaded blue and red. Their overlap forms a 2$\times$1 cluster, containing bond $M$.}
    \label{fig:overlap}
\end{figure}
\par Due to the possibility of cluster overlaps, the procedure for mapping cluster quantities to lattice quantities is, in general, not unique.
However, in the simplest scheme where we use only $2\times1$ clusters, overlaps between cluster will contain at most a single site.
If none of the constrained and representative correlators are purely local, then there is no issue, and the mapping between the lattice and the cluster is unambiguous.
On the other hand, when using 2$\times$2 clusters, also the nearest-neighbor bonds will generally belong to two nonequivalent overlapping clusters (Fig. \ref{fig:overlap}). Then one must avoid the double counting of the overlapping bonds and there are multiple ways this can be done.
We propose two schemes. In the first, we apply the nested cluster scheme\cite{Vucicevic2018} \textemdash additional 2$\times$1 clusters formed by overlaps of 2$\times$2 clusters are taken into account. Contributions from these 2$\times$1 clusters are subtracted to avoid double counting the bond M,
\begin{equation}
    \hat{B}_M^\mathrm{nest} = \hat{B}_M^{2\times2,\mathrm{L}} + \hat{B}_M^{2\times2,\mathrm{R}} - \hat{B}_M^{2\times1,\mathrm{O}}
\end{equation}
where $\hat{B}_M$ is a tensor whose components $\bf r, r', r'' \ldots$ are restricted to the bond M, i.e.
\begin{eqnarray}
    O^\mathrm{nest}_M(c^{2\times2,\mathrm{L/R}}) = +1 \\
    O^\mathrm{nest}_M(c^{2\times1,\mathrm{O}}) = -1.
\end{eqnarray}
where $O(c)$ is the overlap function defined in Eq.~\ref{eq:clustmap}.
In the second scheme, we simply average the contributions from overlapping clusters without introducing additional overlap clusters,
\begin{equation}
    \hat{B}_M^\mathrm{avg} = \frac{1}{2} (\hat{B}_M^{2\times2,\mathrm{L}} + \hat{B}_M^{2\times2,\mathrm{R}})
\end{equation}
\begin{eqnarray}
    O^\mathrm{avg}_M(c^{2\times2,\mathrm{L/R}}) = +\frac{1}{2}.
\end{eqnarray}
The two schemes are further discussed in Section~\ref{sec:clustsize}.

\subsection{OQCET for the square lattice Hubbard model}
\label{sec:OQCETHubb}
\par We will describe the application of OQCET to the Hubbard model \cite{Hubbard1963} with the aim to calculate the density response function $\chi_{\textbf{q}}(\omega)$. We start with the square lattice Hamiltonian of the Hubbard model with a time-dependent non-uniform scalar potential $\phi_{\bf r}(t)$
\begin{equation}
    \label{eq:hubb_model}
    \begin{split}
        H (t) = &-J \sum_{\textbf{r},\mathbf{u} \in\{\pm\mathbf{e}_{x},\pm\mathbf{e}_{y} \}, \sigma} c^{\dag}_{\sigma, \textbf{r}} c_{\sigma, \textbf{r}+\textbf{u}} 
        \\& - \sum_{\textbf{r}, \sigma} ( \mu - \phi_\textbf{r}(t)) n_{\sigma, \textbf{r}}
         + U \sum_{\textbf{r}} n_{\uparrow,\textbf{r}} n_{\downarrow,\textbf{r}} 
    \end{split}
\end{equation}
where $\bf r$ denotes position vectors of lattice sites, $ \mathbf{e}_{x}$ and $\mathbf{e}_{y}$ are lattice vectors and $\sigma$ denotes spin. $c/c^{\dag}$ are annihilation/creation operators and $n_{\sigma,\textbf{r}}=  c^{\dag}_{\sigma, \textbf{r}} c_{\sigma, \textbf{r}}$ denotes the particle-number operator. $J$, $\mu$ and $U$ denote the hopping amplitude, chemical potential and interaction strength, respectively. We set as the energy unit the bare half-bandwidth $D=4J=1$.
We will restrict our calculations to homogeneous paramagnetic phases, so we assume that in equilibrium (in the absence of external fields $\phi$) the system has both translation invariance and $SU(2)$ spin symmetry. Since we are interested in the density response, we will have the density operator $c^{\dag}_{\sigma, i} c_{\sigma, i}$ in the set of constraining operators $\{\hat{A}_\lambda\}$. If we wish for ensure total energy conservation, we must also constrain the bilinear $c^{\dag}_{\sigma, i} c_{\sigma, j}$ and the double occupancy operator $n_{i\uparrow} n_{i\downarrow}$. Since the total energy is given by $E = \langle H \rangle$, in Appendix \ref{sec:conservationNE} we show that constraining the bilinear and the double occupancy gives us both conservation of energy and conservation of density. We do not expect our theory to obey conservation of momentum, as embedding with finite clusters generally breaks momentum conservation.
\par The equations of motion for expectation values can be derived from the Heisenberg equation (Eq.~\ref{eq:heisenberg_eq}). 
For the expectation value of the fermionic bilinear $\X{\sigma}{r}{r'}= \left\langle c^{\dag}_{\sigma, \textbf{r}} c_{\sigma, \textbf{r}'}\right\rangle$, working out the commutator in Eq.~\ref{eq:heisenberg_eq} gives
\begin{equation} \label{eq:bilinear_eom}
    \begin{aligned}
        \partial_{t} X_{\sigma, \mathbf{r r}^{\prime}}  = i \bigg(&-J \sum_{\mathbf{u} \in\{\pm\mathbf{e}_{x},\pm\mathbf{e}_{y} \}} ( X_{\sigma, \mathbf{r}+\mathbf{u}, \mathbf{r}^{\prime}}-X_{\sigma, \mathbf{r}, \mathbf{r}^{\prime}+\mathbf{u}} )\\
        &+\left(\phi_\textbf{r} - \phi_{\textbf{r}'}\right) X_{\sigma, \textbf{r},\textbf{r}'}\\
        &+U \left(\left(X_{\sigma,\textbf{r},\textbf{r}}-X_{\sigma,\textbf{r}',\textbf{r}'}\right)X_{\sigma,\textbf{r},\textbf{r}'}+Y_{\sigma,\textbf{r},\textbf{r}'}\right) \bigg) 
    \end{aligned}
\end{equation}
where we have introduced
\begin{equation}
    {Y_{\sigma, \bf {r,r}'}=\left\langle c^\dag_{\bar{\sigma},\textbf{r}}c_{\bar{\sigma},\textbf{r}}c^\dag_{\sigma,\textbf{r}}c_{\sigma,\textbf{r}'}\right\rangle^\mathrm{conn} - \left\langle c^\dag_{\bar{\sigma},\textbf{r}'}c_{\bar{\sigma},\textbf{r}'}c^\dag_{\sigma,\textbf{r}}c_{\sigma,\textbf{r}'}\right\rangle^\mathrm{conn}}
    \label{eq:Y}
\end{equation}
to denote the connected part of the four-point correlator.
For the the expectation value of the density-density operator $ \left\langle n_{\sigma, \bf r} n_{\sigma', \bf r'} \right\rangle$ we have
\begin{equation} \label{eq:n_eom}
\begin{aligned}
&\partial_{t}  \left\langle n_{\sigma, \bf r} n_{\sigma', \bf r'} \right\rangle \\&
=-i J \sum_{\mathbf{u} \in\{\pm\mathbf{e}_{x},\pm\mathbf{e}_{y} \}} \biggl\{ \left(X_{\sigma,\bf r + u,r'}-X_{\sigma,\bf r,r+u}\right)X_{\sigma,\bf r',r'}\\&
+ X_{\sigma,\bf r,r}(X_{\sigma,\bf r'+u,r'}-X_{\sigma,\bf r',r'+u}) \\&
+\delta_{\sigma,\sigma'} X_{\sigma,\bf r,r'}(X_{\sigma,\bf r',r+u}-X_{\sigma,\bf r'+u,r}) \\&+\delta_{\sigma,\sigma'} X_{\sigma,\bf r',r}(X_{\sigma,\bf r,r'+u}-X_{\sigma,\bf r+u,r'}) \\&
+ \delta_{\sigma,\sigma'} \delta_{\bf r,r'}(X_{\sigma,\bf r+u,r'}-X_{\sigma,\bf r,r'+u}) \\&+ \delta_{\sigma,\sigma'}(\delta_{\bf r,r'+u}-\delta_{\bf r+u,r'})X_{\sigma,\bf r,r'}\biggr\} - iJ W_{\sigma,\sigma',\textbf{r},\textbf{r}'}
\end{aligned}
\end{equation}
where
\begin{equation}
    \begin{split}
        W_{\sigma,\sigma',\textbf{r},\textbf{r}'}=&\Biggl\langle \sum_{\textbf{u}} \biggl\{  \left( c^\dag_{\sigma,\textbf{r}+\textbf{u}} c _{\sigma,\textbf{r}} - c^\dag_{\sigma,\textbf{r}} c _{\sigma,\textbf{r}+\textbf{u}}\right)n_{\sigma',\textbf{r}'} \\
        &+  n_{\sigma,\textbf{r}} \left( c^\dag_{\sigma',\textbf{r}'+\textbf{u}} c _{\sigma',\textbf{r}'} - c^\dag_{\sigma',\textbf{r}'} c _{\sigma',\textbf{r}'+\textbf{u}}\right)\biggr\} \Biggr\rangle^{\mathrm{conn}}.
    \end{split}
    \label{eq:W}
\end{equation}
The double occupancy expectation value, $d_{\bf r}=  \left\langle n_{\uparrow, \bf r} n_{\downarrow, \bf r} \right\rangle$ is a special case of the above expression,
\begin{equation} \label{eq:d_eom}
\partial_t d_\textbf{r}= -2iJ \sum_{\textbf{u}}\ \left(X_{\sigma,\textbf{r}+\textbf{u},\textbf{r}}-X_{\sigma,\textbf{r},\textbf{r}+\textbf{u}}\right)X_{\sigma,\textbf{r},\textbf{r}} -iJ W_{\uparrow,\downarrow,\textbf{r},\textbf{r}}.
\end{equation}

\begin{widetext}

Eqs.~\ref{eq:bilinear_eom} and \ref{eq:n_eom} are solved using the second order Taylor method
\begin{align}
X_{\sigma,\textbf{r}\textbf{r}'}(t_{i+1})&  \approx  X_{\sigma,\textbf{r}\textbf{r}'}(t_i) + \Delta t \, \partial_t X_{\sigma,\textbf{r}\textbf{r}'}(t_i) + \frac{1}{2} (\Delta t)^2 \partial_t^2 X_{\sigma,\textbf{r}\textbf{r}'}(t_i)\\
\left\langle n_{\sigma, \bf r} n_{\sigma', \bf r'} \right\rangle(t_{i+1})& \approx \left\langle n_{\sigma, \bf r} n_{\sigma', \bf r'} \right\rangle(t_i) + \Delta t \, \partial_t \left\langle n_{\sigma, \bf r} n_{\sigma', \bf r'} \right\rangle(t_i) + \frac{1}{2} (\Delta t)^2 \partial_t^2 \left\langle n_{\sigma, \bf r} n_{\sigma', \bf r'} \right\rangle(t_i)
\end{align}

\end{widetext}

The local correlator $\Y{\sigma}{r}{r}$ is zero by definition (Eq.~\ref{eq:Y}), while $\W{\sigma}{\sigma'}{r}{r}$ 
contains no purely local terms (Eq.~\ref{eq:W}). This means that the smallest clusters that give nontrivial contributions to these correlators are of size 2$\times$1.

A derivation of the above equations can be found in Appendix \ref{sec:AppendixEOM}, including the explicit expressions for $\partial_t^2 X_{\sigma,\textbf{r}\textbf{r}'}$ and $\partial_t^2 \left\langle n_{\sigma, \bf r} n_{\sigma', \bf r'} \right\rangle(t_i)$, as well as a proof that $\Y{\sigma}{r}{r'}$ and $\W{\sigma}{\sigma'}{r}{r'}$ are zero in equilibrium for a homogeneous system by symmetry.
\par In the following sections we will outline the different variants of OQCET, depending on the choices of field probe protocol, cluster Hamiltonian, the initial density matrix, and choices of constraining and jump operators. 

\subsection{Inverse linear response theory}
\label{sec:invlinearresponse}
Linear response theory provides a connection between dynamic correlation functions and system response to a weak external perturbation. For density response to a perturbing scalar field $\phi_{\bf r} (t)$, the linear response equation reads
\begin{equation}
   \langle \delta n_{\bf r}(t)\rangle =  \sum_{\bf r'}\int_0^t dt' \chi_{\bf r-r'} (t-t') \phi_{\bf r'}(t')
    \label{eq:linsusc_int}
\end{equation}
where
\begin{equation}
    \langle \delta n_{\bf r}(t)\rangle = \langle n_{\bf r}(t)\rangle - \langle n_{\bf r}(0)\rangle
\end{equation}
and
\begin{equation}
    \chi_{\bf r-r'} (t) =  - i \theta(t) \langle [n_{\sigma,\bf r}(t), n_{\sigma,\bf r'}(0)] \rangle_{0}
    \label{eq:chirrp}
\end{equation}
is the retarded charge susceptibility.

The subscript $0$ denotes that the expectation value is calculated at zero external field ($\phi_\textbf{r}(t)\equiv 0$).
If the field $\phi_{\bf r} (t)$ is of an appropriate form (a delta-like function, for example) then Eq.~\ref{eq:linsusc_int} can be inverted easily, and the susceptibility can be calculated from the non-equilibrium response. This approach is referred to as \textit{inverse linear response theory} \cite{Kovacevic2025}.

\par
\begin{figure*}[htbp!]
    \centering
    \includegraphics[width=12cm]{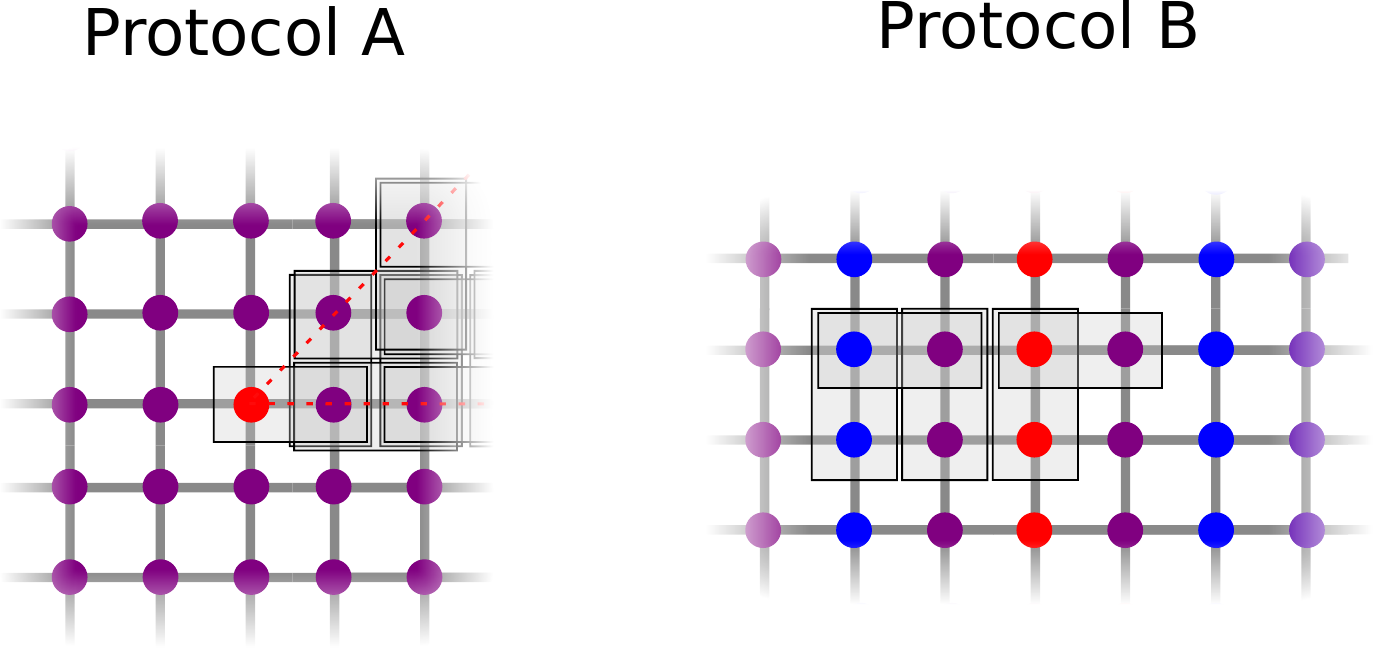}
    \caption{Schematic representation of non-equilibrium protocols. Boxes represent symmetry-irreducible 2$\times$1 clusters. Colors correspond to values of the potential $\phi_{\bf r}$: $+V$ (red), $0$ (purple), $-V$ (blue). The dashed lines in protocol A indicate the symmetry-irreducible part of the lattice. In protocol B, the wavelength represented is $\lambda=4$.}
    \label{fig:protocols}
\end{figure*}

 In our implementation we apply two different non-equilibrium protocols: one with the potential localized in real space (protocol A) and the other localized in momentum space (protocol B) (Fig. \ref{fig:protocols}).

The spatial Fourier transform of Eq.~\ref{eq:chirrp} is
\begin{eqnarray}
    \chi_{\bf q} (t) &&= \sum_{\sigma,\bf r} e^{-i\bf q \cdot r} \chi_{\sigma,\bf r-r'} (t)\\
    \chi_{\bf q} (t) &&= - i \theta(t) \langle [n_{\bf q}(t), n_{\bf -q}(0)] \rangle_{0}
    \label{eq:chiq_FT}
\end{eqnarray}
where
\begin{equation}
    n_{\bf q}(t) = \sum_{\sigma,\bf k} c^\dag_{\sigma,\bf k}(t)c_{\sigma,\bf k+q}(t).
\end{equation}
In protocol A we introduce a potential at $\bf r=0$
\begin{equation}
    \phi^{(\mathrm{A})}_{\bf r}(t) = \alpha \delta_{\bf r,0} \delta(t)  
\end{equation}
where $\delta(t)$ is a Dirac delta function in time. As the lattice is discrete, $\delta_{\bf r,0}$ is the Kronecker delta in space.
Using Eq.~\ref{eq:linsusc_int} we get
\begin{equation}
    \begin{split}
        \langle \delta n_{\bf r}(t)\rangle&= \sum_{\bf r'}\int_0^t dt' \chi_{\bf r-r'} (t-t') \alpha \delta_{\textbf{r}',\mathbf{0}} \delta(t')  \\
        \langle \delta n_{\bf r}(t)\rangle&= \alpha \chi_{\bf r} (t) .
    \end{split}
\end{equation}
$ \chi_{\bf q} (t) $ is then obtained as the Fourier transform of
\begin{eqnarray}
    \chi_{\bf r} (t) = \frac{1}{\alpha} \langle \delta n_{\bf r}(t)\rangle
\end{eqnarray}
For protocol B, we probe the system with a delta potential at wavevector $\bf q=q^*$,
\begin{equation}
    \phi^{(\mathrm{B})}_{\bf q}(t) = \alpha \left(\delta_{\textbf{q}^*,\bf q}+\delta_{-\textbf{q}^*,\bf q} \right)\delta(t)  .
\end{equation}
To ensure the potential in real space has no imaginary component and to preserve inversion symmetry we keep both $\bf q^*$ and $\bf -q^*$ terms. In real space, this is equivalent to
\begin{equation}
    \phi^{(\mathrm{B})}_{\bf r}(t) = 2 \alpha \cos{\left(\bf q^* \cdot r\right)}\delta(t)  .
\end{equation}
Inverting the linear response equations
\begin{equation}
    \begin{split}
        \langle \delta n_{\bf q}(t)\rangle &= \int_0^t dt' \chi_{\bf q} (t-t') \phi^{(\mathrm{B})}_{\bf -q}(t)(t') \\
        \langle \delta n_{\bf q}(t)\rangle&= \int_0^t dt' \chi_{\bf q} (t-t') \frac{\alpha}{2} \left(\delta_{\textbf{q}^*,\bf q}+\delta_{-\textbf{q}^*,\bf q} \right)\delta(t')   \\
        \langle \delta n_{\bf \textbf{q}}(t)\rangle&=  \alpha (\chi_{\bf q^*}(t)\delta_{\textbf{q}^*,\bf q} + \chi_{\bf -q^*}(t)\delta_{-\textbf{q}^*,\bf q})\\
        \langle \delta n_{\bf \textbf{q}^*}(t)\rangle&= \alpha \chi_{\bf q^*}(t).
    \end{split}
\end{equation}
This gives us $\chi_{\bf q^*}(t)$ as a function of the density response $\langle \delta n_{\bf q^*}(t)\rangle$, which can be obtained as a Fourier transform of $\langle \delta n_{\bf r}(t)\rangle$. Numerically, both protocols are implemented in real space.
\par In protocol A, we obtain the entire $\chi_{\bf q}(t)$ in one calculation, but the potential $\phi^{(\mathrm{A} )}_{\bf r'}(t)$ breaks all translation symmetries. This means we need to solve many independent clusters. In protocol B one calculation only gives us $\chi_{\bf q}(t)$ at $\bf q = q^*$, but we preserve translation symmetry in the direction orthogonal to $\bf q^*$. The presence of additional symmetries means that we have fewer independent clusters, making the individual calculations significantly less computationally expensive. The number of clusters needed for protocol A scales as $O(L^2)$, where L is the linear size of the system, while for protocol B the scaling is $O(L)$. In the special case where $\bf q^*$ is collinear with a lattice unit vector and the wavelength $\lambda=2\pi/q$ is an integer, the system also retains translation symmetry along $\bf q^*$ with a step of $\lambda$ lattice spacings. This means that we then require $O(\lambda)$ clusters, and we lose dependence on $L$, making calculations on large lattices easier.
\begin{figure}[htbp!]
    \centering
    \includegraphics[width=0.95\columnwidth]{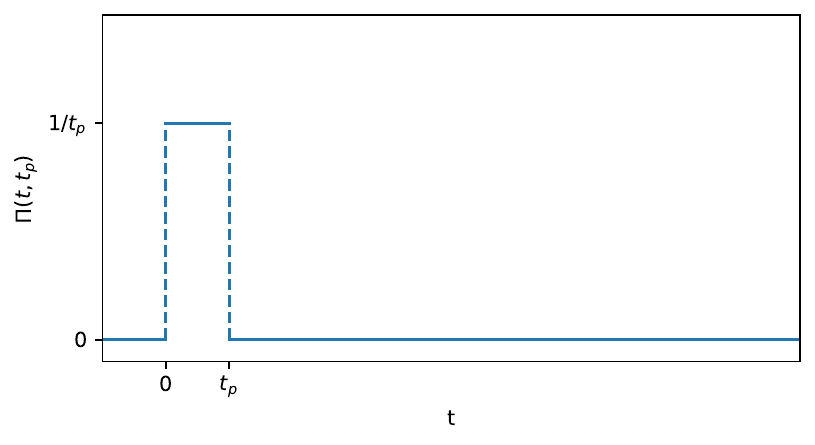}
    \caption{Boxcar function $\Pi(t,t_p)$}
    \label{fig:boxcar}
\end{figure}
\par In our implementation, since we work with finite time steps, we approximate the Dirac delta $\delta(t)$ with a boxcar function $\Pi(t,t_p)$
\begin{equation}
    \begin{split}
        \delta(t) &\approx \Pi(t,t_p)\\
        \Pi(t,t_p)&= 
        \begin{cases}
            \frac{1}{t_p},& \text{if } 0<t<t_p\\
            0,              & \text{otherwise}
        \end{cases}
    \end{split}
    \label{eq:boxcar}
\end{equation}
where $t_p$ is the pulse duration. Throughout the paper we use $t_p=0.001$ (in units of inverse bare half-bandwidth, $(4J)^{-1}$).
To ensure that the boxcar function is implemented correctly, we need to choose the time step $\Delta t$ such that $t_p$ is an integer multiple of $\Delta t$. This way, the field is constant during each time step, and we can use the equations of motion for a piecewise constant field ($\partial_t \phi = 0$ within each time step).
\subsection{Preparing the initial state}
\par To prepare the initial state for OQCET, we must obtain lattice expectation values of our constraining operators $\{\langle \hat{A}^\mathrm{latt}_\lambda (t=0) \rangle\}$ in an equilibrium calculation and initialize a cluster density matrix $\rho_\mathrm{clust}$ satisfying
\begin{equation} \label{eq:initial_self_consistency_condition}
\mathrm{Tr} \left[\rho_\mathrm{clust}(t=0) \hat{A}^{\mathrm{clust}}_\lambda\right]= \{\langle \hat{A}^\mathrm{latt}_\lambda (t=0) \rangle\}
\end{equation}
with $\partial_t \rho_\mathrm{clust}=0$.
\subsubsection{Thermal equilibrium lattice initial state}
\label{sec:lattice_eq_avg}
\par We obtain $\{\langle \hat{A}^\mathrm{latt}_\lambda (t=0) \rangle\}$ from CTINT\cite{Rubtsov2005}, a numerically exact quantum Monte Carlo method, implemented in the TRIQS library. \cite{Parcollet2015}.
\par
\begin{figure}[htbp!]
    \centering
    \includegraphics[width=\columnwidth]{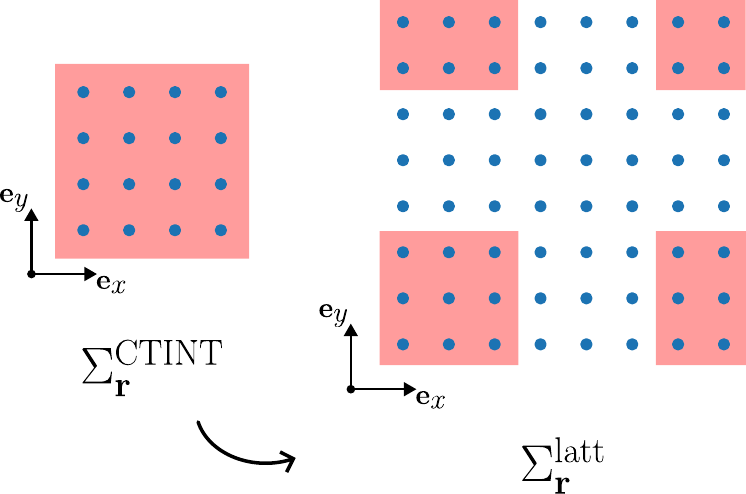}
    \caption{Mapping of CTINT self-energy (red) onto a larger lattice. The self energy for remaining sites is set to zero.}
    \label{fig:symm_proj}
\end{figure}

 The correlators we are interested in, the instantaneous fermionic bilinear $\langle c^{\dag}_{\sigma, \textbf{r}} c_{\sigma, \textbf{r}'}\rangle$ and double occupancy $\langle n_{\uparrow, \textbf{r}} n_{\downarrow, \textbf{r}}\rangle$ can both be obtained directly from the Green's function and self-energies in the imaginary-time formalism, without analytical continuation. The Matsubara Green's function is defined as
\begin{equation}
    G_{\sigma,ij}(\tau)= -\left\langle \mathcal{T}c_{\sigma,i}(\tau)c^\dag_{\sigma,j}(0) \right\rangle.
\end{equation}
The bilinear is simply the appropriate component of the Green's function at time $\tau=0^-$:
\begin{equation}
    \begin{split}
        G_{\sigma,ij}(\tau=0^-)&= -\left\langle \mathcal{T}c_{\sigma,i}(0^-)c^\dag_{\sigma,j}(0) \right\rangle \\&
        = \left\langle c^\dag_{\sigma,j}(0)c_{\sigma,i}(0^-) \right\rangle.
    \end{split}
\end{equation}
The equilibrium double occupancy $d \equiv \langle n_{\uparrow, \textbf{r}}(0) n_{\downarrow, \textbf{r}}(0)\rangle$ can be calculated from the well-known Migdal-Galitskii formula \cite{Krien2017}
\begin{equation}\label{eq:Migdal-Galitskii}
    \begin{split}
        d&=\frac{1}{U} \sum_{\textbf{k},i \omega_n}G_\textbf{k}(i \omega_n)\Sigma_\textbf{k}(i \omega_n)
        \\&= \lim_{\eta\rightarrow 0^-} \frac{1}{U} \sum_{\textbf{k},i \omega_n}G_\textbf{k}(i \omega_n)\Sigma_\textbf{k}(i \omega_n) e^{i \omega_n \eta}
    \end{split}
\end{equation}
In practice, the above sum is evaluated as an inverse Fourier transform $\mathcal{F}$ of the function $f(i\omega_n)=\sum_{\bf k}G_\textbf{k}(i \omega_n)\Sigma_\textbf{k}(i \omega_n)$ at time $\tau=0^-$,
\begin{equation}
    d = \mathcal{F}\left[\sum_{\bf k}G_\textbf{k}(i \omega_n)\Sigma_\textbf{k}(i \omega_n)\right](\tau=0^-).
\end{equation}
We start from numerically exact CTINT results for the self-energy. The CTINT data is then symmetrized to reduce Monte Carlo statistical noise by enforcing lattice symmetries. To study long wavelength response, we need to work with lattices of size $> 64$$\times$64. We cannot obtain the self-energy for such large lattices directly, as CTINT is limited to $\sim$8$\times$8 lattices. However, at high temperatures the self-energy is sufficiently short-ranged and the 8$\times$8 lattice self-energy is a good approximation. The CTINT self-energy can be mapped onto the larger lattice (Fig. \ref{fig:symm_proj}) by the formula
\begin{equation}   
    \Sigma_{\bf r}^\mathrm{latt}= 
    \begin{cases}
        \Sigma_{\bf r}^\mathrm{CTINT}& \text{if } |x| \leq \frac{1}{2} L^\mathrm{CTINT}  \text{ and } |y| \leq \frac{1}{2} L^\mathrm{CTINT} \\
        0,              & \text{otherwise}
    \end{cases}
\end{equation}
for a CTINT lattice of size $L^\mathrm{CTINT} \times L^\mathrm{CTINT}$.
\par We perform a Fourier transform in space to obtain $\Sigma^\mathrm{latt}_\mathbf{k}(i\omega_n)$. The full lattice Green's function is then calculated as
\begin{equation}
    G_\textbf{k}(i\omega_n)=\frac{1}{i\omega_n+\mu-\varepsilon_\textbf{k}-\Sigma_\textbf{k}(i\omega_n)}
\end{equation}
This procedure is somewhat analogous to the periodization step in cellular DMFT \cite{Vucicevic2018}, but the analogy is not complete as the cellular DMFT cluster is not cyclic.
\par Two-particle quantities such as the nearest-neighbor density-density correlator $n_{\sigma,\textbf{r}}n_{\sigma',\bf r+u}$ cannot be obtained just from the self energy, but can be measured directly in CTINT. In this case, we use the CTINT data directly, assuming that these quantities do not significantly depend on lattice size.

\subsubsection{Thermal equilibrium cluster initial state}
\label{sec:clust_init}

The initial cluster expectation values $\langle \hat{A}^\mathrm{clust}_\lambda (t=0) \rangle$ are given by the cluster density matrix $\rho_\mathrm{clust}$ as
\begin{equation} 
    \langle \hat{A}^\mathrm{clust}_\lambda (t=0) \rangle = \mathrm{Tr} \left[\rho_{\mathrm{clust}}(t=0) \hat{A}^\mathrm{clust} \right].
\end{equation} 
\par A natural way to prepare the cluster density matrix would be to create a reduced density matrix by tracing all lattice degrees of freedom outside the cluster (from the environment)
\begin{equation} \label{eq:redrho}
    \rho_{\mathrm{clust}}(t=0)=\mathrm{Tr}_{\mathrm{env}} [\rho_\mathrm{latt}(t=0)]
\end{equation}
where $\mathrm{Tr}_{\mathrm{env}}$ represents a partial trace of the environment degrees of freedom. This guarantees that cluster expectation values will match lattice expectation values for all possible single- and many-body operators in the initial state. The reduced density matrix is however computationally very difficult to obtain for large lattices. Another problem is that the reduced density matrix will generally not commute with the projected cluster Hamiltonian $H_\mathrm{clust}$, meaning that the EOM for $\rho_{\mathrm{clust}}$ will not be stationary even in absence of external fields. One possible way to address this is to introduce additional jump operators $\{\Tilde{L}_i\}$ and coupling $\{\Tilde{\Gamma}_i\}$ in Eq.~\ref{eq:lindblad_clust} such that $\rho_{\mathrm{clust}}$ presents a stationary solution for the corresponding Lindblad equation. We present a simplified implementation of such a scheme in Appendix \ref{sec:redrho}.
\par As it is not immediately clear how to choose $\Tilde{L}_i$ and $\Tilde{\Gamma}_i$ to satisfy stationarity, and the computation of a reduced density matrix on large lattices is difficult, we propose two additional schemes for setting $\rho_\mathrm{clust}(t=0)$. 

In the first scheme, we take that the denisty matrix represents a Boltzmann ensemble, i.e. a thermal state at the same temperature $T=1/\beta$ as the lattice, i.e.
\begin{equation}
 \rho_\mathrm{clust} = \frac{e^{-\beta H_\mathrm{clust}(\mu,J,U,...)}}{\mathrm{Tr} e^{-\beta H_\mathrm{clust}(\mu,J,U,...)}}
\end{equation}
The parameters of the cluster Hamiltonian $\mu,J,U,...$ are tuned so that we satisfy the self-consistency condition at the initial time, Eq.~\ref{eq:initial_self_consistency_condition}, i.e. the cluster expectation values of constraining operators must match the corresponding lattice expectation values. If we are only constraining $X_{i,j}$ and $d_i$, it is certain that by tuning $\mu$, $J$, $U$ we will be able to satisfy Eq.~\ref{eq:initial_self_consistency_condition}, as these parameters affect the density, as well as the kinetic and potential energy.
The cluster Hamiltonian then has the same terms as the lattice Hamiltonian, but the parameters are different.
If additional constraining operators are used, additional terms also have to be introduced into the cluster Hamiltonian (with additional parameters) to ensure enough flexibility of the representative model, as needed to satisfy the self-consistency criterion.
\par In the second scheme we start with an unmodified, projected lattice Hamiltonian. The density matrix is then constructed as a modified Boltzmann ensemble
\begin{equation}\label{eq:modified_boltzmann_ensemble}
    \begin{split}
        &\rho_{\mathrm{clust}}(t=0)=\\
        &\frac{1}{Z'}\sum_{\lambda,k \in \lambda} \ \sum_{\alpha,\{i,j\} \in \alpha} e^{-\beta E_\alpha \delta_{i,j} - \gamma_{\lambda,k} \bra{\Psi_{\alpha_i}} A_{\lambda,k} \ket{\Psi_{\alpha_j}}} \ket{\Psi_{\alpha_j}}\bra{\Psi_{\alpha_i}}    \\
        &Z' = \mathrm{Tr}\left[\sum_{\lambda,k \in \lambda} \ \sum_{\alpha,\{i,j\} \in \alpha} e^{-\beta E_\alpha \delta_{i,j} - \gamma_{\lambda,k} \bra{\Psi_{\alpha_i}} A_{\lambda,k} \ket{\Psi_{\alpha_j}}} \right]
    \end{split}
\end{equation}
where $\alpha$ are eigenspaces of the Hamiltonian, $E_\alpha$ are the associated eigenvalues and $\Psi_{\alpha_i}$ are vectors in the eigenspace $\alpha$. The index $k$ represents a sum over all components of the constraining operator tensors $\{ \hat{A}^\mathrm{latt}_\lambda \}$, so $A_{\lambda,k}$ is a scalar in the space of cluster indices (it is of course still a matrix in the cluster Hilbert space). The parameters $ \gamma_{\lambda,k}$ are again chosen to reproduce lattice expectation values for constraining operators. 
\par Projecting onto the eigenspaces guarantees that $[\rho,H]=0$. In the Hamiltonian eigenbasis, the density matrix has a block diagonal form $\rho = \bigoplus_\alpha \rho_\alpha$, with the blocks corresponding to eigenspaces $\alpha$. In each block $\alpha$ the Hamiltonian has a scalar form, $H_\alpha= E_\alpha I_{\mathrm{dim}(\alpha)}$, where $I_{\mathrm{dim}(\alpha)}$ is the identity matrix with dimension of the eigenspace $\alpha$. Since a scalar matrix commutes with every matrix, we have
\begin{equation}
    [H_\alpha,\rho_\alpha]=0.
\end{equation}
The commutator of block matrices is simply a block matrix of commutators of individual blocks, so
\begin{equation}
    [H,\rho]=\bigoplus_\alpha [H_\alpha,\rho_\alpha] = 0.
\end{equation}

Perhaps a more obvious approach might have been to try to modify the weights $w(\Psi)$ in the ensemble
\begin{equation}
    \rho_{\mathrm{clust}}= \sum_\Psi w(\Psi) \ket{\Psi}\bra{\Psi}
\end{equation}
where $\ket{\Psi}$ are system eigenstates. However, these eigenstates must be obtained numerically, and we cannot make sure to get $\ket{\Psi}$ that obey spatial or other symmetries. This means that tuning $w(\Psi)$ can break these symmetries, and that additional steps would have to be implemented to rotate the basis of each eigenspace so as to impose those symmetries.
The approach we take (as laid out in Eq.~\ref{eq:modified_boltzmann_ensemble}) respects symmetries by construction and is independent of the choice of basis for degenerate eigenstates. Our approach is therefore simpler, yet equally physically justified.

\subsection{Cluster dynamics}
\label{sec:ClusterDynamics}
\par As previously introduced, cluster dynamics are governed by the Lindblad equation
\begin{equation}
    \begin{split}
        \frac{d\rho(t)}{dt}=&-i[H_\mathrm{clust},\rho(t)]\\&
         +\sum_{l} \Gamma_l(t)\left(  L_l\rho(t)L^\dag_l-\frac{1}{2}\left\{      L^\dag_l L_l,\rho(t)        \right\}\right)
    \end{split}
    \label{eq:lindblad_eq}
\end{equation}
where we use the notation $\rho = \rho_\mathrm{clust}$ for brevity.
For some choices of jump operators, the first nonzero contributions to constrained operator expectation values appear at the second order.
For example, if we take $L=c^\dag_i$, the first order contribution to the current is zero at $t=0$. In equilibrium, $\frac{d\rho(t=0)}{dt}$ is purely real. $[H_\mathrm{clust},\rho(t=0)]$ is zero by definition, and both $\rho(t=0)$ and $L=c^\dag_0$ are real (in the cluster index basis $\ket{i}$). From this it follows that the first order contributions to the current, which are of the form $\mathrm{Im}\left(\mathrm{Tr}\left[ \Delta t \cdot \frac{d \rho}{dt} c^\dag_{i,\sigma}c_{j,\sigma}\right]\right)$, will be zero at $t=0$.
To account for this, Eq.~\ref{eq:lindblad_eq} is solved using the second order Taylor method
\begin{equation}
\rho(t_{i+1}) \approx \rho(t_i) + \Delta t \cdot \frac{d \rho}{dt}\Bigr|_{t=t_i} +\frac{1}{2} (\Delta t)^2 \cdot \frac{d^2 \rho}{dt^2}\Bigr|_{t=t_i} .
\label{eq:lindblad_euler_order_2}
\end{equation}
The coupling coefficients $\Gamma_l(t)$ are determined at each time step. Details of the implementation of the Lindblad equation and the optimization procedure can be found in Appendix \ref{sec:AppendixClust}.
\subsection{Choice of jump operators}
\label{sec:Lchoice}
\par In general, there are many possible choices for the set of jump operators $\{L_l\}$. There are, however, a few guiding principles for making this choice. Firstly, the solutions to the optimization problem should be unique, so the number of $\Gamma_l$'s should match the number of constraining operators (counting the real and imaginary parts of complex-valued expectation values separately).
\par Secondly, the jump operators and their $\Gamma_l$'s must respect symmetries of the system. For example, if sites $i$ and $j$ are equivalent by symmetry and the operator $L_i=c_{\sigma,i}$ is present, then $L_j=c_{\sigma,j}$ must also be included, with $\Gamma_j=\Gamma_i$. This reasoning also applies for spin symmetry (if present).
\par In the Lindblad equation, $\Gamma_l$ are non-negative real numbers. To make the optimization problem simpler, we can extend the domain of $\Gamma_l$ to the real axis ($\Gamma_l \in \mathbf{R}$) by pairing up conjugated operators $L_l$ and $\bar{L}_i$:
\begin{equation}
    \begin{split}
    \frac{d\rho(t)}{dt}&=-i[H_\mathrm{clust},\rho(t)] \\
    & +\sum_{l} \biggl[ |\Gamma_l|\theta(\Gamma_l) \left(  L_l\rho(t)L^\dag_l-\frac{1}{2}\left\{      L^\dag_l L_l,\rho(t)\right\}\right)  \\
    &+ |\Gamma_l|\theta(-\Gamma_l) \left(  \bar{L}_l\rho(t)\bar{L}^\dag_l-\frac{1}{2}\left\{      \bar{L}^\dag_l \bar{L}_l,\rho(t)        \right\}\right) \biggr]\\
    \end{split}
\end{equation}

\par In the simple case of 2$\times$1 clusters with constraints on $X_{\sigma,i,j}$ we have one complex and two purely real constraints (four unknowns). A good starting point for jump operators are the annihilation (and creation) operators of the single particle single-site and plane-wave states
\begin{equation}
    \left\{  c_{\sigma,0},\quad  c_{\sigma,1} , \quad c_{\sigma,k=0}, \quad c_{\sigma,k=\frac{\pi}{2}} \right\}_{\sigma\in \{\uparrow,\downarrow\}}
    \label{eq:2x1_X_Lset}
\end{equation} 
where 
\begin{equation*}
    c_{\sigma,k=0}\equiv c_{\sigma,0} + c_{\sigma,1} ,\quad  c_{\sigma,k=\frac{\pi}{2}}\equiv c_{\sigma,0} +i c_{\sigma,1}
\end{equation*}

\begin{figure}[htbp!]
    \centering
    \includegraphics[width=4cm]{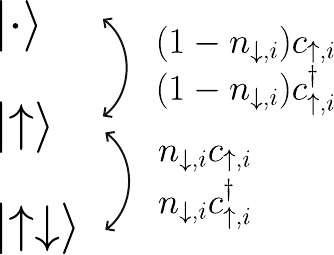}
    \caption{Effect of $n_{\bar{\sigma},i} c_{\sigma,i}$ and $(1-n_{\bar{\sigma},i}) c_{\sigma,i}$ on a lattice site.}
    \label{fig:n_jump_action}
\end{figure}

After introducing additional constraints in the form of the double occupancy $d_i$, it might be tempting to simply add the operators $c_{\sigma,i} c_{\sigma,i}$ to Eq.~\ref{eq:2x1_X_Lset}. In practice we find that such choice leads to problems; at some point in time evolution, a solution to the self-consistency condition no longer seems to exist, and no choice of $\{\Gamma_l\}$ can be found to satisfy Eq.~\ref{eq:gamma}. We interpret this as being related to the operators being insufficiently independent. For example, adding a particle to increase the density will at the same time increase the double occupancy, which we might not want. To try and alleviate the problem, we replace $c_{\sigma,i}$ operators and their conjugates $c^\dagger_{\sigma,i}$ by the following operators
\begin{equation}
    n_{\bar{\sigma},i} c_{\sigma,i}, \quad (1-n_{\bar{\sigma},i}) c_{\sigma,i}.
\end{equation}
The effect of these operators is illustrated in Fig. \ref{fig:n_jump_action}.
This choice of operators allows us to increase density while not affecting the double occupancy and \emph{vice versa}.

After some trial and error for the other operators, we arrive at the set
\begin{equation}
    \begin{split}
        \{ & n_{\bar{\sigma},0}c_{\sigma,0},\quad n_{\bar{\sigma},1} c_{\sigma,1} , \quad (1-n_{\bar{\sigma},0})c_{\sigma,0},\quad (1-n_{\bar{\sigma},1}) c_{\sigma,1} , \\
         & (1-n_{\bar{\sigma},0})(1-n_{\bar{\sigma},1})c_{\sigma,k=0}, \quad c_{\sigma,k=\frac{\pi}{2}} \}_{\sigma\in \{\uparrow,\downarrow\}}
    \end{split}
    \label{eq:2x1_d_Lset}
\end{equation} 
If we wish to introduce additional non-local density-density constraints ($n_{\sigma,i}n_{\sigma',j}$), we use the same jump operators as Eq.~\ref{eq:2x1_d_Lset}, adding
\begin{equation}
    \{(1-n_{i,\bar{\sigma}})c^\dag_{j,\sigma} \quad i\neq j \}_{\sigma\in \{\uparrow,\downarrow\}}
\end{equation}
\par Jump operators for other cluster sizes and constraint choices can be constructed in an analogous manner. In Appendix \ref{sec:appendixLset} we give the jump operator sets used for $2\times2$ clusters.
\subsection{Limits in which OQCET becomes exact}
\par OQCET is exact in the atomic limit $J = 0$, as the problem is reduced to a sum of individual Hubbard atoms. The solution to the optimization problem is then trivially $\{\Gamma_l=0 \quad \forall l\}$, as the individual atoms are decoupled.
\par In the non-interacting limit $U =0$, the situation is more complicated. Even at $U=0$, for the cluster evolution to follow the lattice EOMs nonzero $\{\Gamma_l\}$ are needed. This coupling introduces correlations into clusters, meaning that $Y_{\sigma,\bf r,r'}$ and $W_{\sigma,\sigma',\bf r,r'}$ are nonzero. $Y_{\sigma,\bf r,r'}$ no longer factors into the equations of motion for $X_{\sigma,\bf r,r'}$ (as it is multiplied by U), so Eq.~\ref{eq:bilinear_eom} reduces to the exact EOM for the non-interacting problem. 
This means that $X_{\sigma,\bf r,r'}$ is exact in this limit. In the lattice EOM for $d_{\bf r}$ (Eq.~\ref{eq:d_eom}), the $W_{\sigma,\sigma',\bf r,r'}$ term does not vanish since it is not multiplied by $U$. The nonzero contributions from $W_{\sigma,\sigma',\bf r,r'}$ mean that the results for the double occupancy are not exact even at $U=0$. However, in this limit the average Hamiltonian only depends on $X_{\sigma,\bf r,r'}$, giving us exact results for the total energy. Since we are ultimately interested in the susceptibility $\chi_{\bf q}(t) \propto \langle \delta n_{\bf q}(t)\rangle$, the fact that  $d_{\bf r}$ is not exact does not affect the final results.
\par OQCET also becomes exact in the limit of infinite cluster size, as the lattice EOM become equivalent to the Lindblad equation for the single remaining cluster with $\{\Gamma_l=0 \quad \forall l \}$. We note that in this limit, the Lindblad equations of motion for all three initial cluster state schemes (Section \ref{sec:clust_init}) trivially reduce to the exact solution, with $\rho_\mathrm{clust} = \rho_\mathrm{latt}$, $H_\mathrm{clust}=H_\mathrm{latt}$ and $[H,\rho_\mathrm{clust}(t=0)]=0$.
\begin{figure*}[t]
    \centering
    \includegraphics[width=0.9\linewidth]{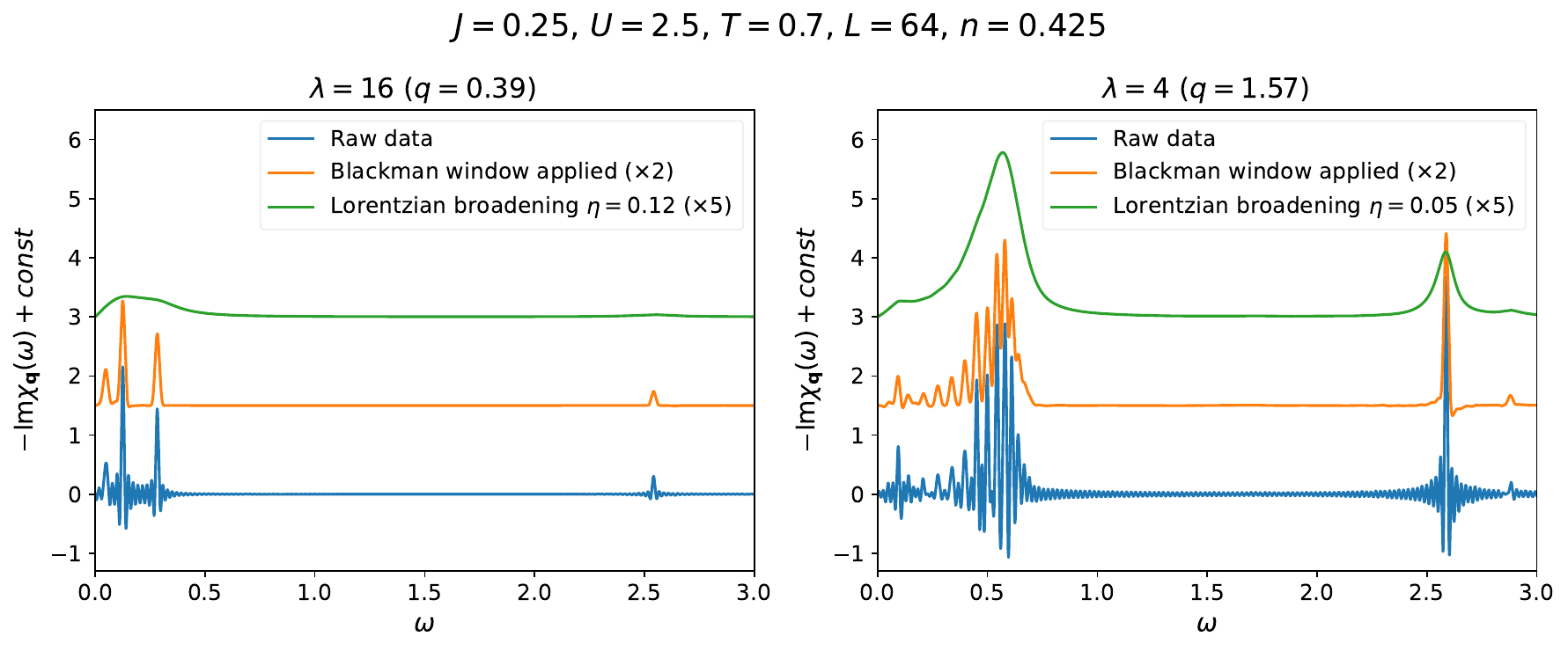}
    \caption{Comparison of raw and processed OQCET response for two values of wavelength $\lambda=2\pi/q$. To reduce noise in the raw data (blue) and recover the positions of peaks, a Blackman window is applied (orange, result multiplied by 2 for visibility). To obtain a spectrum for comparison with experiment, the raw data is broadened with a Lorentzian kernel (green, result multiplied by 5). At shorter wavelengths the peaks become more dense and smaller broadening is necessary. Results obtained with protocol B}
    \label{fig:postprocess}
\end{figure*}
\subsection{Post-processing}
\label{sec:postprocess}

\par It turns out that the response calculated in OQCET has roughly a discrete spectrum. Our understanding is that this is due to the fact that both the clusters and the lattice have finite size, meaning that the response has no long time decoherence. The coupling of the cluster to the environment can in principle provide the decoherence, but this effect turns out to be weak, at least in the non-equilibrium protocols we employ.
As we are never far away from equilibrium, $\Gamma_l$ tend to be small, and should in principle decay with time, as we approach equlibrium; yet, as $\Gamma_l$ become smaller, one gets less decoherence and this in turn precludes thermalization.
Nevertheless, this becomes a significant issue only when the system is probed at long wavelengths: the dynamics is then slow, the coupling to environment particularly weak, and we only get a response in a very narrow range of low frequencies; thus, the discrete structure of the spectrum becomes apparent.
A possible solution to this problem would be to use the reduced density matrix as the initial density matrix (the approach we have discussed in Section~\ref{sec:clust_init}). In this approach, one includes additional dissipative terms in the Linblad equation that do not decay with time providing a steady amount of decoherence. As this approach is currently infeasible for large lattices, we deal with the discretness of the spectra in a post-processing step.

\par The time evolution is necessarily performed up to a finite final time $t_\mathrm{max}$, and Fourier windowing with a Blackman window is used to obtain the frequency spectra~\cite{Harris1978}.
The window cutoff $t_{max}$ is taken to be long enough to resolve all peaks in the spectrum.
\par When comparing with experimental data, we instead apply Lorentzian broadening, as is standard in ED-based methods~\cite{Vucicevic2019}. To do so, we perform a convolution of the raw frequency response with the Lorentzian kernel
\begin{equation}
    f(\omega) = \frac{1}{\pi \eta \left[1+\left(\frac{\omega}{\eta}\right)^2\right]}
\end{equation}
where $\eta$ is the broadening width. An example of the post-processing steps can be seen in Fig. \ref{fig:postprocess}. As a consequence of particle number conservation $\lim_{\mathbf{q} \to 0} \chi_{\bf q} (\omega \neq 0) = 0$, so in the $\mathbf{q} \to 0$ limit, the broadening must go to zero. To ensure this we choose $\eta(q)$ as 
\begin{equation}
    \eta(q) = \min\left\{ \eta_{max} \sqrt{q}, \eta_{max} \right\}
\end{equation}
where $\eta_{max}$ is the maximum value of the broadening.

\section{Results}
\label{sec:Results}
In this section we present the results of OQCET calculations for the charge response in the square-lattice Hubbard model, as defined by Eq. \ref{eq:hubb_model}. We first benchmark different versions of OQCET, depending on choice of initial cluster state, constraining operators and experimental protocol, as well as cluster size. Next, we perform comparisons with other numerical methods. Finally, we compare OQCET results to a recent cold atom experiment by Brown et al. \cite{Brown2019}.

\subsection{Comparison with exact diagonalization}
\par The 2x2 Hubbard lattice is small enough to be solvable by exact diagonalization (ED). While results obtained from it cannot tell us much about long-wavelength response, it allows us to compare OQCET with numerically exact results.
\par
\begin{figure}
    \centering
    \includegraphics[width=\linewidth]{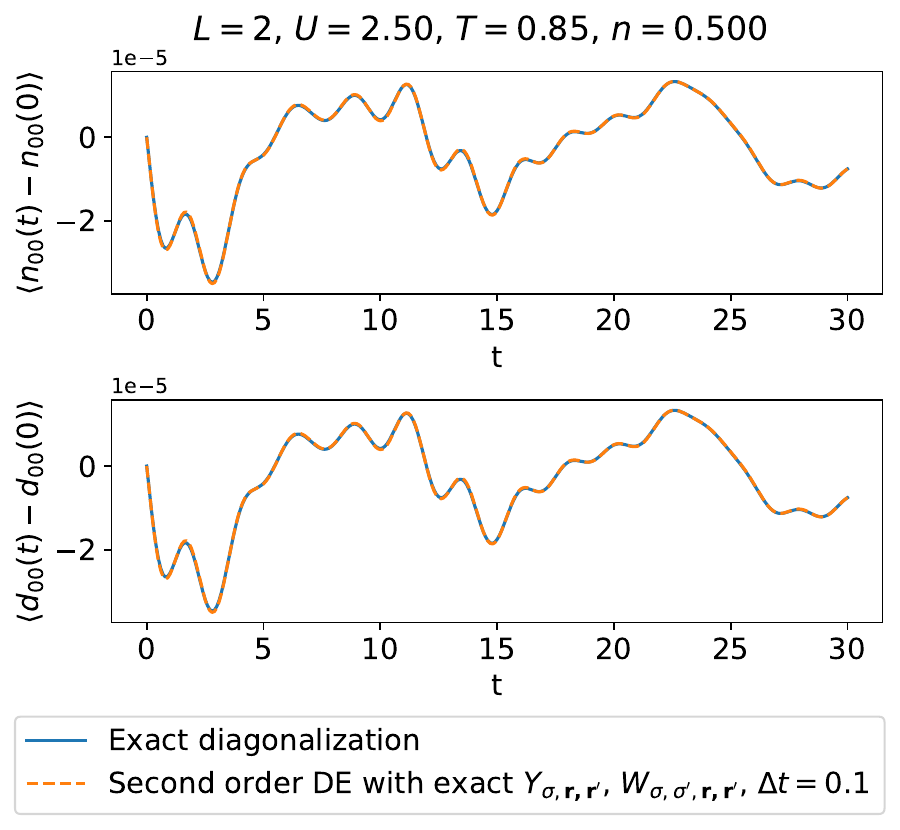}
    \caption{Comparison of density and double occupancy evolution for an interacting 2x2 lattice.}
    \label{fig:DE_ED_comparison}
\end{figure}

 To verify that the differential equations are correct in the interacting case, we compare results obtained by exact diagonalization on a 2x2 lattice with results from our implementation of the equations of motion, given the exact $Y_{\sigma,\bf r,r'}$ and $W_{\sigma,\sigma',\bf r,r'}$ extracted from ED (Fig. \ref{fig:DE_ED_comparison}). 
\par
\begin{figure}
    \centering
    \includegraphics[width=\linewidth]{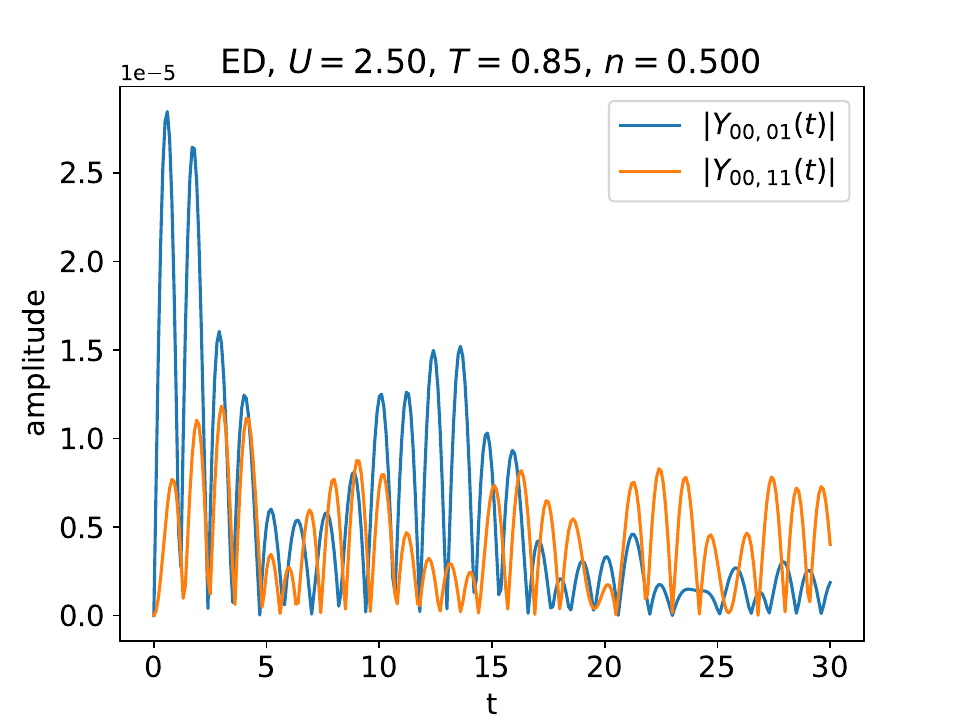}
    \caption{Values of nearest and next-nearest neighbor $Y_{\sigma,\textbf{r},\textbf{r}'}(t)$ on a 2x2 lattice}
    \label{fig:2x2_yrange}
\end{figure}

\begin{figure}
    \centering
    \includegraphics[width=\linewidth]{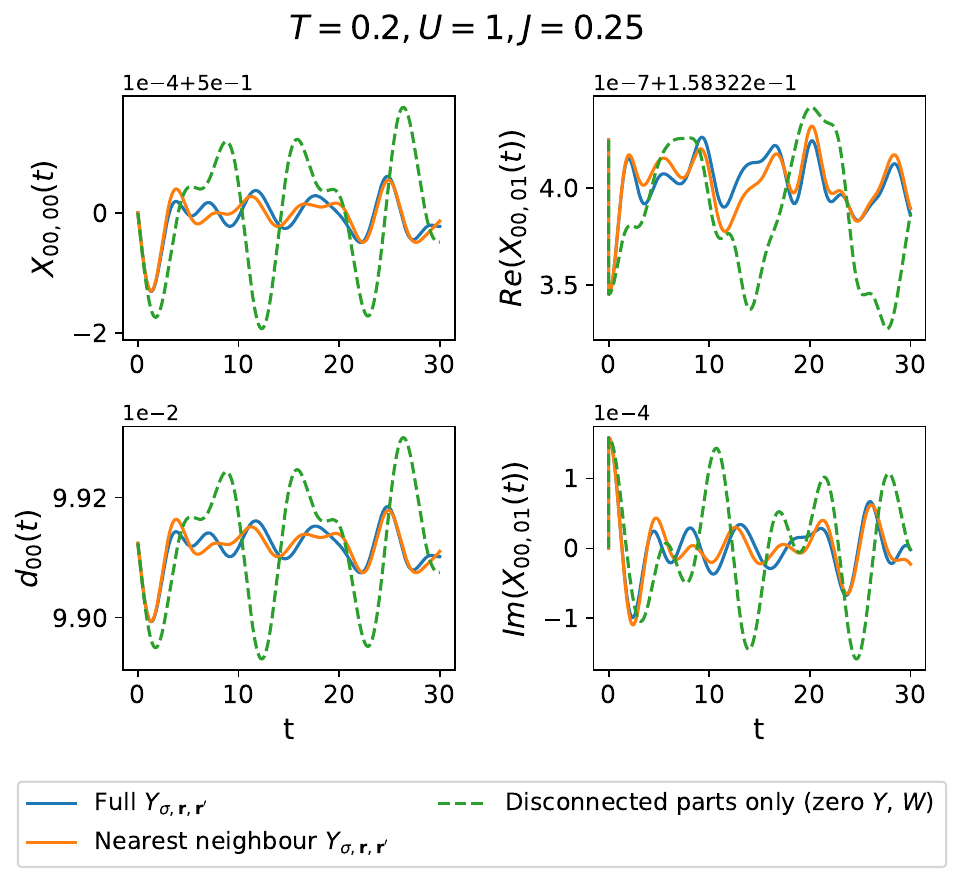}
    \caption{Comparison of expectation values between exact diagonalization results and results using truncated $Y_{\sigma,\textbf{r},\textbf{r}'}(t)$ on a 2x2 lattice. $X_{\sigma,\textbf{r},\textbf{r}'}$ and $d_{\bf r}$ are obtained from the lattice equations of motion.} 
    \label{fig:2x2_bench}
\end{figure}

 When using 2x1 clusters, $Y_{\sigma,\textbf{r},\textbf{r}'}(t)$ is truncated to its nearest-neighbor components. To check the validity of this approximation, we compare results for evolving $X_{\sigma,\textbf{r},\textbf{r}'}$ and $d_{\bf r}$ equations of motion using truncated $Y_{\sigma,\textbf{r},\textbf{r}'}(t)$ with exact results (Figs. \ref{fig:2x2_yrange} and \ref{fig:2x2_bench}). We see that the data are qualitatively similar, while the $Y_{\sigma,\textbf{r},\textbf{r}'}=0$ approximation gives significantly worse results. We see that the next-nearest neighbor Y is not always smaller than the nearest neighbor Y, but its effect on the EOM is apparently significantly less important.

\subsection{Benchmarking initial cluster state methods}

\par In Section \ref{sec:clust_init} we introduced three methods for preparing the initial density matrix and the cluster Hamiltonian: 
\begin{enumerate}
    \item a reduced density matrix with additional environment coupling needed to ensure stationarity in absence of external fields (see Appendix \ref{sec:redrho})
    \item a modified Hamiltonian with a thermal state
    \item a density matrix with a modified Boltzmann ensemble.
\end{enumerate}
We will refer to these three methods as \textit{reduced}, \textit{thermal} and \textit{weighted}, respectively.
\par
\begin{figure}
    \centering
    \includegraphics[width=\linewidth]{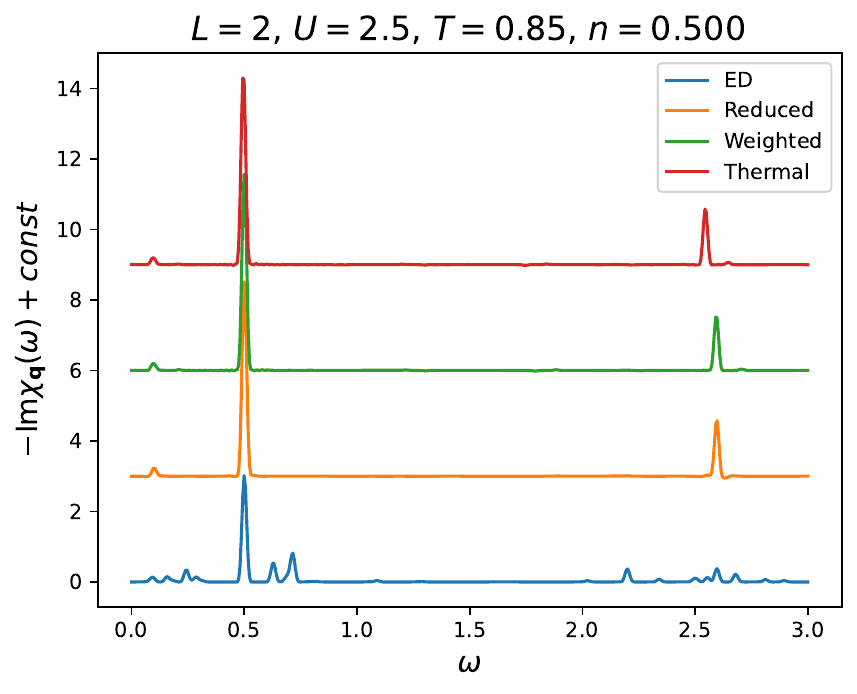}
    \caption{$-\mathrm{Im}\chi_{\bf q}(\omega)$ for different initial state schemes, obtained with protocol A. A Blackman window with $t_{max}=200$ has been applied to all curves, including ED.}
    \label{fig:init_comparions}
\end{figure}

 In Fig. \ref{fig:init_comparions} we compare the three methods on a 2$\times$2 lattice, which allows us to benchmark them against numerically exact results obtained from ED. We see that $\chi_q(\omega)$ does not significantly depend on the choice of method. In the remainder of the paper, we will use the weighted method to prepare initial states. 
\par The response spectrum for both ED and OQCET is discrete, as is expected for a small lattice. However, the OQCET spectrum features fewer peaks as the clusters used are smaller (2$\times$1). Nevertheless, OQCET captures the position of the main peak, as well as the presence of a Hubbard resonance at $\omega \approx U$.

\subsection{Bechmarking effects of constraints and cluster size}
\label{sec:clustsize}
\begin{figure}
    \centering
    \includegraphics[width=\linewidth]{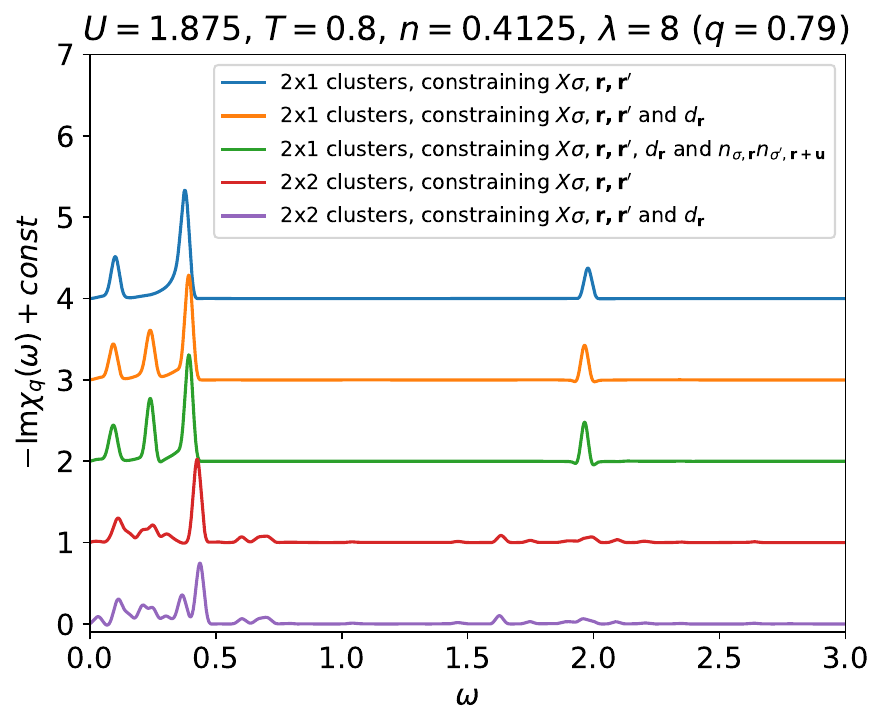}
    \caption{$-\mathrm{Im}\chi_{\bf q}(\omega)$ for different cluster sizes and choices of constraining operators, obtained with protocol B.}
    \label{fig:2x1_2x2_X_d_nn_comparison}
\end{figure}
\par Cluster evolution in OQCET is constrained by operator expectation values $\{\langle \hat{A}_\lambda \rangle \}$. The choice of operators $\{\hat{A}_\lambda\}$ is arbitrary to a degree\textemdash charge response is given as $\langle n_i(t) - n_i(t=0)\rangle$, so it is necessary to place a constraint on the density $\langle n_i(t) \rangle$. If we want conservation of energy in our theory it is necessary to include constraints on both $X_{\sigma,i,j}$ and $d_{i}$. In Fig. \ref{fig:2x1_2x2_X_d_nn_comparison} we compare the charge response for three sets of constraining operators. In the first, we only constrain $X_{\sigma,i,j}$. For the second, we constrain both $X_{\sigma,i,j}$ and $d_{i}$. We see that the latter choice introduces an additional peak in the spectrum. Finally, we include additional constraints on the nearest neighbor density-density correlators $n_{\sigma,\bf r}n_{\sigma',\bf r+u}$, but this does not qualitatively change the spectrum.

\begin{figure}
    \centering
    \includegraphics[width=\linewidth]{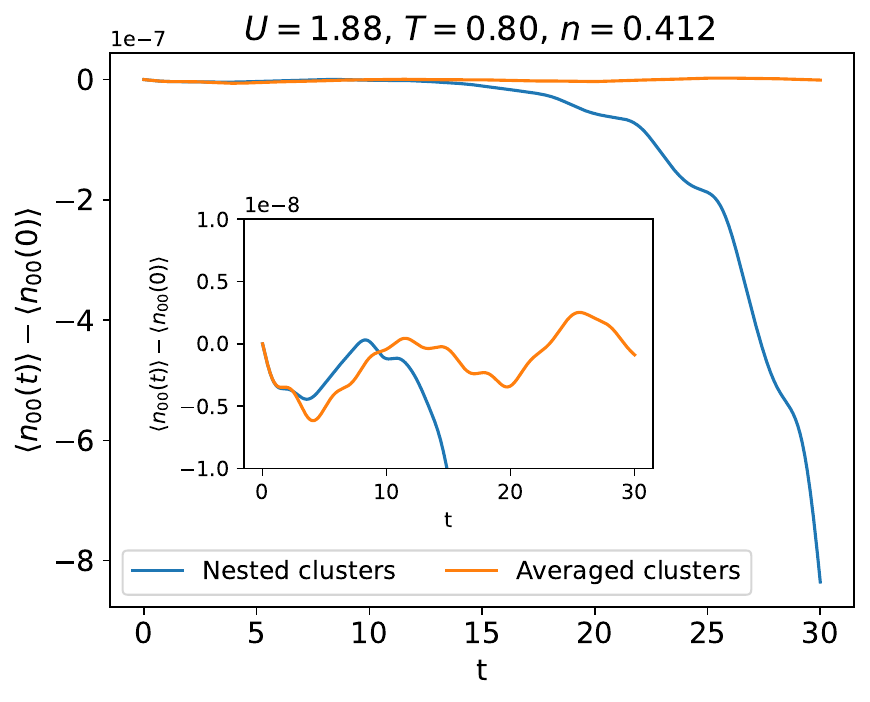}
    \caption{Density response for different 2$\times$2 overlap schemes, obtained by protocol B. The response for the nested scheme starts to diverge away from equilibrium at $t\approx10$.}
    \label{fig:nested_avg_comp}
\end{figure}

\par To see the effect of cluster size on the charge response, we compare results for 2x1 and 2x2 clusters with constraints on $X_{\sigma,i,j}$. If we try to apply a nested cluster scheme, the system response very quickly diverges (Fig. \ref{fig:nested_avg_comp}), giving unphysical results. A possible reason for this is that $Y_{\sigma,\bf r, r'}(\omega)$ has a discrete spectrum, the contributions from the 2x2 and 2x1 clusters do not cancel out properly; in general, subtracting two discrete spectra with peaks whose positions do not perfectly match results in a non-causal spectrum.  This non-causality might drive the system further and further out of equilibrium. If instead we average contributions from overlapping 2x2 clusters, we get a well-behaved response (Figs. \ref{fig:nested_avg_comp} and \ref{fig:2x1_2x2_X_d_nn_comparison}), although, as already mentioned, we do not observe thermalization at long times, which is related to the discreteness of the frequency spectrum of the response.
\begin{figure}
    \centering
    \includegraphics[width=\linewidth]{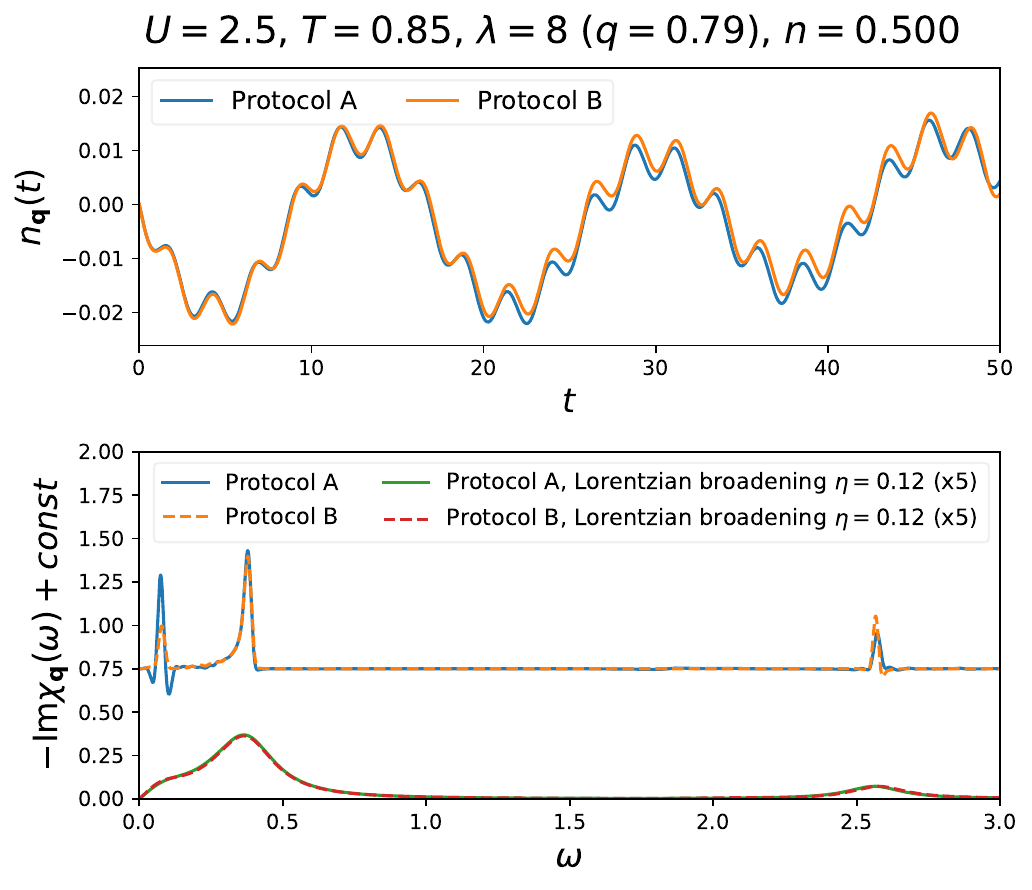}
    \caption{$n_{\bf q}(t)$ and $-\mathrm{Im}\chi_{\bf q}(\omega)$ for non-equilibrium protocols using 2$\times$1 clusters. }
    \label{fig:protocolAvsB}
\end{figure}

\subsection{Bechmarking non-equilibrium protocols}

\begin{figure*}[ht]
    \centering
    \includegraphics[width=0.9\linewidth]{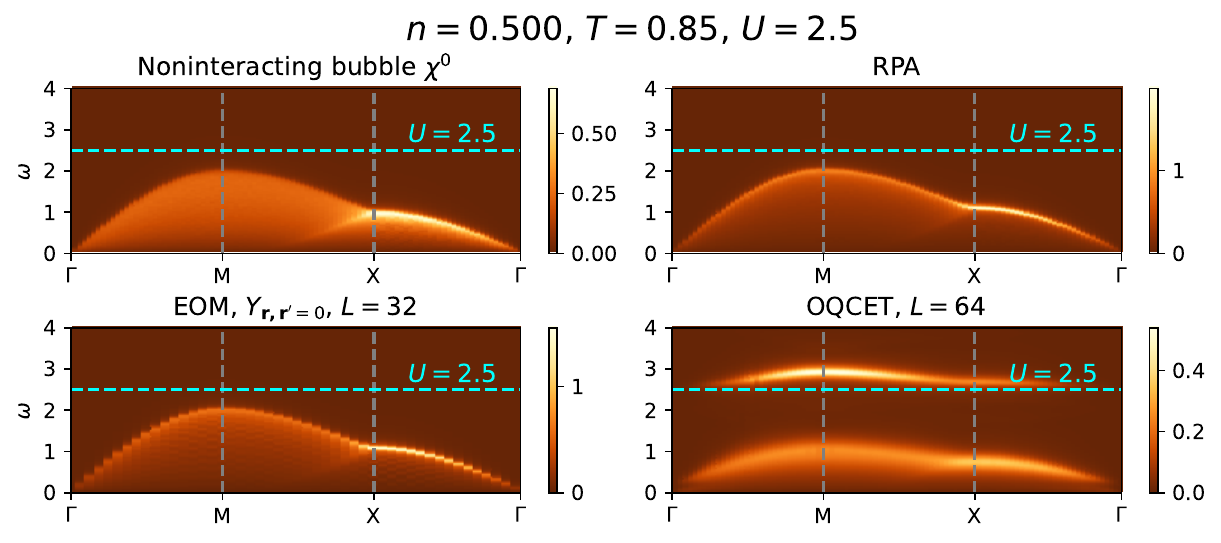}
    \caption{$-\mathrm{Im}\chi_{\bf q}(\omega)$ at half filling for several methods. EOM refers to disconnected equations of motion (Eq. \ref{eq:bilinear_eom} with connected parts set to zero). To calculate the entire spectrum OQCET results were obtained by protocol A using 2x1 clusters.}
    \label{fig:Hafermann_comp}
\end{figure*}
\begin{figure*}[!ht]
    \centering
    \includegraphics[width=\linewidth]{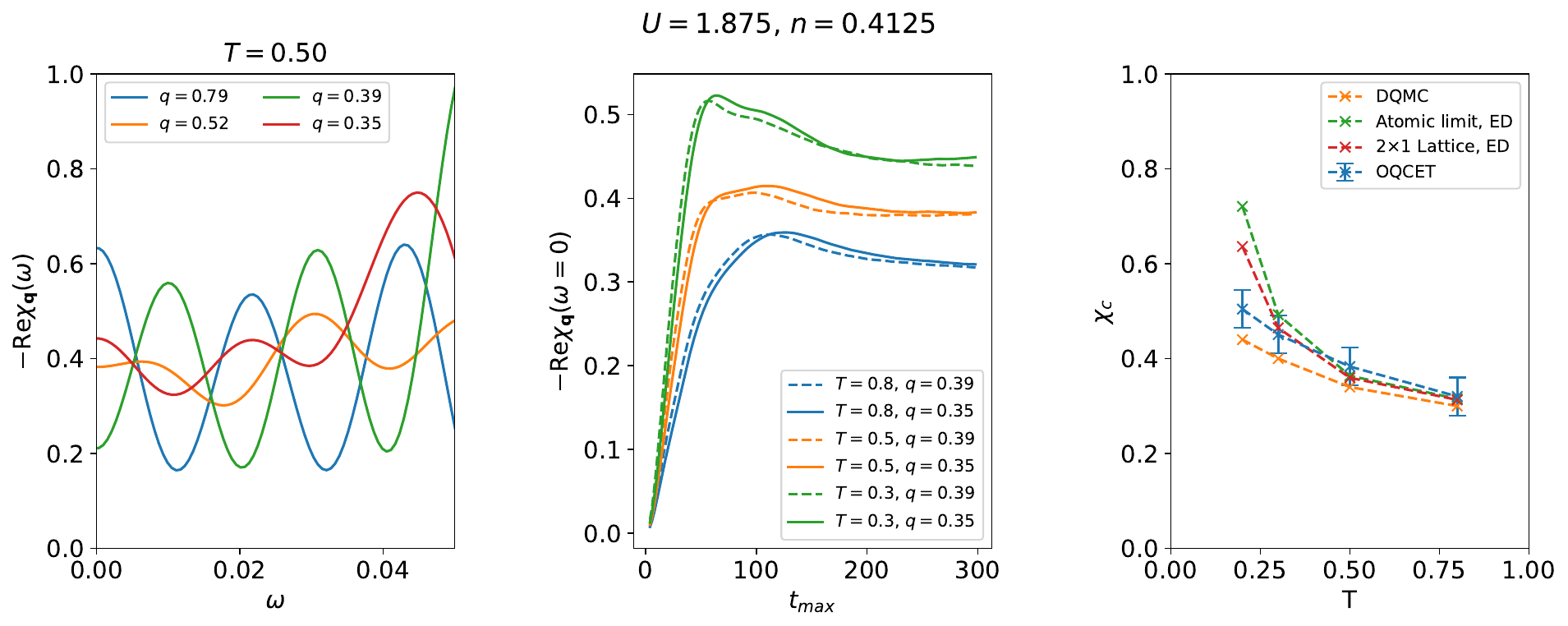}
    \caption{(a) Real part of charge susceptibility for different values of $\textbf{q}$ in the raw data. (b) $- \chi_\textbf{q} (\omega=0)$ as a function of Blackman window size (cutoff $t_{max}$) for different values of $T$ and $\bf q$. (c) Comparison of $\chi_c(T)$ between OQCET, numerically exact DQMC results, and exact diagonalization results for a 2$\times$1 lattice and the atomic limit. DQMC results have been taken from Brown et al.\cite{Brown2019}. All OQCET results were obtained for 2$\times$1 clusters using protocol B.}
    \label{fig:chic}
\end{figure*}

In Fig.~\ref{fig:protocolAvsB} we see that protocols A and B give qualitatively similar results. The initial density response ($\chi_{\bf q}(t)$ for small values of $t$) is nearly identical; after Lorentzian broadening, so are the spectra.
Since embedded cluster theories break momentum conservation (as discussed in Section \ref{sec:OQCETHubb}),
the response at any the given wave vector is expected to depend on the spatial distribution of the external probe. The presence of peaks with negative spectral weight is discussed in Appendix \ref{sec:negweight}.

\subsection{Comparison with other numerical methods}
\par In Fig. \ref{fig:Hafermann_comp} we compare results for $\mathrm{Im} \chi_\textbf{q} (\omega)$ between several methods.
The noninteracting (retarded) bubble susceptibility $\chi_{0,\bf q}(\omega)$ can be obtained by analytical continuation of the expression\cite{Hafermann2014}
\begin{equation}
    \chi_{0,\bf q}(i\Omega_\nu) = 2\sum_{\mathbf{k}, i\omega_n} G_{0, \bf k+q} (i\omega_n) G_{0, \bf k} (i\omega_n + i\Omega_\nu)
\end{equation}
where $i\Omega_\nu$ and $i\omega_n$ are bosonic and fermionic Matsubara frequencies, respectively. The RPA susceptibility is given as
\begin{equation}
    \chi_{\mathrm{RPA},\bf q}(\omega)= \frac{\chi_{0,\bf q}(\omega) }{1-U \chi_{0,\bf q}(\omega) }
\end{equation}

If we set the connected correlators $Y_{\sigma,\bf r,r'}$ and $W_{\sigma,\sigma',\bf r,r'}$ to zero, the equations of motion Eq. \ref{eq:bilinear_eom} reduce to the time-dependent Hartree-Fock EOM
\begin{equation} \label{eq:hf_eom}
    \begin{aligned}
        \partial_{t} X_{\sigma, \mathbf{r r}^{\prime}}  = i \bigg(&-J \sum_{\mathbf{u} \in\{\pm\mathbf{e}_{x},\pm\mathbf{e}_{y} \}} ( X_{\sigma, \mathbf{r}+\mathbf{u}, \mathbf{r}^{\prime}}-X_{\sigma, \mathbf{r}, \mathbf{r}^{\prime}+\mathbf{u}} )\\
        &+\left(\phi_\textbf{r} - \phi_{\textbf{r}'}\right) X_{\sigma, \textbf{r},\textbf{r}'}\\
        &+U \left(\left(X_{\sigma,\textbf{r},\textbf{r}}-X_{\sigma,\textbf{r}',\textbf{r}'}\right)X_{\sigma,\textbf{r},\textbf{r}'}\right) \bigg) .
    \end{aligned}
\end{equation}
In this case, the susceptibility most resembles RPA results, which is expected, as RPA can be obtained as the linear response of time-dependent Hartree-Fock~\cite{Nakatsukasa2007}. The results do not match exactly, due to the fact that the equilibrium $X_{\sigma,\textbf{r},\textbf{r}'}$ used in the EOM is not the same as the noninteracting $G_{0, \bf r,r'}$ used in the RPA calculation.
In the full OQCET solution we see a renormalization of the lower band energy and the presence of a Hubbard resonance at $\omega \approx U$. The Hubbard resonance is expected to be present, as can be seen in DMFT results from Hafermann et al. \cite{Hafermann2014}. The lower band is absent in the DMFT results, but this could be explained by the significantly higher temperature in OQCET calculations. Away from half-filling however, we expect to see both bands, even for large values of $U$.

\begin{figure*}[t]
    \centering
    \includegraphics[width=0.80\linewidth]{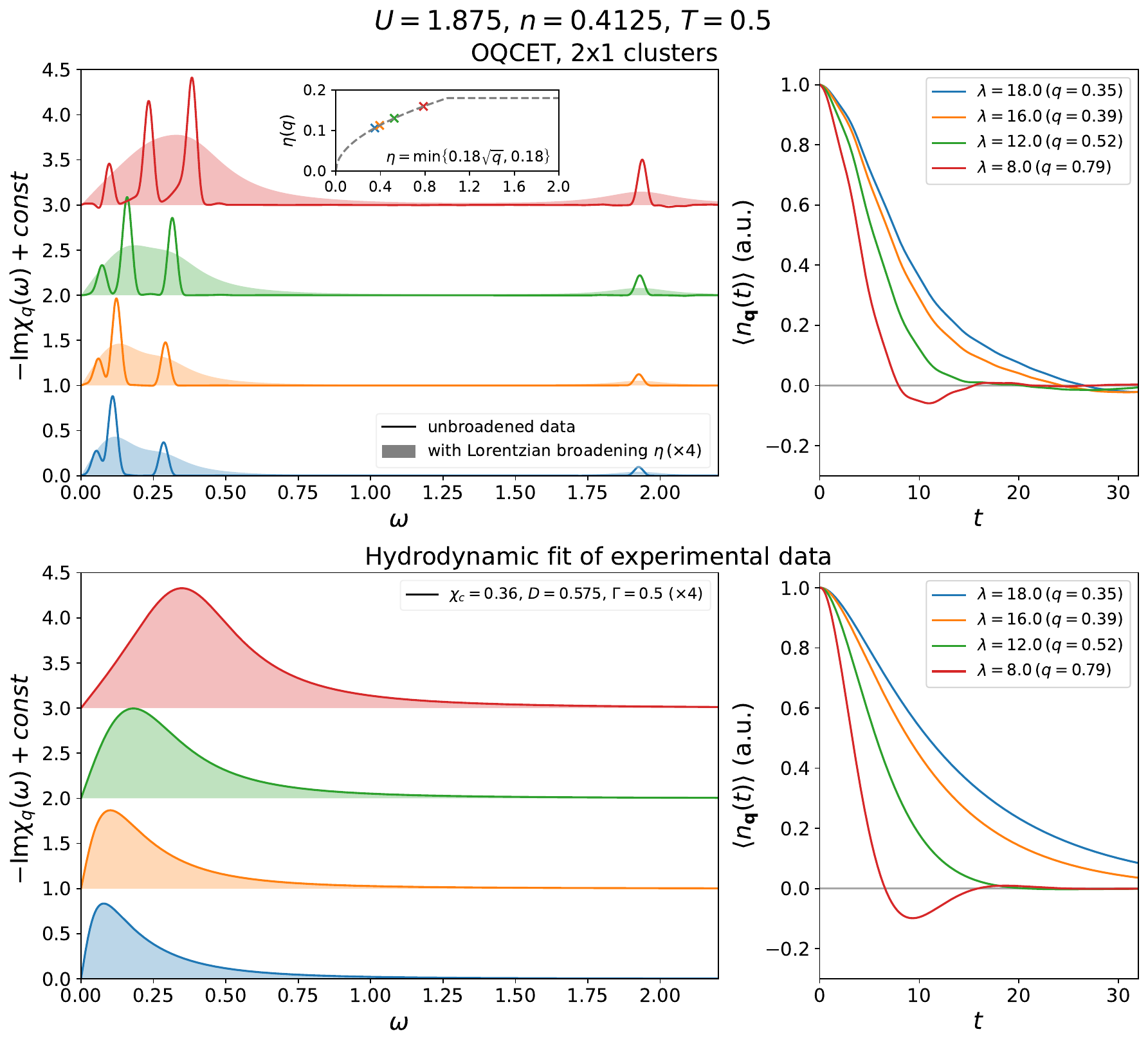}
    \caption{$-\mathrm{Im}\chi_{\bf q}(\omega)$ (left panels) and experimental protocol response (right panels) for different values of the wavelength $\lambda$ (wavevector $q$). Hydrodynamic ansatz data is taken from Brown et al.\cite{Brown2019}. The Lorentzian broadening $\eta$ as a function of wavevector $q$ is shown in the inset. Results obtained with protocol B.}
    \label{fig:imchiqw_q}
\end{figure*}
The charge compressibility $\chi_c = \partial n / \partial \mu$ can be obtained as the limit of the retarded susceptibility
\begin{equation}
    \chi_c =- \lim_{\textbf{q} \to 0}  \chi_\textbf{q} (\omega=0)
\end{equation}

Since we stop the calculations at a finite time, the raw response is oscillatory (Fig. \ref{fig:chic}), and is as such unsuitable for calculating $\chi_c$. Instead, the value of $\chi_c$ (Fig. \ref{fig:chic}) is taken from Blackman $\chi_{\textbf{q}}(\omega=0)$, converged with $t_{max}$ and $\bf q$. Results for $2\times2$ clusters are not shown as the response does not behave well enough at the long times necessary to get convergence with window size. The error is estimated from the dependence of $\chi_{\textbf{q}}(\omega=0)$ on $t_\mathrm{max}$ and $\bf q$.
\par The compressibility $\chi_c$ is a static quantity, so it can be calculated numerically exactly by Monte Carlo methods such as DQMC. In Fig. \ref{fig:chic} we see good agreement between OQCET and DQMC results. OQCET overestimates $\chi_c$ somewhat, but the results present an improvement compared to the corresponding isolated cluster ED calculation; with increasing temperature, results tend to the atomic limit, as expected.

\begin{figure*}[!ht]
    \centering
    \includegraphics[width=\linewidth]{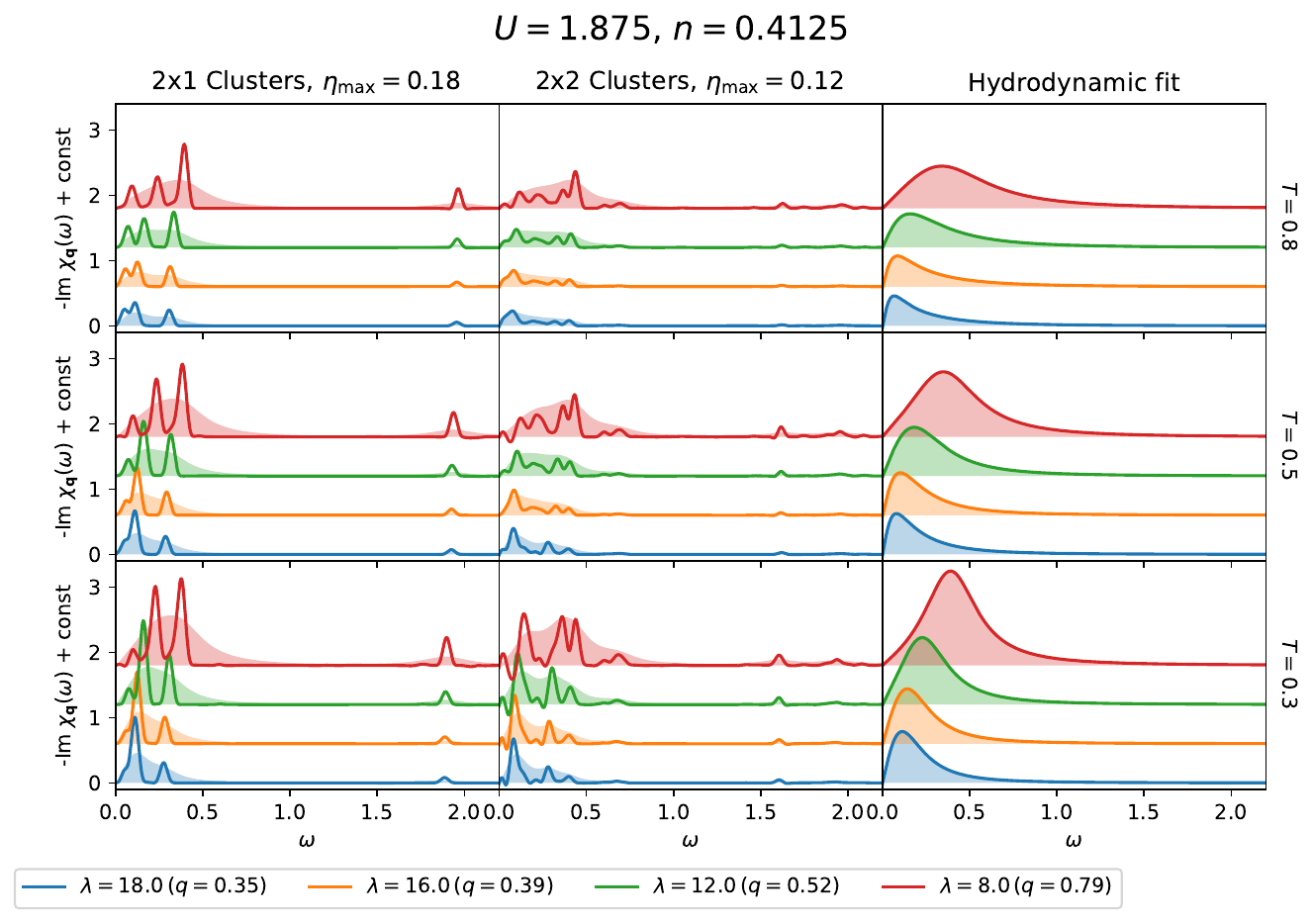}
    \caption{$-\mathrm{Im}\chi_{\bf q}(\omega)$ as a function of the wavelength $\lambda$ (wavevector $q$) and cluster size. OQCET results are compared with the hydrodynamic ansatz fit from Brown et al.\cite{Brown2019}. For OQCET results, Blackman spectra ($t_{\mathrm{max}} = 150$) are represented by a solid line, while Lorentzian broadened spectra are shaded. The Lorentzian broadened and hydrodynamic $\chi_{\bf q}$ have been multiplied by a factor of 3 for visibility. The 2$\times$2 spectrum is denser, so a smaller $\eta_\mathrm{max}$ is used. Results obtained with protocol B.}
    \label{fig:imchiqw_q_full}
\end{figure*}
\subsection{Comparison with experimental results}
\par In the experiment performed by Brown et al.~\cite{Brown2019}, lithium atoms are placed in an optical lattice simulating the Hubbard model with an external sinusoidal potential
\begin{equation}
    H \rightarrow H + V \int d\textbf{r} \sin{\left(x q^*\right) n\left(\textbf{r}\right)} \theta(-t)
\end{equation}
After first thermalizing the system in the external potential, the potential is switched off at $t=0$, and the response in the experimental protocol is given by 
\begin{equation}
    \begin{split}
        n_\textbf{q}(t) &= \int_{-\infty}^t dt' \chi_\textbf{q}(t-t') \theta(-t') \\
        &= \int_t^\infty \chi_\textbf{q}(t')dt'
    \end{split}
    \label{eq:hydro_ansatz}
\end{equation}
Certain assumptions about hydrodynamic behavior at the longest length and time scales yield an ansatz for the charge susceptibility at long wavelengths and low frequencies \cite{Brown2019,Vucicevic2023}
\begin{equation}
    \chi_\mathbf{q}(\omega)=\frac{\chi_c}{1-\frac{i \omega}{q^2D}-\frac{\omega^2}{q^2D\Gamma}}
    \label{eq:hydro}
\end{equation}

where $\chi_c$ is the charge compressibility, $D$ is the diffusion constant and $\Gamma$ is the momentum relaxation rate. The experiment was performed for a range of temperatures and wavelengths of the external potential, and the response was fit to the ansatz Eq. \ref{eq:hydro_ansatz} to obtain the parameters $\chi_c$, $D$ and $\Gamma$. The experiment found excellent agreement between experimental data and the hydrodynamic ansatz.

\par In Figs. \ref{fig:imchiqw_q} and \ref{fig:imchiqw_q_full} we compare OQCET to experimental results from Brown et al. \cite{Brown2019}. The raw experimental data is not available, so OQCET results are compared with the hydrodynamic ansatz Eq. \ref{eq:hydro} using $\chi_c$, $D$ and $\Gamma$ extracted from experiment. The trends with wavelength and temperature in OQCET are in qualitative agreement with the hydrodynamic theory. The Hubbard peak at $\omega \approx U$ present in OQCET cannot be described by the ansatz. In the time domain, this peak corresponds to high frequency oscillations that cannot be seen in the experiment due to the finite time resolution and relatively large error bars. However, this peak is expected to be present, as it can be seen in the optical conductivity spectrum \cite{Kokalj2017} and the spectral function \cite{Vucicevic2015}.

We note that, on the OQCET side, using larger clusters yields a larger number of peaks; for this reason we have used a smaller maximum broadening $\eta_{\mathrm{max}}$ when using $2\times 2$ clusters than when using $2\times 1$ clusters.

\section{Conclusions}
\label{sec:Conclusions}
\par We have formulated, developed and tested OQCET, an embedded cluster method, based on the exact solution of the Lindbladian dynamics of small open quantum clusters. OQCET becomes formally exact in the atomic, noninteracting and infinite cluster size limits, obeys conservation laws, allows computations for large lattices, and avoids analytical continuation.
Different variants of the method can be formulated depending on the specifics of the impurity and lattice equations (the choice of constrained and jump operators, the initial density matrix and the impurity Hamiltonian, and the choice of non-equilibrium protocol). We benchmark the differences in the results obtained by the different variants of the method and identify the most optimal choices for the computation of the quantity of our interest, which is the charge-charge correlation function.
\par Due to finite lattice and cluster sizes, OQCET yields a discrete charge-fluctuation spectrum, which is especially evident at long wavelengths. Discrete spectrum means a non-decaying response in real time. Our calculation is formulated in real-time and performed up to a finite time; frequency dependence is obtained by a Fourier transform. To compare with experimental results, we employ Lorentzian broadening, as is standard in ED-based methods. We are able to probe large lattices and response at wavelengths of order 20 lattice spacings, which is not possible with any straightforward application of ED (or even FTLM). Qualitative agreement with the experiment shows that OQCET is a promising method for the computation of dynamical response functions in strongly correlated lattice models.
\par Further work is needed to develop a way to ensure decoherence of the response at long times, which is necessary to obtain a continuous spectrum. One possible route is to try and use embedded clusters of increased size, but this would likely require further algorithmic optimizations (or even approximations) at the level of the Lindblad equations. Nevertheless, OQCET can be readily generalized for the computation of the dynamic spin susceptibilities and other two-particle response functions. It can be used to compute even the single-particle spectral function, by applying an appropriate non-equilibrium protocol and formulating the equations in Nambu space (as needed for an induced breaking of particle-number conservation by external field). Finally, OQCET can be straightforwardly formulated in momentum space to directly investigate the response to uniform electric fields (e.g. using protocols outlined in \cite{Kovacevic2025}), but we leave this for future work.

\begin{acknowledgments}
We acknowledge useful discussions with Michel Ferrero. The CTINT data was provided by Michel Ferrero. Computations were performed on the PARADOX supercomputing facility (Scientific Computing Laboratory, Center for the Study of Complex Systems, Institute of Physics Belgrade). J.V.\ acknowledges funding provided by the Institute of Physics Belgrade, through the grant by the Ministry of Science, Technological Development and Innovation of the Republic of Serbia. J.V.\ and P.B.\ acknowledge funding by the European Research Council, grant ERC-2022-StG: 101076100.
H.U.R.S.\ acknowledges financial support from the Swedish Research Council (Vetenskapsrådet, VR) grant number 2024-04652 and funding from the European Research Council (ERC) under the European Union’s Horizon 2020 research and innovation programme (grant agreement No. 854843-FASTCORR).
\end{acknowledgments}

\appendix

\section{Lattice equations of motion}
\label{sec:AppendixEOM}
In this section we will be working in the Heisenberg picture, and we omit time-dependence in notation for brevity, $c^{\dag}_{\sigma, \textbf{r}}\equiv c^{\dag}_{\sigma, \textbf{r}}(t)$. The Heisenberg equation for $c^{\dag}_{\sigma, \textbf{r}} c_{\sigma, \textbf{r}'}$ reads
\begin{equation}
\partial_t (c^{\dag}_{\sigma, \textbf{r}} c_{\sigma, \textbf{r}'}) = i \left[H,c^{\dag}_{\sigma, \textbf{r}} c_{\sigma, \textbf{r}'}\right]
\end{equation}
The Hamiltonian in Eq.~\ref{eq:hubb_model} can be split into three terms, $H=H_\mathrm{kin}+H_\mathrm{pot}+H_\mathrm{int}$. The kinetic energy term gives us
\begin{equation}
    \begin{split}
        &i \left[H_\mathrm{kin},  c^{\dag}_{\sigma, \textbf{r}} c_{\sigma, \textbf{r}'}\right] \\
    &= -iJ \sum_{\textbf{r}'',\mathbf{u} \in\{\pm\mathbf{e}_{x},\pm\mathbf{e}_{y}\}, \sigma'}\left[ c^{\dag}_{\sigma', \textbf{r}''} c_{\sigma', \textbf{r}''+\textbf{u}},c^{\dag}_{\sigma, \textbf{r}} c_{\sigma, \textbf{r}'}\right]
    \end{split}
\end{equation}
applying the identity $\left[a^\dag b, c^\dag d\right]=\delta_{bc} a^\dag d - \delta_{ad} c^\dag b$
\begin{equation}
    \begin{split}
        i &\left[H_\mathrm{kin},c^{\dag}_{\sigma, \textbf{r}} c_{\sigma, \textbf{r}'}\right] \\
        &= -iJ \sum_{\textbf{r}'',\mathbf{u}, \sigma'}\delta_{\sigma,\sigma'}\left( \delta_{\textbf{r}''+\textbf{u},\textbf{r}} c^{\dag}_{\sigma', \textbf{r}''} c_{\sigma, \textbf{r}'} - \delta_{\textbf{r}'',\textbf{r}'} c^{\dag}_{\sigma, \textbf{r}} c_{\sigma', \textbf{r}''+\textbf{u}}\right)\\
        &= -iJ \sum_{\mathbf{u}}\left(  c^{\dag}_{\sigma, \textbf{r}-\textbf{u}} c_{\sigma, \textbf{r}'} -  c^{\dag}_{\sigma, \textbf{r}} c_{\sigma, \textbf{r}'+\textbf{u}}\right)\\
    \end{split}
\end{equation}
For the potential we have
\begin{equation}
    \begin{split}
        &i \left[H_\mathrm{pot},c^{\dag}_{\sigma, \textbf{r}} c_{\sigma, \textbf{r}'}\right]\\
        &= -i\sum_{\textbf{r}'', \sigma'}(\mu - \phi_{\textbf{r}''})\left[ c^{\dag}_{\sigma', \textbf{r}''} c_{\sigma', \textbf{r}''},c^{\dag}_{\sigma, \textbf{r}} c_{\sigma, \textbf{r}'}\right] \\
        &= -i\sum_{\textbf{r}'', \sigma'}(\mu - \phi_{\textbf{r}''}) \delta_{\sigma,\sigma'}\left( \delta_{\textbf{r}'',\textbf{r}} c^{\dag}_{\sigma', \textbf{r}''} c_{\sigma, \textbf{r}'} - \delta_{\textbf{r}'',\textbf{r}'} c^{\dag}_{\sigma, \textbf{r}} c_{\sigma', \textbf{r}''}\right) \\
        &= -i\left((\mu - \phi_{\textbf{r}}) - \left(\mu - \phi_{\textbf{r}'}\right)\right) c^{\dag}_{\sigma, \textbf{r}} c_{\sigma, \textbf{r}'}\\
        &= i\left(\phi_{\textbf{r}}- \phi_{\textbf{r}'}\right) c^{\dag}_{\sigma, \textbf{r}} c_{\sigma, \textbf{r}'}
    \end{split}
\end{equation}
Finally, for the interacting part
\begin{equation}
    \begin{split}
        &i \left[H_\mathrm{int},c^{\dag}_{\sigma, \textbf{r}} c_{\sigma, \textbf{r}'}\right]\\
        &= iU\sum_{\textbf{r}''} \left[c^{\dag}_{\sigma, \textbf{r}''} c_{\sigma, \textbf{r}''} c^{\dag}_{\bar{\sigma}, \textbf{r}''} c_{\bar{\sigma}, \textbf{r}''},c^{\dag}_{\sigma, \textbf{r}} c_{\sigma, \textbf{r}'} \right]\\
        &= iU\sum_{\textbf{r}''} c^{\dag}_{\bar{\sigma}, \textbf{r}''} c_{\bar{\sigma}, \textbf{r}''} \left( \delta_{\textbf{r}'',\textbf{r}} c^{\dag}_{\sigma, \textbf{r}''} c_{\sigma, \textbf{r}'} - \delta_{\textbf{r}'',\textbf{r}'} c^{\dag}_{\sigma, \textbf{r}} c_{\sigma, \textbf{r}''}\right) \\
        &= iU (c^{\dag}_{\bar{\sigma}, \textbf{r}} c_{\bar{\sigma}, \textbf{r}} - c^{\dag}_{\bar{\sigma}, \textbf{r}'} c_{\bar{\sigma}, \textbf{r}'})  c^{\dag}_{\sigma, \textbf{r}} c_{\sigma, \textbf{r}'}
    \end{split}
\end{equation}
Combining everything together we have
\begin{equation}
\begin{aligned}
    \partial_t (c^{\dag}_{\sigma, \textbf{r}} c_{\sigma, \textbf{r}'}) ={} & i \biggl(-J \sum_{\mathbf{u} } \left( c^\dag_{\sigma,\textbf{r}-\textbf{u}} c_{\sigma \textbf{r}'}-c^\dag_{\sigma,\textbf{r}} c_{\sigma \textbf{r}'+\textbf{u}}\right) \\
    &+ \left(\phi_\textbf{r} - \phi_{\textbf{r}'}\right)c^\dag_{\sigma,\textbf{r}}c_{\sigma,\textbf{r}'}\\ &
    + U \left(c^\dag_{\bar{\sigma},\textbf{r}}c_{\bar{\sigma},\textbf{r}}-c^\dag_{\bar{\sigma},\textbf{r}'}c_{\bar{\sigma},\textbf{r}'}\right)c^\dag_{\sigma,\textbf{r}}c_{\sigma,\textbf{r}'} \biggr)
\end{aligned}
\end{equation}
We now take the expectation value of the bilinear, $X_{\sigma,\textbf{r},\textbf{r}'}=\left\langle c^{\dag}_{\sigma, \textbf{r}} c_{\sigma, \textbf{r}'}\right\rangle$
\begin{equation}
    \begin{split}
        \partial_{t} X_{\sigma, \textbf{r},\textbf{r}'} \ =\ i \Bigg(&-J \sum_{\mathbf{u}} ( X_{\sigma, \mathbf{r}-\mathbf{u}, \mathbf{r}^{\prime}}-X_{\sigma, \mathbf{r}, \mathbf{r}^{\prime}+\mathbf{u}} )\\
        &+\left(\phi_\textbf{r} - \phi_{\textbf{r}'}\right) X_{\sigma, \textbf{r},\textbf{r}'}\\
        &+U \langle( c^{\dag}_{\bar{\sigma}, \mathbf{r}} c_{\bar{\sigma}, \mathbf{r}}-c_{\bar{\sigma}, \mathbf{r}^{\prime}}^{\dag} c_{\bar{\sigma}, \mathbf{r}^{\prime}} ) c_{\sigma, \mathbf{r}}^{\dag} c_{\sigma, \mathbf{r}^{\prime}} \rangle\Bigg) 
    \end{split}
\end{equation}
The 4-operator average can be decomposed into connected and disconnected components
\begin{equation}
    \begin{split}
        &\left\langle\left( c^{\dag}_{\bar{\sigma}, \mathbf{r}} c_{\bar{\sigma}, \mathbf{r}}-c_{\bar{\sigma}, \mathbf{r}^{\prime}}^{\dag} c_{\bar{\sigma}, \mathbf{r}^{\prime}} \right) c_{\sigma, \mathbf{r}}^{\dag} c_{\sigma, \mathbf{r}^{\prime}} \right\rangle \\ &= \left( \left\langle c^{\dag}_{\bar{\sigma}, \mathbf{r}} c_{\bar{\sigma}, \mathbf{r}} \right\rangle -\left\langle c_{\bar{\sigma}, \mathbf{r}^{\prime}}^{\dag} c_{\bar{\sigma}, \mathbf{r}^{\prime}} \right\rangle \right) \left\langle c_{\sigma, \mathbf{r}}^{\dag} c_{\sigma, \mathbf{r}^{\prime}} \right\rangle + Y_{\sigma, \textbf{r},\textbf{r}'}
    \end{split}
\end{equation}
where we have introduced
\begin{equation}
    Y_{\sigma, \bf {r,r}'}=\left\langle c^\dag_{\bar{\sigma},\textbf{r}}c_{\bar{\sigma},\textbf{r}}c^\dag_{\sigma,\textbf{r}}c_{\sigma,\textbf{r}'}\right\rangle^\mathrm{conn} - \left\langle c^\dag_{\bar{\sigma},\textbf{r}'}c_{\bar{\sigma},\textbf{r}'}c^\dag_{\sigma,\textbf{r}}c_{\sigma,\textbf{r}'}\right\rangle^\mathrm{conn}
    \label{eq:Y_app}
\end{equation}
    By using $X_{\sigma,\textbf{r},\textbf{r}'}=X_{\bar{\sigma},\textbf{r},\textbf{r}'}$ and substituting $\textbf{u}\rightarrow -\textbf{u}$ in the first term we get
\begin{equation}
    \begin{aligned}
        \partial_{t} X_{\sigma, \mathbf{r r}^{\prime}}  = i \bigg(&-J \sum_{\mathbf{u}} ( X_{\sigma, \mathbf{r}+\mathbf{u}, \mathbf{r}^{\prime}}-X_{\sigma, \mathbf{r}, \mathbf{r}^{\prime}+\mathbf{u}} )\\
        &+\left(\phi_\textbf{r} - \phi_{\textbf{r}'}\right) X_{\sigma, \textbf{r},\textbf{r}'}\\
        &+U \left(\left(X_{\sigma,\textbf{r},\textbf{r}}-X_{\sigma,\textbf{r}',\textbf{r}'}\right)X_{\sigma,\textbf{r},\textbf{r}'}+Y_{\sigma,\textbf{r},\textbf{r}'}\right) \bigg) 
    \end{aligned}
\end{equation}
For the density-density operator we have

\begin{equation}
    \partial_{t} ( n_{\sigma, \bf r} n_{\sigma', \bf r'} )  = i \left[H, n_{\sigma, \bf r} n_{\sigma', \bf r'}\right]
\end{equation}
The operator $ n_{\sigma, \bf r} n_{\sigma', \bf r'} $ commutes with the density operator  $n_{\sigma', \bf r''}$, so the commutator $[H_\mathrm{pot} + H_\mathrm{int}, n_{\sigma, \bf r} n_{\sigma', \bf r'} ]$ vanishes and $\left[H, n_{\sigma, \bf r} n_{\sigma', \bf r'}\right]=\left[H_\mathrm{kin}, n_{\sigma, \bf r} n_{\sigma', \bf r'}\right]$
\begin{equation}
    \begin{split}
        &\partial_{t} ( n_{\sigma, \bf r} n_{\sigma', \bf r'} ) \\
        &= i \left[H_\mathrm{kin},n_{\sigma, \bf r} n_{\sigma', \bf r'}\right] \\
        &= -iJ \sum_{\textbf{r}'',\mathbf{u}, \sigma''}\left[ c^{\dag}_{\sigma'', \textbf{r}''} c_{\sigma'', \textbf{r}''+\textbf{u}},n_{\sigma, \bf r} n_{\sigma', \bf r'}\right] \\
        &= -iJ \sum_{\textbf{r}'',\mathbf{u}, \sigma''} \biggl( n_{\sigma,\bf r} \left[ c^{\dag}_{\sigma'', \textbf{r}''} c_{\sigma'', \textbf{r}''+\textbf{u}}, n_{\sigma', \bf r'}\right] \\
        &\qquad \ \ \qquad \qquad + \left[ c^{\dag}_{\sigma'', \textbf{r}''} c_{\sigma'', \textbf{r}''+\textbf{u}}, n_{\sigma, \bf r}\right] n_{\sigma', \bf r'} \biggr)
    \end{split}
\end{equation}
Following the same procedure as for the bilinear, we get
\begin{equation}
\begin{split}
    &\partial_{t} ( n_{\sigma, \bf r} n_{\sigma', \bf r'} )  \\
    & =-i J \sum_{\textbf{u}} \biggl\{ \left( c^\dag_{\sigma,\textbf{r}-\textbf{u}} c _{\sigma,\textbf{r}} - c^\dag_{\sigma,\textbf{r}} c _{\sigma,\textbf{r}+\textbf{u}}\right)n_{\sigma',\textbf{r}'} \\
    & \qquad \qquad \quad \ +n_{\sigma,\textbf{r}} \left( c^\dag_{\sigma',\textbf{r}'-\textbf{u}} c _{\sigma',\textbf{r}'} - c^\dag_{\sigma',\textbf{r}'} c _{\sigma',\textbf{r}'+\textbf{u}}\right) \biggr\}
\end{split}
\end{equation}
By substituting $\textbf{u}\rightarrow-\textbf{u}$ in the first terms in brackets we arrive
\begin{equation}
\begin{split}
    &\partial_{t} ( n_{\sigma, \bf r} n_{\sigma', \bf r'} )  \\
    & =-i J \sum_{\textbf{u}} \biggl\{  \left( c^\dag_{\sigma,\textbf{r}+\textbf{u}} c _{\sigma,\textbf{r}} - c^\dag_{\sigma,\textbf{r}} c _{\sigma,\textbf{r}+\textbf{u}}\right)n_{\sigma',\textbf{r}'} \\
    & \qquad \qquad \quad \ + n_{\sigma,\textbf{r}} \left( c^\dag_{\sigma',\textbf{r}'+\textbf{u}} c _{\sigma',\textbf{r}'} - c^\dag_{\sigma',\textbf{r}'} c _{\sigma',\textbf{r}'+\textbf{u}}\right)\biggr\}
\end{split}
\end{equation}
We now take the expectation value, and decompose the averages into connected and disconnected components
\begin{equation} \label{eq:nn_therm_ev}
\begin{aligned}
&\partial_{t}  \left\langle n_{\sigma, \bf r} n_{\sigma', \bf r'} \right\rangle  =-i J \sum_{\textbf{u}} \\
& \times \Bigl\{  \left.  \left(\left\langle c^\dag_{\sigma,\textbf{r}+\textbf{u}} c _{\sigma,\textbf{r}} \right\rangle-\left\langle c^\dag_{\sigma,\textbf{r}} c _{\sigma,\textbf{r}+\textbf{u}}\right\rangle\right) \left\langle n_{\sigma',\textbf{r}'} \right\rangle+ \right. \\&
\left. \left\langle n_{\sigma,\textbf{r}}\right\rangle \left( \left\langle c^\dag_{\sigma',\textbf{r}'+\textbf{u}} c _{\sigma',\textbf{r}'} \right\rangle- \left\langle c^\dag_{\sigma',\textbf{r}'} c _{\sigma',\textbf{r}'+\textbf{u}}\right\rangle \right) \right.+ \\&
\left. \delta_{\sigma,\sigma'}  \left\langle c^\dag_{\sigma,\textbf{r}} c _{\sigma',\textbf{r}'} \right\rangle \left( \left\langle c^\dag_{\sigma',\textbf{r}'} c _{\sigma,\textbf{r}+\textbf{u}}\right\rangle -\left\langle c^\dag_{\sigma',\textbf{r}'+\textbf{u}} c _{\sigma,\textbf{r}} \right\rangle  \right)\right.+\\&
\left.\delta_{\sigma,\sigma'}\left\langle c^\dag_{\sigma',\textbf{r}'} c _{\sigma,\textbf{r}} \right\rangle \left( \left\langle c^\dag_{\sigma,\textbf{r}} c _{\sigma',\textbf{r}'+\textbf{u}}\right\rangle-\left\langle c^\dag_{\sigma,\textbf{r}+\textbf{u}} c _{\sigma',\textbf{r}'} \right\rangle  \right) \right.+\\&
\left.\delta_{\sigma,\sigma'} \delta_{\textbf{r},\textbf{r}'} \left(\left\langle c^\dag_{\sigma,\textbf{r}+\textbf{u}} c _{\sigma',\textbf{r}'} \right\rangle- \left\langle c^\dag_{\sigma,\textbf{r}} c _{\sigma',\textbf{r}'+\textbf{u}}\right\rangle\right) \right.+\\&
\left. \delta_{\sigma,\sigma'} \left( \delta_{\textbf{r},\textbf{r}'+\textbf{u}}-\delta_{\textbf{r}+\textbf{u},\textbf{r}'} \right)\left\langle c^\dag_{\sigma,\textbf{r}} c _{\sigma',\textbf{r}'} \right\rangle \right.\Bigl\} -iJ W_{\sigma,\sigma',\textbf{r},\textbf{r}'}
\end{aligned}
\end{equation} 
Where
\begin{equation}
    \begin{split}
        W_{\sigma,\sigma',\textbf{r},\textbf{r}'}=&\Biggl\langle \sum_{\textbf{u}} \biggl\{  \left( c^\dag_{\sigma,\textbf{r}+\textbf{u}} c _{\sigma,\textbf{r}} - c^\dag_{\sigma,\textbf{r}} c _{\sigma,\textbf{r}+\textbf{u}}\right)n_{\sigma',\textbf{r}'} \\
        &+  n_{\sigma,\textbf{r}} \left( c^\dag_{\sigma',\textbf{r}'+\textbf{u}} c _{\sigma',\textbf{r}'} - c^\dag_{\sigma',\textbf{r}'} c _{\sigma',\textbf{r}'+\textbf{u}}\right)\biggr\} \Biggr\rangle^{\mathrm{conn}}
    \end{split}
    \label{eq:W_app}
\end{equation}
Assuming spin symmetry in our system, we can rewrite Eq. \ref{eq:nn_therm_ev} using $X_{\sigma,\bf r,\bf r'}=X_{\sigma',\bf r,\bf r'}$
\begin{equation}
\begin{aligned}
&\partial_{t}  \left\langle n_{\sigma, \bf r} n_{\sigma', \bf r'} \right\rangle \\&
=-i J \sum_{\textbf{u}} \biggl\{ \left(X_{\sigma,\bf r + u,r'}-X_{\sigma,\bf r,r+u}\right)X_{\sigma,\bf r',r'}\\&
+ X_{\sigma,\bf r,r}(X_{\sigma,\bf r'+u,r'}-X_{\sigma,\bf r',r'+u}) \\&
+\delta_{\sigma,\sigma'} X_{\sigma,\bf r,r'}(X_{\sigma,\bf r',r+u}-X_{\sigma,\bf r'+u,r}) \\&+\delta_{\sigma,\sigma'} X_{\sigma,\bf r',r}(X_{\sigma,\bf r,r'+u}-X_{\sigma,\bf r+u,r'}) \\&
+ \delta_{\sigma,\sigma'} \delta_{\bf r,r'}(X_{\sigma,\bf r+u,r'}-X_{\sigma,\bf r,r'+u}) \\&+ \delta_{\sigma,\sigma'}(\delta_{\bf r,r'+u}-\delta_{\bf r+u,r'})X_{\sigma,\bf r,r'}\biggr\} - iJ W_{\sigma,\sigma',\textbf{r},\textbf{r}'}
\end{aligned}
\end{equation}
The double occupancy $d_{\bf r}=  \left\langle n_{\uparrow, \bf r} n_{\downarrow, \bf r} \right\rangle$ is a special case of the above expression,
\begin{equation}
\partial_t d_\textbf{r}= -2iJ \sum_{\textbf{u}}\ \left(X_{\sigma,\textbf{r}+\textbf{u},\textbf{r}}-X_{\sigma,\textbf{r},\textbf{r}+\textbf{u}}\right)X_{\sigma,\textbf{r},\textbf{r}} -iJ W_{\uparrow,\downarrow,\textbf{r},\textbf{r}}
\end{equation}
The second order term $\partial_t^2 X_{\sigma,\textbf{r}\textbf{r}'}(t_i)$ reads
\begin{equation}
\begin{aligned}
\partial_t^2 X_{\sigma,\textbf{r}\textbf{r}'}(t_i) = & i \biggl(-J \sum_{\mathbf{u}} ( \partial_{t} X_{\sigma, \mathbf{r}-\mathbf{u}, \mathbf{r}^{\prime}}-\partial_{t} X_{\sigma, \mathbf{r}, \mathbf{r}^{\prime}+\mathbf{u}} )\\&
+\left(\phi_\textbf{r} - \phi_{\textbf{r}'}\right) \partial_{t} X_{\sigma, \mathbf{r r}^{\prime}}\\&
+\left(\partial_{t}\phi_\textbf{r} - \partial_{t}\phi_{\textbf{r}'}\right)  X_{\sigma, \mathbf{r r}^{\prime}}\\&
+ U \bigl(\left(\partial_{t} X_{\bar{\sigma},\textbf{r,r}}-\partial_{t} X_{\bar{\sigma},\textbf{r}'\textbf{r}'}\right)X_{\sigma,\textbf{r}\textbf{r}'}\\&
+ \left(X_{\bar{\sigma},\textbf{r},\textbf{r}}-X_{\bar{\sigma},\textbf{r}',\textbf{r}'}\right)\partial_t X_{\sigma,\textbf{r},\textbf{r}'}+\partial_{t} Y_{\sigma,\textbf{r},\textbf{r}'}\bigr) \biggr).
\end{aligned}
\end{equation}
Analogously,
\begin{equation}
\begin{aligned}
&\partial_{t}^2  \left\langle n_{\sigma, \bf r} n_{\sigma', \bf r'} \right\rangle \\&
=-i J \sum_{\textbf{u}} \biggl\{ \left(\partial_{t} X_{\sigma,\bf r + u,r'}-\partial_{t} X_{\sigma,\bf r,r+u}\right)X_{\sigma,\bf r',r'}\\&
+ \left(X_{\sigma,\bf r + u,r'}-X_{\sigma,\bf r,r+u}\right)\partial_{t} X_{\sigma,\bf r',r'}\\&
+ \partial_{t} X_{\sigma,\bf r,r}(X_{\sigma,\bf r'+u,r'}-X_{\sigma,\bf r',r'+u}) \\&
+ X_{\sigma,\bf r,r}(\partial_{t} X_{\sigma,\bf r'+u,r'}-\partial_{t} X_{\sigma,\bf r',r'+u}) \\&
+\delta_{\sigma,\sigma'} \partial_{t} X_{\sigma,\bf r,r'}(X_{\sigma,\bf r',r+u}-X_{\sigma,\bf r'+u,r}) \\&
+\delta_{\sigma,\sigma'} X_{\sigma,\bf r,r'}(\partial_{t} X_{\sigma,\bf r',r+u}-\partial_{t} X_{\sigma,\bf r'+u,r}) \\&
+\delta_{\sigma,\sigma'} \partial_{t} X_{\sigma,\bf r',r}(X_{\sigma,\bf r,r'+u}-X_{\sigma,\bf r+u,r'}) \\&
+\delta_{\sigma,\sigma'} X_{\sigma,\bf r',r}(\partial_{t} X_{\sigma,\bf r,r'+u}-\partial_{t} X_{\sigma,\bf r+u,r'}) \\&
+ \delta_{\sigma,\sigma'} \delta_{\bf r,r'}(\partial_{t} X_{\sigma,\bf r+u,r'}-\partial_{t} X_{\sigma,\bf r,r'+u}) \\&
+ \delta_{\sigma,\sigma'}(\delta_{\bf r,r'+u}-\delta_{\bf r+u,r'})\partial_{t} X_{\sigma,\bf r,r'}\biggr\} - iJ \partial_{t} W_{\sigma,\sigma',\textbf{r},\textbf{r}'}.
\end{aligned}
\end{equation}
The time-derivatives $\partial_{t}Y_{\sigma,\textbf{r,r}'}$ and $\partial_{t} W_{\sigma,\sigma',\textbf{r},\textbf{r}'}$ are approximated as 
\begin{eqnarray}
    \partial_{t}Y_{\sigma,\textbf{r,r}'}(t_i) =& \frac{Y_{\sigma,\textbf{r,r}'}(t_i)-Y_{\sigma,\textbf{r,r}'}(t_{i-1})}{t_i-t_{i-1}}\\
    \partial_{t}W_{\sigma,\sigma',\textbf{r},\textbf{r}'}(t_i) =& \frac{W_{\sigma,\sigma',\textbf{r},\textbf{r}'}(t_i)-W_{\sigma,\sigma',\textbf{r},\textbf{r}'}(t_{i-1})}{t_i-t_{i-1}}.
\end{eqnarray}
By taking the complex conjugate of Eq. \ref{eq:Y_app}, we get
\begin{equation}
    \begin{split}
        Y^*_{\sigma, \bf {r,r}'}&=\left\langle c^\dag_{\bar{\sigma},\textbf{r}}c_{\bar{\sigma},\textbf{r}}c^\dag_{\sigma,\textbf{r}'}c_{\sigma,\textbf{r}}\right\rangle^\mathrm{conn} - \left\langle c^\dag_{\bar{\sigma},\textbf{r}'}c_{\bar{\sigma},\textbf{r}'}c^\dag_{\sigma,\textbf{r}'}c_{\sigma,\textbf{r}}\right\rangle^\mathrm{conn}   \\
        &=-Y_{\bar{\sigma}, \textbf{r}',\textbf{r}}
    \end{split}
\end{equation}
Analogously for Eq. \ref{eq:W_app}
\begin{equation}
    \begin{split}
        W^*_{\sigma,\sigma',\textbf{r},\textbf{r}'}&=-\Biggl\langle \sum_{\textbf{u}} \biggl\{  n_{\sigma',\textbf{r}'}\left( c^\dag_{\sigma,\textbf{r}+\textbf{u}} c _{\sigma,\textbf{r}} - c^\dag_{\sigma,\textbf{r}} c _{\sigma,\textbf{r}+\textbf{u}}\right) \\
        &\qquad +  \left( c^\dag_{\sigma',\textbf{r}'+\textbf{u}} c _{\sigma',\textbf{r}'} - c^\dag_{\sigma',\textbf{r}'} c _{\sigma',\textbf{r}'+\textbf{u}}\right) n_{\sigma,\textbf{r}}\biggr\} \Biggr\rangle^{\mathrm{conn}}\\
        & = -W_{\sigma',\sigma,\textbf{r}',\textbf{r}}
    \end{split}
\end{equation}
Since at equilibrium $ Y_{\sigma, \textbf{r},\textbf{r}'} = Y_{\bar{\sigma}, \textbf{r}'\textbf{r}}$ and $W_{\sigma,\sigma',\textbf{r},\textbf{r}'} = W_{\sigma',\sigma,\textbf{r}',\textbf{r}}$ by symmetry, we conclude $ Y_{\sigma, \textbf{r},\textbf{r}'} (t=0)= 0$ and $W_{\sigma,\sigma',\textbf{r},\textbf{r}'} (t=0)=0$.

\section{Proof of energy and particle number conservation}
\label{sec:conservationNE}
When evaluating the lattice equations of motion with exact correlators $X$, $Y$ and $W$, the density and energy conversion is trivial since $\partial_t \langle \hat{N} \rangle = i\langle [H, \hat{N}] \rangle = 0$ and $\partial_t \langle H \rangle = i\langle [H, H] \rangle = 0$, i.e. the Hamitonian $H$ commutes with the total density operator $\hat{N}$ and itself.

On the other hand, to prove that OQCET conserves particle number $N= \sum_{\sigma,\bf r} X_{\sigma,\bf r,r'}$ and energy $E=\langle H\rangle$, we will have to show that $\partial_t N = 0$ and $\partial_t E = 0$ for any values of $Y_{\sigma, \textbf{r}\textbf{r}'},W_{\sigma,\sigma',\textbf{r},\textbf{r}'}$, since these connected correlators are approximated by their corresponding cluster expectation values.
  
Starting from equations Eq. \ref{eq:bilinear_eom} conservation of total particle number is easy to show
\begin{equation} \label{eq:cons_N}
\partial_t N = \sum_{\textbf{r}, \sigma} \partial_t X_{\sigma, \textbf{r},\textbf{r}} = -iJ\sum_{\textbf{r},\textbf{u},\sigma} \left(X_{\sigma,\textbf{r}-\textbf{u},\textbf{r}} - X_{\sigma,\textbf{r},\textbf{r}+\textbf{u}}\right)
\end{equation}
Making the substitution $\textbf{r} \rightarrow \textbf{r} - \textbf{u}$ in the second term we see that the terms cancel out and we have
\begin{equation}
\partial_t N=0
\end{equation}
To show conservation of energy we will prove
\begin{equation} \label{eq:dt_E}
    \begin{split}
        \partial_t E & = -J \sum_{\textbf{r},\textbf{u},\sigma} \partial_t X_{\sigma,\textbf{r},\textbf{r+u}} - \mu \sum_{\textbf{r},\sigma} \partial_t X_{\sigma, \textbf{r},\textbf{r}} + U \sum_r \partial_t d_\textbf{r}
        \\& = 0   
    \end{split}
\end{equation}
The second term is zero from the conservation of particle number. From Eq. \ref{eq:bilinear_eom} we get
\begin{equation}
\begin{aligned}
&\sum_{\textbf{r},\textbf{u},\sigma} \partial_t X_{\sigma, \textbf{r},\textbf{r}+\textbf{u}} \\
& = \sum_{\textbf{r},\textbf{u},\sigma}  i \Biggl( -J \sum_{\textbf{u}'} \left( X_{\sigma,\textbf{r}-\textbf{u}',\textbf{r}+\textbf{u}} - X_{\sigma,\textbf{r},\textbf{r}+\textbf{u}+\textbf{u}'}\right)\\
&\quad + U \left( X_{\bar{\sigma},\textbf{r},\textbf{r}} - X_{\bar{\sigma},\textbf{r}+\textbf{u},\textbf{r}+\textbf{u}} \right) X_{\sigma,\textbf{r},\textbf{r}+\textbf{u}} + Y_{\sigma,\textbf{r},\textbf{r}+\textbf{u}} \Biggr) 
\end{aligned}
\end{equation}
By substituting $\textbf{r}\rightarrow \textbf{r}-\textbf{u}'$ in the term $\sum_{\textbf{r},\textbf{u},\textbf{u}',\sigma}X_{\sigma,\textbf{r},\textbf{r}+\textbf{u}+\textbf{u}'}$ the hopping terms cancel out, leaving
\begin{equation} 
\begin{split}
    &\sum_{\textbf{r},\textbf{u},\sigma} \partial_t X_{\sigma, \textbf{r},\textbf{r}+\textbf{u}} \\&
= iU   \sum_{\textbf{r},\textbf{u},\sigma} \left(  \left( X_{\bar{\sigma},\textbf{r},\textbf{r}} - X_{\bar{\sigma},\textbf{r}+\textbf{u},\textbf{r}+\textbf{u}} \right) X_{\sigma,\textbf{r},\textbf{r}+\textbf{u}} + Y_{\sigma,\textbf{r},\textbf{r}+\textbf{u}}\right)
\end{split}
\end{equation}
A further substitution $\textbf{r}+\textbf{u}\rightarrow \textbf{r}$, followed by substituting $\bf u \rightarrow -u$ in the term $\sum_{\textbf{r},\textbf{u},\sigma}X_{\bar{\sigma},\textbf{r}+\textbf{u},\textbf{r}+\textbf{u}}  X_{\sigma,\textbf{r},\textbf{r}+\textbf{u}}$ gives us the form
\begin{equation} \label{eq:hopping_eom}
    \begin{split}
        &\sum_{\textbf{r},\textbf{u},\sigma} \partial_t X_{\sigma, \textbf{r},\textbf{r}+\textbf{u}}
        \\& = iU   \sum_{\textbf{r},\textbf{u},\sigma} \left(  \left( X_{\bar{\sigma},\textbf{r},\bf r+u} - X_{\bar{\sigma},\textbf{r}+\textbf{u},\textbf{r}} \right) X_{\sigma,\textbf{r},\textbf{r}} + Y_{\sigma,\textbf{r},\textbf{r}+\textbf{u}}\right)
    \end{split}
\end{equation}
Substituting Eq. \ref{eq:hopping_eom} and Eq. \ref{eq:d_eom} into Eq. \ref{eq:dt_E}
\begin{equation} \label{eq:dt_E_WY}
\partial_t E = -iJU\sum_{\bf r}\left(  W_{\textbf{r},\textbf{r},\uparrow,\downarrow} + \sum_{\sigma,\bf u} Y_{\sigma,\textbf{r},\textbf{r}+\textbf{u}} \right)
\end{equation}
From Eq. \ref{eq:W}, $W_{\uparrow,\downarrow,\textbf{r},\textbf{r}}$ is given as
\begin{equation}
    \begin{split}
        W_{\uparrow,\downarrow,\textbf{r},\textbf{r}}=\Biggl\langle \sum_{\textbf{u}} \biggl\{&  \left( c^\dag_{\uparrow,\textbf{r}+\textbf{u}} c _{\uparrow,\textbf{r}} - c^\dag_{\uparrow,\textbf{r}} c _{\uparrow,\textbf{r}+\textbf{u}}\right)n_{\downarrow,\textbf{r}} \\
        &+  n_{\uparrow,\textbf{r}} \left( c^\dag_{\downarrow,\textbf{r}+\textbf{u}} c _{\downarrow,\textbf{r}} - c^\dag_{\downarrow,\textbf{r}} c _{\downarrow,\textbf{r}+\textbf{u}}\right)\biggr\} \Biggr\rangle^{\mathrm{conn}}
    \end{split} 
\end{equation}
which can be written as
\begin{equation}
    W_{\uparrow,\downarrow,\textbf{r},\textbf{r}} = \sum_{\textbf{u},\sigma} \Biggl\langle \left( c^\dag_{\sigma,\textbf{r}+\textbf{u}} c _{\sigma,\textbf{r}} - c^\dag_{\sigma,\textbf{r}} c _{\sigma,\textbf{r}+\textbf{u}}\right)n_{\bar{\sigma},\textbf{r}}  \Biggr\rangle^{\mathrm{conn}}
\end{equation}
If we introduce
\begin{equation}
\tilde{W}_{\sigma,\bf r,u}=\left\langle \left( c^\dag_{\sigma,\textbf{r}+\textbf{u}} c _{\sigma,\textbf{r}} - c^\dag_{\sigma,\textbf{r}} c _{\sigma,\textbf{r}+\textbf{u}}\right)n_{\bar{\sigma},\bf r}\right\rangle^\mathrm{conn}
\label{eq:W_tild}
\end{equation}
the above can be written as 
\begin{equation}
    W_{\uparrow,\downarrow,\textbf{r},\textbf{r}}=\sum_{\sigma,\bf u} \tilde{W}_{\sigma,\bf r,u}
\end{equation}
We can then write Eq. \ref{eq:dt_E_WY} as
\begin{equation}
    \begin{split}
        \partial_t E &= -\frac{1}{2}iJU\sum_{\sigma, \bf r,u} \\
        & \times \left(  \tilde{W}_{\sigma,\bf r,u} + \tilde{W}_{\sigma,\bf r+u,-u} +  Y_{\sigma,\textbf{r},\textbf{r}+\textbf{u}} +Y_{\sigma,\bf r+u,r} \right)
    \end{split}
\end{equation}
where we have made the substitutions $\bf r \to r +u$ and $\bf u \to -u$ to obtain the second $\tilde{W}$ and $Y$ terms.
Applying Eq. \ref{eq:Y} and Eq. \ref{eq:W_tild} we write out the terms in brackets
\begin{equation}
    \begin{split}
        &\tilde{W}_{\sigma,\bf r,u} + \tilde{W}_{\sigma,\bf r+u,-u} +  Y_{\sigma,\textbf{r},\textbf{r}+\textbf{u}} +Y_{\sigma,\bf r+u,r} \\
        &=
        \left\langle \left( c^\dag_{\sigma,\textbf{r}+\textbf{u}} c _{\sigma,\textbf{r}} - c^\dag_{\sigma,\textbf{r}} c _{\sigma,\textbf{r}+\textbf{u}}\right)n_{\bar{\sigma},\bf r}\right\rangle^\mathrm{conn}  \\
        &\quad + \left\langle \left( c^\dag_{\sigma,\textbf{r}} c _{\sigma,\textbf{r}+\textbf{u}} - c^\dag_{\sigma,\textbf{r}+\textbf{u}} c _{\sigma,\textbf{r}}\right)n_{\bar{\sigma},\bf r +u}\right\rangle^\mathrm{conn}  \\
        &\quad +\left\langle c^\dag_{\bar{\sigma},\bf{r}}c_{\bar{\sigma},\bf{r}}c^\dag_{\sigma,\bf{r}}c_{\sigma,\bf r+u}\right\rangle^\mathrm{conn} \\
        &\quad - \left\langle c^\dag_{\bar{\sigma},\bf{r+u}}c_{\bar{\sigma},\bf{r+u}}c^\dag_{\sigma,\bf{r}}c_{\sigma,\bf r+u}\right\rangle^\mathrm{conn}\\
        &\quad + \left\langle c^\dag_{\bar{\sigma},\bf r+u}c_{\bar{\sigma},\bf r+u}c^\dag_{\sigma,\bf r+u}c_{\sigma,\bf{r}}\right\rangle^\mathrm{conn} \\
        &\quad - \left\langle c^\dag_{\bar{\sigma},\bf{r}}c_{\bar{\sigma},\bf{r}}c^\dag_{\sigma,\bf r+u}c_{\sigma,\bf{r}}\right\rangle^\mathrm{conn}
    \end{split}
\end{equation}

Single-particle operators with opposite spins commute, so all the terms in the above equation cancel out and
\begin{equation}
    \tilde{W}_{\sigma,\bf r,u} + \tilde{W}_{\sigma,\bf r+u,-u} +  Y_{\sigma,\textbf{r},\textbf{r}+\textbf{u}} +Y_{\sigma,\bf r+u,r} =  0
\end{equation}
Since $\tilde{W}_{\sigma,\bf r,u}$ and $Y_{\sigma,\textbf{r},\textbf{r}+\textbf{u}}$ lie on the same bond (in the same cluster), the identity is valid for all $\tilde{W}_{\sigma,\bf r,u}$, $Y_{\sigma,\textbf{r},\textbf{r}+\textbf{u}}$ calculated on clusters. This finally leads us to
\begin{equation}
    \partial_t E=0
\end{equation}
which we set out to prove.
\newpage

\begin{widetext}
\section{Cluster dynamics}
\label{sec:AppendixClust}
\par As previously introduced, cluster dynamics are governed by the Lindblad equation
\begin{equation}
    \frac{d\rho_\mathrm{clust}(t)}{dt}=-i[H_\mathrm{clust},\rho_\mathrm{clust}(t)] +\sum_{l} \Gamma_l(t)\left(  L_l\rho_\mathrm{clust}(t)L^\dag_l-\frac{1}{2}\left\{      L^\dag_l L_l,\rho_\mathrm{clust}(t)        \right\}\right)
    \label{eq:lindblad_eq_app}
\end{equation}
For brevity, we will use $H=H_\mathrm{clust}$ and $\rho=\rho_\mathrm{clust}$ in the remainder of the section.

The second derivative in Eq.~\ref{eq:lindblad_euler_order_2} $\frac{d^2 \rho}{dt^2}$ can be expanded as follows:
\begin{equation}
\begin{aligned}
\frac{d^2 \rho}{dt^2} = &- i \left[H, \frac{d \rho(t)}{dt}\right] + \sum_l \Gamma_l (t) \left( L_l \frac{d \rho(t)}{dt} L_l^{\dag}- \frac{1}{2} \left\{ \frac{d \rho(t)}{dt}, L_l^{\dag} L_l \right\} \right)\\ &+ \sum_l \frac{d \Gamma_l(t)}{dt} \left( L_l \rho(t) L_l^{\dag}- \frac{1}{2} \left\{ \rho(t), L_l^{\dag} L_l \right\} \right)
\end{aligned}
\label{eq:lindblad_2nd_deriv}
\end{equation}
Substituting Eq. \ref{eq:lindblad_eq_app} and Eq. \ref{eq:lindblad_2nd_deriv} into Eq. \ref{eq:lindblad_euler_order_2} we get
\begin{equation}
\begin{aligned}
    \rho(t_{i+1}) ={} & \rho(t_i) + \Delta t \cdot \left(- i \left[H, \rho(t_i)\right] + \sum_l \left( \Gamma_l (t_i) + \frac{1}{2} \Delta t \frac{d \Gamma_l}{dt}\Bigr|_{t=t_i}\right) \left( L_l \rho(t_i) L_l^{\dag}- \frac{1}{2} \left\{ \rho(t_i), L_l^{\dag} L_l \right\} \right)\right) \\
&+ \frac{1}{2} (\Delta t)^2 \left( - i \left[H, \frac{d \rho}{dt}\Bigr|_{t=t_i}\right] + \sum_l \Gamma_l (t_i) \left( L_l \frac{d \rho}{dt}\Bigr|_{t=t_i} L_l^{\dag}- \frac{1}{2} \left\{ \frac{d \rho}{dt}\Bigr|_{t=t_i}, L_l^{\dag}  L_l\right\} \right) \right)
\end{aligned}
\end{equation}
The time derivative with respect to $\Gamma_l$ can be removed by observing that $\Gamma_l (t) + \frac{1}{2} \Delta t \frac{d \Gamma_l(t)}{dt} \approx \Gamma_l \left(t+\frac{\Delta t}{2} \right)$, i.e. using the system-environment coupling evaluated at half the time step (reminiscent of the leapfrog integration method). Ignoring terms of order $(\Delta t)^3$ we arrive at
\begin{equation}
\begin{aligned}
    \rho(t_{i+1}) &= \rho(t_i) + \Delta t \cdot \left(- i \left[H, \rho(t_i)\right] + \sum_l  \Gamma_l \left(t_i+\frac{\Delta t}{2}\right) \left( L_l \rho(t_i) L_l^{\dag}- \frac{1}{2} \left\{ \rho(t_i), L_l ^{\dag} L_l \right\} \right)\right) \\
&+ \frac{1}{2} (\Delta t)^2 \left( - i \left[H, \frac{d \rho}{dt}\Bigr|_{t=t_i}\right] + \sum_l \Gamma_l \left(t_i+\frac{\Delta t}{2}\right) \left( L_l \frac{d \rho}{dt}\Bigr|_{t=t_i} L_l^{\dag}- \frac{1}{2} \left\{ \frac{d \rho}{dt}\Bigr|_{t=t_i}, L_l^{\dag} L_l \right\} \right) \right)
\end{aligned}
\label{eq:lindblad_euler_order_2_exp}
\end{equation}
The derivative $\frac{d \rho}{dt}$ in Eq.~\ref{eq:lindblad_euler_order_2_exp} is calculated as:
\begin{equation}
\begin{split}
    \frac{d \rho}{dt}\Bigr|_{t=t_i} \approx - i \left[H, \rho(t_i)\right] + \sum_l \Gamma_l \left(t_i+\frac{\Delta t}{2}\right) &\left( L_l \rho(t_i) L_l^{\dag}- \frac{1}{2} \left\{ \rho(t_i), L_l^{\dag} L_l \right\} \right) \\&
    - \sum_l \frac{1}{2} \Delta t \frac{d \Gamma}{dt} \Bigr|_{t=t_i} \left( L_l \rho(t_i) L_l^{\dag}- \frac{1}{2} \left\{ \rho(t_i), L_l^{\dag} L_l \right\} \right)
\end{split}
\label{eq:lindblad_1st_derivative}
\end{equation}
Where we have substituted $\Gamma_l (t+\frac{\Delta t}{2}) \approx \Gamma_l (t) + \frac{1}{2} \Delta t \frac{d \Gamma_l(t)}{dt}$ into Eq. \ref{eq:lindblad_eq_app}.  The $\sim\Delta t$ term on the right-hand side is discarded, as its contribution to Eq. \ref{eq:lindblad_euler_order_2_exp} would be of the order $(\Delta t)^3$.
Expanding each term in Eq. \ref{eq:lindblad_euler_order_2_exp} we get:
\begin{equation}
-i \left[ H,\frac{d \rho}{dt}\right]= - \left\{ HH,\rho\right\} + 2 H \rho H -i \sum_l \Gamma_l \left(\left[ H,L_l \rho L_l^\dag\right] - \frac{1}{2} \left[H,\left\{\rho,L_l^\dag L_l\right\}\right]\right)
\end{equation}
\begin{equation}
\sum_l \Gamma_l L_l \frac{d \rho}{dt} L_l^\dag = -i \sum_l \Gamma_l L_l \left[H,\rho \right]L_l^\dag + \sum_{l,k} \Gamma_l \Gamma_k \left( L_l L_k \rho L_k^\dag L_l^\dag - \frac{1}{2} L_l\left\{\rho,L_k^\dag L_k\right\}L_l^\dag\right)
\end{equation}
\begin{equation}
\begin{aligned}
-\frac{1}{2} \sum_l  \Gamma_l\left\{ \frac{d \rho}{dt}, L^\dag_l L_l \right\}= & \frac{i}{2}\sum_l  \Gamma_l \left\{\left[H,\rho\right],L^\dag_l L_l\right\} - \frac{1}{2}\sum_{l,k} \Gamma_l \Gamma_k \left\{L_k \rho L^\dag_k,L^\dag_l L_l\right\}  +  \frac{1}{4}\sum_{l,k} \Gamma_l \Gamma_k \left\{\left\{\rho,L^\dag_k L_k\right\},L^\dag_l L_l\right\}
\end{aligned}
\end{equation}
where $\rho=\rho(t)$ and $\Gamma_l=\Gamma_l(t+\frac{\Delta t}{2})$
We can rewrite Eq. \ref{eq:lindblad_euler_order_2_exp} in terms of sums over $\{\Gamma_l\}$ by introducing the following:
\begin{equation}
P = \Delta t \cdot \left(- i \left[H, \rho\right]\right) + \frac{1}{2} (\Delta t)^2 \left( - \left\{ HH,\rho\right\}  + 2 H \rho H \right)
\end{equation}
\begin{equation}
\begin{aligned}
Q_l =  \Delta t \bigg( L_l \rho L_l^{\dag}- \frac{1}{2}&\left\{ \rho, L_l ^{\dag} L_l \right\} \bigg) + \frac{1}{2} (\Delta t)^2 \biggl(-i\left[ H,L_l \rho L_l^\dag\right] + 
 \frac{i}{2} \left[H,\left\{\rho,L_l^\dag L_l\right\}\right] - i  L_l \left[H,\rho \right]L_l^\dag+\frac{i}{2} \Bigl\{[H,\rho],L^\dag_l L_l\Bigr\} \biggr) 
\end{aligned}
\end{equation}
\begin{equation}
\begin{split}
    R_{lk} = \frac{1}{2} (\Delta t)^2 \biggl( L_l L_k \rho L_k^\dag L_l^\dag - \frac{1}{2} \biggl(L_l\Bigl\{\rho,L_k^\dag L_k\Bigr\}L_l^\dag& + \Bigl\{L_k \rho L^\dag_k,L^\dag_l L_l\Bigr\} -\frac{1}{2} \Bigl\{\Bigl\{\rho,L^\dag_k L_k\Bigr\},L^\dag_l L_l\Bigr\} \biggr)
    \biggr)
\end{split}
\end{equation}
Now Eq. \ref{eq:lindblad_euler_order_2_exp} can be written more compactly as:
\begin{equation}
\begin{aligned}
\rho(t_{i+1})= \rho(t_i) + P(t_i) + \sum_l \Gamma_l Q_l(t_i) + \sum_{l,k} \Gamma_l \Gamma_k R_{lk}(t_i)
\end{aligned}
\label{eq:lindblad_2nd_final}
\end{equation}
At the beginning of each step we calculate $P$, $Q_l$ and $R_{lk}$, then we find the appropriate $\{\Gamma_l\}$ such that we match the desired set of expectation values.

\end{widetext}
\begin{equation} \label{eq:sc_cond}
    \langle \hat{A}^\mathrm{clust}_\lambda (t+\Delta t) \rangle = \mathrm{Tr} \left[\rho_{\mathrm{clust}}(t+\Delta t) \hat{A}^\mathrm{clust} \right].
\end{equation} 
Determining the appropriate $\{\Gamma_l\}$ such that Eq.~\ref{eq:sc_cond} is satisfied represents a root-finding problem. However, due to a discontinuity in the derivative of $\rho_\mathrm{clust}(t)$ with respect to $\Gamma_l$ at $\Gamma_l=0$, reframing the root-finding problem as a minimization problem results in better numerical stability.
We define the merit function as
\begin{equation} \label{eq:meritfunc}
r(t_i) = \sqrt{\frac{1}{N_A}\sum_\mu \left\lvert \left\langle \hat{A}_{\mu}\right\rangle_\mathrm{latt}(t_{i+1})-\left\langle \hat{A}_{\mu}\right\rangle_\mathrm{clust}(t_{i+1})\right\rvert^2}
\end{equation}
Where $N_A$ is the number of operators $\{\hat{A}_{\mu}\}$. The average $\left\langle \hat{A}_{\mu}(t_{i+1})\right\rangle_\mathrm{clust}$ is given by
\begin{equation}
\left\langle \hat{A}_{\mu}(t_{i+1}) \right\rangle_\mathrm{clust} = \mathrm{Tr}(\rho(t_{i+1}) \cdot \hat{A}_{\mu})
\end{equation}
The norm in Eq. \ref{eq:meritfunc} is defined as
\begin{equation}
    \begin{split}
        &\left\lvert \left\langle \hat{A}_{\mu}\right\rangle_\mathrm{latt}(t_{i+1})-\left\langle \hat{A}_{\mu}\right\rangle_\mathrm{clust}(t_{i+1})\right\rvert = \\&
        \frac{1}{N_k}\sum_{k} \left\lvert \left\langle A_{\mu,k}\right\rangle_\mathrm{latt}(t_{i+1})-\left\langle A_{\mu,k}\right\rangle_\mathrm{clust}(t_{i+1})\right\rvert
    \end{split}
\end{equation}
where $k$ denotes a sum over all cluster tensor indices.
We perform the minimization numerically using the Nelder-Mead algorithm implemented in the SciPy Python package. \cite{2020SciPy-NMeth, Nelder1965}

\section{Reduced density matrix scheme for the initial state} \label{sec:redrho}

\par Using a reduced density matrix (Eq.~\ref{eq:redrho}) for the initial state guarantees that the cluster expectation values of all operators will match the lattice expectation values at $t=0$. To ensure stationarity of the initial state with respect to the cluster Hamiltonian $H_\mathrm{clust}$, we introduce additional jump operators $\{\Tilde{L}_i\}$ and time-independent coupling $\{\Tilde{\Gamma}_i\}$ into Eq.~\ref{eq:lindblad_clust}. 

Assuming $\Gamma_l (t=0)=0$, by applying the stationarity condition $\partial \rho_{0,\mathrm{clust}}/\partial t =0$ ($\rho_{0,\mathrm{clust}} \equiv \rho_{\mathrm{clust}}(0)$) to Eq. \ref{eq:lindblad_clust} we get
\begin{equation}
    i [H_\mathrm{clust},\rho_{0,\mathrm{clust}}] =  \sum_{l} \Tilde{\Gamma}_l\left(  \Tilde{L}_l\rho_{0} \Tilde{L}^\dag_l-\frac{1}{2}\left\{      \Tilde{L}^\dag_l \Tilde{L}_l,\rho_{0}      \right\}\right)
    \label{eq:equilib_cond}
\end{equation}
\par It is not immediately clear how to choose $\Tilde{L}_i$ and $\Tilde{\Gamma}_i$ to satisfy this condition. If we now write the equations of motion for $\rho$ very close to $\rho_{0,\mathrm{clust}}$; $\rho_\mathrm{clust}(t)=\rho_{0,\mathrm{clust}}+\Delta \rho_\mathrm{clust}(t)$
\begin{widetext}  
\begin{equation}
    \begin{split}
        \frac{d\rho_\mathrm{clust}(t)}{dt}=-i[H_\mathrm{clust},\rho_\mathrm{clust}(t)] &+\sum_{l} \Tilde{\Gamma}_l\left(  \Tilde{L}_l(\rho_{0,\mathrm{clust}}+\Delta \rho_\mathrm{clust}(t)) \Tilde{L}^\dag_l-\frac{1}{2}\left\{      \Tilde{L}^\dag_l \Tilde{L}_l,(\rho_{0,\mathrm{clust}}+\Delta \rho_\mathrm{clust}(t))      \right\}\right)\\
        & +\sum_{l} \Gamma_l(t)\left(  L_l\rho_\mathrm{clust}(t)L^\dag_l-\frac{1}{2}\left\{      L^\dag_l L_l,\rho_\mathrm{clust}(t)        \right\}\right)
    \end{split}
\end{equation}
Assuming that $\Delta \rho_\mathrm{clust} \ll \rho_\mathrm{clust}$ for a weak perturbation, if we drop contributions from the $\Delta \rho_\mathrm{clust}$ term in the second term on the RHS, substituting Eq. \ref{eq:equilib_cond} we get
\begin{equation}
    \begin{split}
        \frac{d\rho_\mathrm{clust}(t)}{dt}&=-i[H_\mathrm{clust},\rho_\mathrm{clust}(t)] + i [H_\mathrm{clust},\rho_{0,\mathrm{clust}}] +\sum_{l} \Gamma_l(t)\left(  L_l\rho_\mathrm{clust}(t)L^\dag_l-\frac{1}{2}\left\{      L^\dag_l L_l,\rho_\mathrm{clust}(t)        \right\}\right)\\
        &= -i[H_\mathrm{clust},\rho_\mathrm{clust}(t)-\rho_{0,\mathrm{clust}}] +\sum_{l} \Gamma_l(t)\left(  L_l\rho_\mathrm{clust}(t)L^\dag_l-\frac{1}{2}\left\{      L^\dag_l L_l,\rho_\mathrm{clust}(t)        \right\}\right)
    \end{split}
\end{equation}

At $t=0$, this reduces to $\frac{d\rho_\mathrm{clust}(0)}{dt}=-i[H_\mathrm{clust},\rho_\mathrm{clust}(0)]=0$, satisfying the stationarity condition.

The derivation in Appendix~\ref{sec:AppendixClust} can be repeated with the modified Lindblad equation, and we arrive at the same form as in Eq. \ref{eq:lindblad_2nd_final}, but with modified coefficients $P$ and $Q_l$:
\begin{equation}
    P = \Delta t \cdot \left(- i \left[H, \rho-\rho_{0}\right]\right) + \frac{1}{2} (\Delta t)^2 \left( - \left\{ HH,\rho\right\}  + 2 H \rho H \right)
\end{equation}
\begin{equation}
    \begin{aligned}
    Q_l =  \Delta t \bigg( L_l \rho L_l^{\dag}- \frac{1}{2}&\left\{ \rho, L_l ^{\dag} L_l \right\} \bigg) + \frac{1}{2} (\Delta t)^2 \biggl(-i\left[ H,L_l \rho L_l^\dag\right] + 
    \\& \frac{i}{2} \left[H,\left\{\rho,L_l^\dag L_l\right\}\right] - i  L_l \left[H,\rho -\rho_{0}\right]L_l^\dag+\frac{i}{2} \Bigl\{[H,\rho],L^\dag_l L_l\Bigr\} \biggr) 
    \end{aligned}
\end{equation}
The coefficients $R_{lk}$ remain unchanged. This construction ensures that the stationarity condition is satisfied, while the reduced density matrix gives us appropriate expectation values.
\end{widetext}

\section{Jump operators for 2x2 clusters}
\label{sec:appendixLset}
\par
\begin{figure}[htbp!]
    \centering
    \includegraphics[width=3cm]{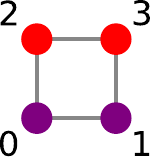}
    \caption{A 2x2 cluster in protocol B with labeled sites. The colors denote symmetry equivalent sites}
    \label{fig:2x2_clust}
\end{figure}

 For 2$\times$2 clusters we select jump operators tailored to the non-equilibrium protocol B, as this allows us significant savings on computation time. Since we have translation symmetry along one direction (e.g. the $x$ direction) the expectation value $X_{\sigma,\bf r,r+e_x}$ is purely real. This means that we need fewer jump operators (the current operator in the form $c_i + i c_j$ is unnecessary). With reasoning analogous to the 2x1 case, for 2x2 clusters with $X_{\sigma,i,j}$ constraints we choose
\begin{equation}
    \begin{split}
    \bigl\{ & c_{\sigma,0},\   c_{\sigma,1} , \  c_{\sigma,2},\   c_{\sigma,3} , \  
    c_{\sigma,0} + c_{\sigma,1},\  c_{\sigma,2} + c_{\sigma,3}, \\& c_{\sigma,1} + c_{\sigma,2}, \  c_{\sigma,0} + c_{\sigma,3}, \  c_{\sigma,1} + i c_{\sigma,2}, \  c_{\sigma,0} + i c_{\sigma,3}, \\& 
    c_{\sigma,0} + c_{\sigma,1} + c_{\sigma,2} + c_{\sigma,3}, \  c_{\sigma,0} + c_{\sigma,1} + i c_{\sigma,2} +i  c_{\sigma,3} \bigr\}_{\sigma\in \{\uparrow,\downarrow\}}
    \label{eq:2x2_X_Lset}
    \end{split}
\end{equation} 
with indices $\{0,1,2,3\}$ as labeled on Fig. \ref{fig:2x2_clust}.
Adding $d_{i}$ constraints requires adding the following jump operators:
\begin{equation}
    \begin{split}
    \bigl\{ & n_{\bar{\sigma},0} c_{\sigma,0},\   n_{\bar{\sigma},1} c_{\sigma,1} , \  n_{\bar{\sigma},2} c_{\sigma,2},\   n_{\bar{\sigma},3} c_{\sigma,3} , \ (1-n_{\bar{\sigma},0}) c_{\sigma,0},\\&
    (1-n_{\bar{\sigma},1}) c_{\sigma,1} , \  (1-n_{\bar{\sigma},2}) c_{\sigma,2},\   (1-n_{\bar{\sigma},3}) c_{\sigma,3} , \\&
    c_{\sigma,0} + c_{\sigma,1},\  c_{\sigma,2} + c_{\sigma,3}, \ c_{\sigma,1} + c_{\sigma,2}, \  c_{\sigma,0} + c_{\sigma,3}, \\&
    c_{\sigma,1} + i c_{\sigma,2}, \  c_{\sigma,0} + i c_{\sigma,3}, \ c_{\sigma,0} + c_{\sigma,1} + c_{\sigma,2} + c_{\sigma,3}, \\&
    c_{\sigma,0} + c_{\sigma,1} + i c_{\sigma,2} +i  c_{\sigma,3} \bigr\}_{\sigma\in \{\uparrow,\downarrow\}}
    \label{eq:2x2_d_Lset}
    \end{split}
\end{equation} 
This set works well at temperature $T=0.8$, but at lower temperatures we find that convergence is improved by using the following jump operators:
\begin{widetext}

\begin{equation}
    \begin{split}
    \bigl\{ & n_{\bar{\sigma},0} c_{\sigma,0},\   n_{\bar{\sigma},1} c_{\sigma,1} , \  n_{\bar{\sigma},2} c_{\sigma,2},\   n_{\bar{\sigma},3} c_{\sigma,3} , \ (1-n_{\bar{\sigma},0}) c_{\sigma,0},\\&
    (1-n_{\bar{\sigma},1}) c_{\sigma,1} , \  (1-n_{\bar{\sigma},2}) c_{\sigma,2},\   (1-n_{\bar{\sigma},3}) c_{\sigma,3} , \\&
    (c_{\sigma,0} + c_{\sigma,1}+ c_{\sigma,2} + c_{\sigma,3}) (1-n_{\bar{\sigma},0}) (1-n_{\bar{\sigma},1}) (1-n_{\bar{\sigma},2}) (1-n_{\bar{\sigma},3})
    , \\& c_{\sigma,0} + c_{\sigma,1},\  c_{\sigma,2} + c_{\sigma,3}, \ c_{\sigma,1} + c_{\sigma,2}, \  c_{\sigma,0} + c_{\sigma,3}, \\&
    c_{\sigma,1} + i c_{\sigma,2}, \  c_{\sigma,0} + i c_{\sigma,3}, \
    c_{\sigma,0} + c_{\sigma,1} + i c_{\sigma,2} +i  c_{\sigma,3} \bigr\}_{\sigma\in \{\uparrow,\downarrow\}}
    \end{split}
\end{equation} 
\begin{equation}
    \begin{split}
    \bigl\{ & n_{\bar{\sigma},0} c_{\sigma,0},\   n_{\bar{\sigma},1} c_{\sigma,1} , \  n_{\bar{\sigma},2} c_{\sigma,2},\   n_{\bar{\sigma},3} c_{\sigma,3} , \ (1-n_{\bar{\sigma},0}) c_{\sigma,0},\\&
    (1-n_{\bar{\sigma},1}) c_{\sigma,1} , \  (1-n_{\bar{\sigma},2}) c_{\sigma,2},\   (1-n_{\bar{\sigma},3}) c_{\sigma,3} , \\&
    (c_{\sigma,0} + c_{\sigma,1}+ c_{\sigma,2} + c_{\sigma,3}) (1-n_{\bar{\sigma},0}) (1-n_{\bar{\sigma},1}) (1-n_{\bar{\sigma},2}) (1-n_{\bar{\sigma},3})
    , \\& 
    (c_{\sigma,0} + c_{\sigma,1}) (1-n_{\bar{\sigma},0}) (1-n_{\bar{\sigma},1}), \  (c_{\sigma,2} + c_{\sigma,3}) (1-n_{\bar{\sigma},2}) (1-n_{\bar{\sigma},3}), \\&
    (c_{\sigma,1} + c_{\sigma,2}) (1-n_{\bar{\sigma},1}) (1-n_{\bar{\sigma},2}), \  (c_{\sigma,0} + c_{\sigma,3}) (1-n_{\bar{\sigma},0}) (1-n_{\bar{\sigma},3}), \\&
    c_{\sigma,1} + i c_{\sigma,2}, \  c_{\sigma,0} + i c_{\sigma,3}, \
    c_{\sigma,0} + c_{\sigma,1} + i c_{\sigma,2} +i  c_{\sigma,3} \bigr\}_{\sigma\in \{\uparrow,\downarrow\}}
    \end{split}
\end{equation} 
We use the above at temperatures $T=0.5$ and $T=0.3$ respectively.
\end{widetext} 

\section{Negative spectral weight and causality}
\label{sec:negweight}
\par
\begin{figure}[!ht]
    \centering
    \includegraphics[width=\linewidth]{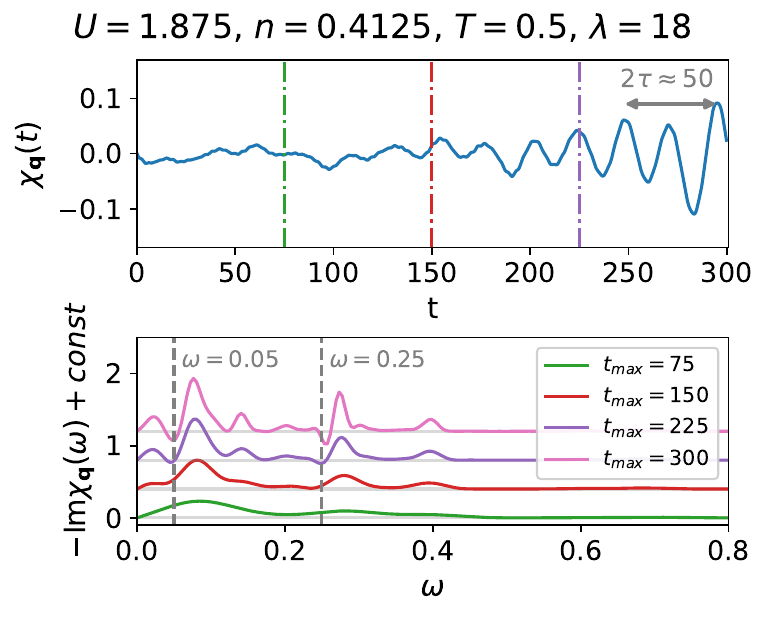}
    \caption{$-\mathrm{Im}\chi_{\bf q}(t)$ and  $-\mathrm{Im}\chi_{\bf q}(\omega)$ for different values of the Blackman window size $t_\mathrm{max}$. The wave of increasing amplitude in the time domain (of period $\tau\approx 25$) corresponds to a negative amplitude peak at $\omega = 2\pi/\tau \approx 0.25$. The negative weight peak at $\omega \approx 0.05$ has a very long period of $\tau \approx 125$ so it is not clearly visible in the time-domain graph above. For windows of size $t_\mathrm{max}=150$ and shorter both components with negative spectral weight are cut off and are not visible in the spectrum.}
    \label{fig:negweight}
\end{figure}

 In Figs. \ref{fig:imchiqw_q_full} and \ref{fig:protocolAvsB} we observe peaks with negative spectral weight, which corresponds to a component of the response increasing in intensity with time. This behaviour breaks causality and is most likely an artifact of our approximation, becoming more prominent at lower temperatures. Solutions of HEOM are also known to suffer from similar instabilities\cite{Jankovic2023}. However, at experimentally relevant times ($t \lesssim 50$), or equivalently, after broadening of the spectral curves, this effect is negligible (Fig.~\ref{fig:negweight}).

\section{Implementation benchmarks and time step convergence}
\label{sec:appendixbenchmark}

\begin{figure}[ht]
    \centering
    \includegraphics[width=\linewidth]{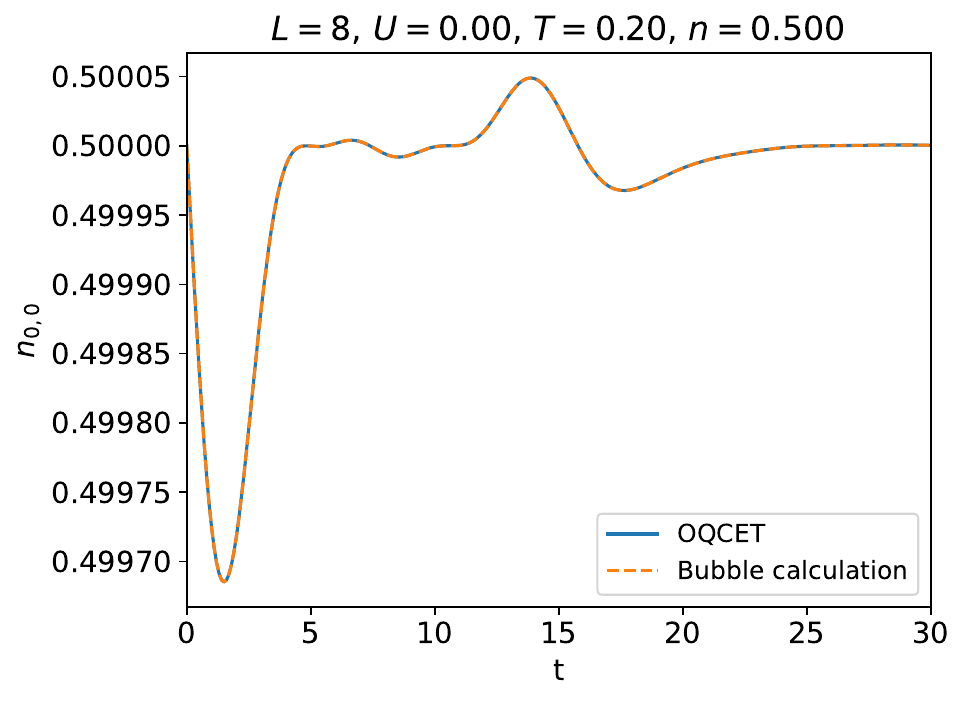}
    \caption{Comparison of density response between OQCET differential equations and bubble calculations.}
    \label{fig:bubble_comp}
\end{figure}
\begin{figure}[ht]
    \centering
    \includegraphics[width=\linewidth]{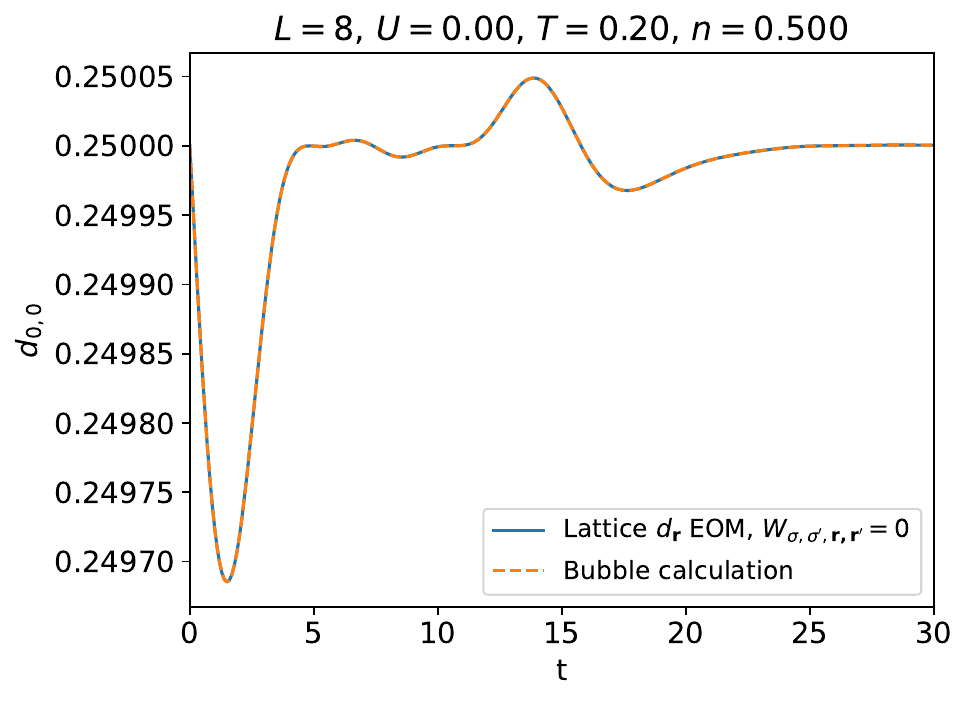}
    \caption{Comparison of double occupancy response between lattice differential equations and bubble calculations. The bubble double occupancy is given as the square of the density shown in Fig. \ref{fig:bubble_comp}.}
    \label{fig:bubble_comp_d}
\end{figure} 
\par We check the correctness of our implementation of OQCET by benchmarking the lattice differential equation and the Lindbladian cluster evolution against known results.
\par In the $U=0$ limit, the lattice differential equations should give exact results for the density. To verify this, we compare the results to bubble calculations which are in this case numerically exact (Fig. \ref{fig:bubble_comp}) \cite{Vucicevic2023}. This also serves to show that our approximation for the delta potential places us in the linear response regime.

\par OQCET does not give us exact $d_\textbf{r}$ in the $U=0$ limit, but if we evolve the double occupancy by equation Eq. \ref{eq:d_eom} with $W_{\sigma,\sigma',\bf r,r'}=0$ we expect to get exact results. For $U=0$, the double occupancy is simply the square of the density, $d_{\bf r}=\langle n_{\bf_r}\rangle ^2$, which we can calculate from the bubble (Fig. \ref{fig:bubble_comp_d}).

\begin{figure}[ht]
    \centering
    \includegraphics[width=0.9\linewidth]{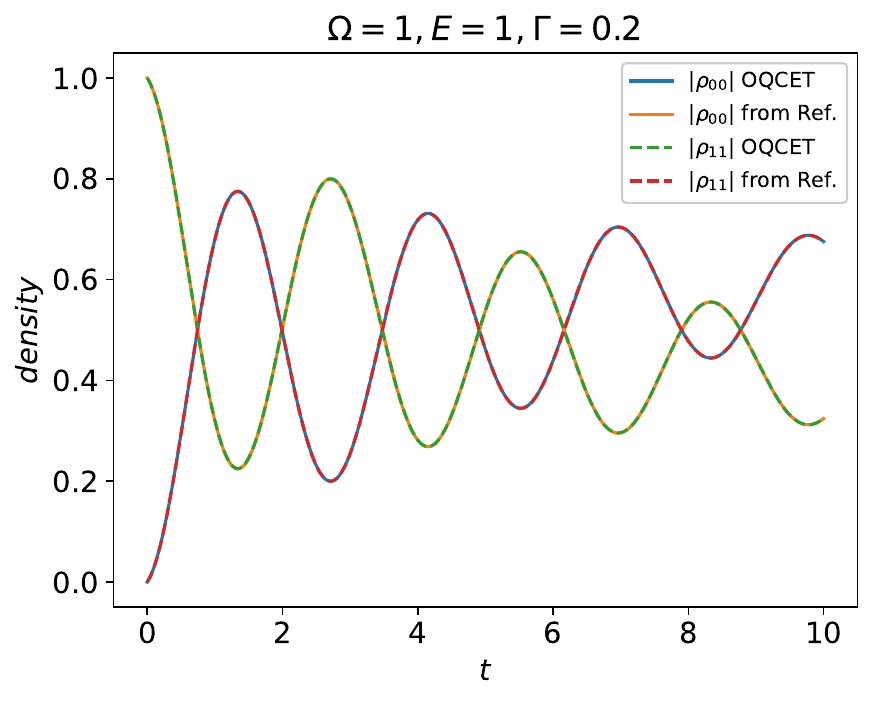}
    \caption{Comparison of density evolution of a two level system between OQCET and implementation provided in Ref.~\onlinecite{Manzano2020}.}
    \label{fig:lindblad_comp}
\end{figure}
\par To check that the equations for Lindbladian evolution (derived in Appendix \ref{sec:AppendixClust}) have been implemented correctly, we reproduce results for the evolution of a simple two-level system with decay shown in Ref.~\onlinecite{Manzano2020}. We compare the OQCET implementation of Eq. \ref{eq:lindblad_2nd_final} with code provided in Ref.~\onlinecite{Manzano2020} (Fig. \ref{fig:lindblad_comp}).
\par The Hamiltonian, initial state and jump operator of the system are given by
\begin{equation}
    H = \begin{bmatrix}
        0 & \Omega \\
        \Omega & E\\
        \end{bmatrix}, \quad
    \rho(t=0) = \begin{bmatrix}
        0 & 0  \\
        0 & 1\\
        \end{bmatrix}, \quad
    L = \sigma^- = \begin{bmatrix}
        0 & 1  \\
        0 & 0\\
        \end{bmatrix}
    \label{eq:twolevel}
\end{equation}

\begin{figure*}[ht]
    \centering
    \includegraphics[width=\linewidth]{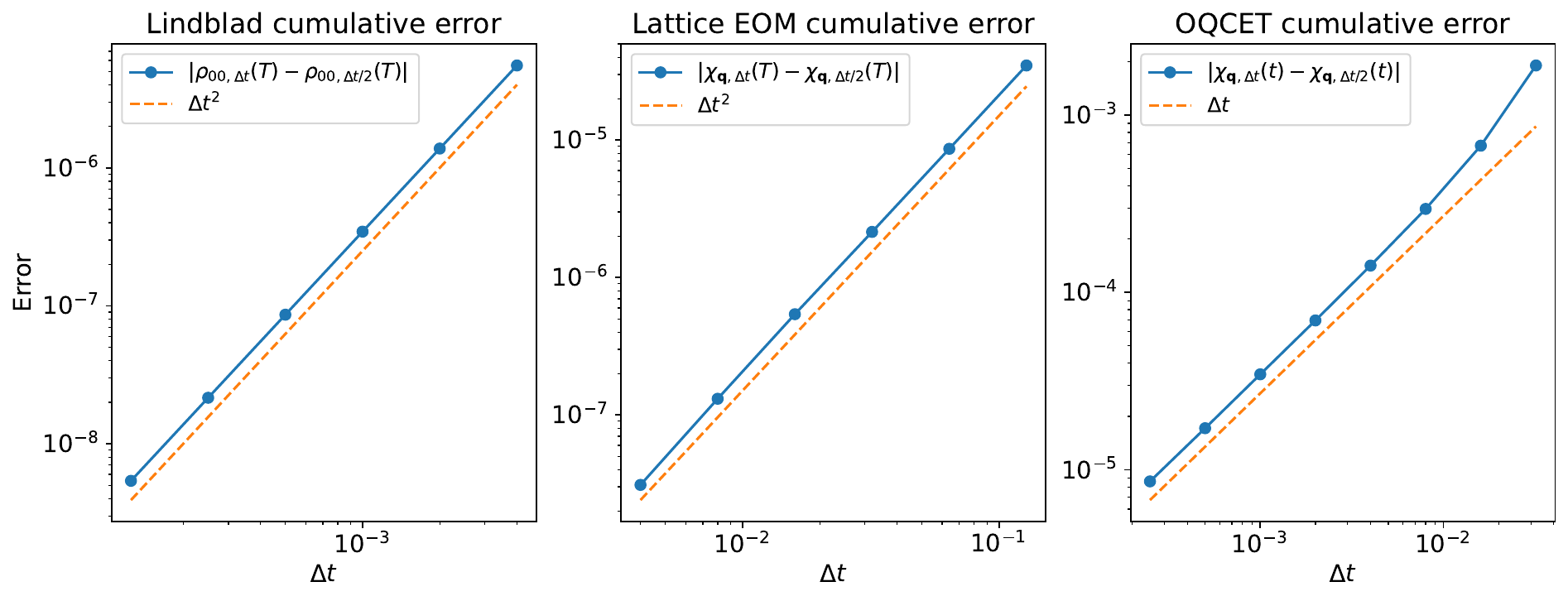}
    \caption{Comparison of cumulative error scaling for the Lindbladian evolution, the lattice EOM, and the full OQCET algorithm. Lindblad results were obtained for the two-level system described by Eq. \ref{eq:twolevel}, with parameters $\Omega=1$, $E=1$ and time-dependent $\Gamma(t)=2\cos(0.1 t)$. Both the Lindbladian evolution and the lattice EOM show the expected $O(\Delta t^2)$ scaling, while the full OQCET algorithm shows $O(\Delta t)$ scaling.}
    \label{fig:dt_err}
\end{figure*}

\par Finally, we check the convergence of our time evolution with respect to the time step $\Delta t$. For a second-order method, we expect the error to scale as $\Delta t^2$. To estimate how the cumulative error scales with $\Delta t$, we write the susceptibility at a fixed time $t$ and step size $\Delta t$ as $\chi_{\bf q}(t,\Delta t) = \chi^*_{\bf q}(t) + E(t,\Delta t)$, where $\chi^*_{\bf q}(t) = \lim_{\Delta t \to 0} \chi_{\bf q}(t,\Delta t)$. Then,
\begin{equation}
    \begin{split}
        \chi^*_{\bf q}(t)  &= \chi_{\bf q}(t,\Delta t) -E (t,\Delta_t) \\
        \chi_{\bf q}(t,\Delta t) -E (t,\Delta_t) &= \chi_{\bf q}(t,\Delta t/2) -E (t,\Delta_t/2) \\
        \chi_{\bf q}(t,\Delta t) -\chi_{\bf q}(t,\Delta t/2) &= E (t,\Delta_t) -E (t,\Delta_t/2) \\
    \end{split}
\end{equation}
If the error can be expressed as $E(t,\Delta t) \approx C(t) (\Delta t)^p$ for some constant $C(t)$, then
\begin{equation}
    \begin{split}
        \chi_{\bf q}(t,\Delta t) -\chi_{\bf q}(t,\Delta t/2) &= C(t) (\Delta t)^p - C(t) (\Delta t/2)^p \\
        &= C(t)  (1 - 2^{-p}) (\Delta t)^p
    \end{split}
\end{equation}
Thus, we can estimate the error from a log-log plot of $\chi_{\bf q}(t,\Delta t) -\chi_{\bf q}(t,\Delta t/2)$ vs $\Delta t$.
\par We have independently verified that our implementation of both the lattice differential equations and the Lindbladian cluster evolution separately give the expected $\Delta t^2$ scaling. However, the error of the full algorithm scales as $O(\Delta t)$ (Fig. \ref{fig:dt_err}), which we believe is due to additional error compounding from the optimization of $\Gamma_l(t)$ in the second step of the algorithm.

\bibliography{refs.bib}
\bibliographystyle{apsrev4-2}

\end{document}